# Photoanodes Based on TiO$_2$ and α-Fe$_2$O$_3$ for Solar Water Splitting – Superior Role of 1D Nanoarchitectures and of Combined Heterostructures

Stepan Kment,[1,]* Francesca Riboni,[2] Sarka Pausova,[3] Lei Wang,[2,#] Lingyun Wang,[1] Hyungkyu Han,[1] Zdenek Hubicka,[1] Josef Krysa,[3] Patrik Schmuki,[1,2,4]* Radek Zboril[1]*

[1]Regional Centre of Advanced Technologies and Materials, Šlechtitelů 27, 78371 Olomouc, Czech Republic

[2]Department of Materials Science and Engineering, University of Erlangen-Nuremberg, Martensstrasse 7, D-91058 Erlangen, Germany

[3]Department of Inorganic Technology, University of Chemistry and Technology Prague, Technicka 5, 166 28 Prague 6, Czech Republic

[4]Department of Chemistry, Faculty of Science, King Abdulaziz University, P.O. Box 80203, Jeddah 21569, Saudi Arabia

# current address: xxx

These authors contributed equally to this work

*Corresponding authors. E-mail: stepan.kment@upol.cz; schmuki@ww.uni-erlangen.de; radek.zboril@upol.cz

Link to the published article:

https://pubs.rsc.org/en/content/articlelanding/2017/cs/c6cs00015k#!divAbstract

**ABSTRACT**

Solar driven photoelectrochemical water splitting (PEC-WS) using semiconductor photoelectrodes represents a promising approach for a sustainable and environmentally friendly production of renewable energy vectors and fuel sources, such as dihydrogen ($H_2$). In this context, titanium dioxide ($TiO_2$) and iron oxide (hematite, $\alpha\text{-}Fe_2O_3$) are among the most investigated candidates as photoanode materials, mainly owing to their resistance to photocorrosion, non-toxicity, natural abundance, and low production cost. Major drawbacks are, however, an inherently low electrical conductivity and a limited hole diffusion length that significantly affect the performance of $TiO_2$ and $\alpha\text{-}Fe_2O_3$ in PEC devices. To this regard, one-dimensional (1D) nanostructuring is typically applied as it provides several superior features such as a significant enlargement of the material surface area, extended contact between the semiconductor and the electrolyte and, most remarkably, preferential electrical transport that overall suppress charge carrier recombination and improve $TiO_2$ and $\alpha\text{-}Fe_2O_3$ photo-electrocatalytic properties. The present review describes various synthetic methods, properties and PEC applications of 1D-photoanodes (nanotubes, nanorods, nanofibers, nanowires) based on titania, hematite, and on $\alpha\text{-}Fe_2O_3/TiO_2$ heterostructures. Various routes towards modification and enhancement of PEC activity of 1D photoanodes are also discussed including doping, decoration with co-catalysts and heterojunction engineering. Finally, the challenges related to the optimization of charge transfer kinetics in both oxides are highlighted.

# Table of contents

## 1. Introduction

The fulfilment of global energy demand still mainly relies on supply and use of fossil fuels such as coal, crude oil, natural gas, *etc*. The *energy content* of fossil fuels is typically released upon their combustion, which in turn generates significant emissions of $CO_2$ into the atmosphere and ultimately promotes global warming and climate change. In 2013 the global primary energy consumption amounted to *ca.* 17 TW (90% of which was generated from fossil fuel).

Taking into account the pace at which the world economy grows, a consequent drastic increase of the global energy consumption is predicted to take place in the near future.[1,2]

Therefore, the urgency to find secure, sustainable, clean and renewable energy sources appears of primary importance and various alternatives to current sources have been introduced with the attempt of limiting $CO_2$ emissions.[3,4]

Dihydrogen ($H_2$), in particular, is one of the most promising energy carriers and fuel sources due to its high energy-per-mass content, a wide range of available storage and transport approaches, and reduced harmful emissions. Indeed, the energy stored into its chemical bond can be used in fuel cells to produce clean electricity. In addition, $H_2$ represents the reactant for well-established industrial processes, such as petroleum refinement, ammonia synthesis, *etc*.

Although industrial mass scale production of dihydrogen is still based on fossil fuel combustion (*i.e.*, the most common method for extracting $H_2$ is still the steam-reforming of methane),[4] green alternatives that utilize only water ($H_2O$) as primary $H_2$ source are currently under development.

One of the most promising solutions is represented by the use of photoelectrochemical cells (PECs) for light-driven water splitting. In these devices, semiconductor materials are used as photoelectrodes that under illumination and an applied electric bias can split $H_2O$ into $H_2$ and $O_2$. Ideally, the goal of this technology is to convert sunlight into clean energy available on demand.

## 1.1 Photoelectrolysis of water – background and basic requirements

After the pioneering work in 1972 by Fujishima and Honda,[5] the photoelectrochemical water splitting (PEC-WS) reaction has been largely investigated as promising energy vector ($H_2$).

PEC-WS consists in the reduction of water into hydrogen (cathodic reaction) accompanied by the anodic reaction of oxygen generation.[5–7] In a conventional PEC device, the reduction and oxidation reactions occur at the surface of photoelectrodes (*i.e.*, the photocathode and the photoanode, respectively) that are immersed in an aqueous electrolyte. The charge carriers that correspondingly promote $H_2$ and $O_2$ evolution (*i.e.*, electrons and holes, respectively) are generated in the semiconductor(s) upon absorption of photons of adequate energy (hν ≥ band-gap, $E_g$).

Different configurations apply to PEC devices: in a so-called tandem cell configuration, both the electrodes are semiconductive materials, while in a Schottky type cell one of the two is a metal electrode. Regardless of the configuration, n-type semiconductors are used as photoanodes while p-type semiconductors as photocathode – this is related to the electronic properties of these materials and the type of majority carriers, *i.e.*, electrons ($e^-$) or holes ($h^+$) produced by photoexcitation.[8]

The focus of the present review is on n-type photoanodes – as the most important class of materials discussed in this review, namely α-$Fe_2O_3$ and $TiO_2$, are in their most abundant forms intrinsically n-type materials.

Overall, the PEC-WS reaction consists of four main sequential steps (Fig. 1.1(a)): (i) light absorption, (ii) separation of photogenerated charge carriers, (iii) transport of charges ($h^+$) towards the electrode surface (electrons to the back contact), and (iv) surface reactions ($h^+$ transfer). In most cases of available n-type semiconductors in contact with a common electrolyte, a space-charge depletion layer is formed at the semiconductor/liquid junction (SCLJ). Upon light illumination, electron-hole ($e^-/h^+$) pairs are generated (1) and separated due to the space-charge field generated by band-bending originating from the intrinsic work-function (red-ox potential) differences between the semiconductor and the electrolyte.[9]

As a consequence, holes in the valence band (VB) diffuse and migrate towards the SCLJ where the oxygen evolution reaction (OER) occurs (3). By contrast, electrons promoted into the semiconductor conduction band (CB) are driven away from the SCLJ, transported through the anode towards the back-contact, and are conveyed through the external circuit to a metallic cathode (*e.g.*, platinum), where the $H_2O$ reduction to $H_2$ (hydrogen evolution reaction, HER – (2)) takes place. Hence, the photoelectrolysis of water can be summarized as follows:

$$semiconductor \xrightarrow{h\nu} e_{cb}^- + h_{vb}^+ \qquad (1)$$

$$2e^- + 2H^+ \rightarrow H_2 \qquad E^0_{H^+/H_2} = 0\ V\ vs.\ RHE \qquad (2)$$

$$4h^+ + 2H_2O \rightarrow O_2 + 4H^+ \qquad E^0_{O_2/H_2O} = 1.23\ V\ vs.\ RHE \qquad (3)$$

$$2H_2O \rightarrow 2H_2 + O_2 \qquad \Delta E = 1.23\ V\ vs.\ RHE \qquad (4)$$

The overall reaction (4) is endoenergetic, with a Gibbs free energy $\Delta_r G^0$ = 273.2 kJ/mol (according to the Nernst equation, $\Delta E$ = 1.23 V). Therefore, the semiconductor used as photoelectrode must absorb light with photon energy greater than 1.23 eV (that is, the semiconductor should exhibit an absorption edge at $\lambda <$ *ca.* 1100 nm).

However, due to the intrinsic kinetic limitations of reactions (2) and (3) and the resulting overpotential needed for oxygen and hydrogen production, $\Delta G > \sim 2.0$ eV (corresponding to a limiting absorption edge $\lambda$ of *ca.* 610 nm) is frequently reported as the threshold energy required to achieve a water splitting reaction at a reasonable rate.[10,11]

Basic requirements for a photoanode material include low cost of production, high chemical stability, high carrier mobility, long carrier lifetime, and rapid interfacial charge transfer. Furthermore, it should feature a conduction band-edge ($E_{cb}$) more negative than the redox potential $E°$ ($H^+/H_2$) and a valence band-edge ($E_{vb}$) more positive than the redox potential $E°$ ($O_2/H_2O$), with the electronic nature of the band edges that also provides photostability to the material itself. Additionally, it is also desired that the semiconductor absorbs a large fraction of visible light in order to enable sunlight-driven water splitting.

Band edges and band gap energies of semiconductors typically explored for photoelectrochemical $H_2O$ splitting are reported in Fig. 1.1(b).

Despite large efforts and hundreds of semiconductors explored, the thermodynamic and kinetic requirements for an efficient PEC-WS reaction have not yet been satisfied by a single semiconductor candidate. For instance, Si, Ge or III – V compounds suffer from severe photocorrosion and instability in water solutions, while transition metal oxides rarely meet the criteria of (i) a band-gap width suitable for sunlight activation, or of (ii) adequate band-edges to spontaneously promote OER and HER.

Nevertheless, metal oxides still represent one of the most viable options for PEC water splitting application, particularly owing to their low processing cost and high stability against photocorrosion even in harsh environments.

The engineering of metal oxide semiconductor photoelectrodes that exhibit long–term stability and a high photoefficiency could be realized using different strategies (applied individually or in combination).

A first approach can be based on theoretical prediction and synthesis of new materials that could intrinsically satisfy the aforementioned criteria.[12]

A second approach, based on experimental attempts, relies on bulk and/or surface modifications of already known/widely investigated semiconductors (*e.g.*, $TiO_2$, α-$Fe_2O_3$,

WO$_3$, BiVO$_4$, *etc.*) in view of improving their properties and functionalities. This translates into strategies such as: (i) doping, that is by introducing other elements (metallic or non-metallic), into the semiconductor lattice, to modify electronic properties such as the donor density and thus the charge mobility, and/or to narrow the band gap of the material and enable visible light photoresponse (this often applies to TiO$_2$); (ii) anchoring a co-catalyst onto the photoelectrode surface to promote charge carrier separation and to facilitate hydrogen and oxygen evolution kinetics; (iii) surface modification by organic dyes, plasmonic materials or quantum dots to enhance visible light absorption; (iv) passivation of surface states or protection of the electrode by deposition of ultra-thin overlayers, to avoid surface charge recombination and enhance charge transfer to the environment (or to suppress photocorrosion).

Finally, another viable strategy is the nanostructuring of photoelectrode material into nanoparticles, nanotubes, nanoplatelets, *etc.* This is an efficient approach in the context of heterogeneous (*i.e.*, solid-liquid and solid-gas) reactions since it typically leads to large surface-to-volume ratio for the semiconductor (large surface area) that facilitates charge transfer at the SCLJ. Also, *e.g.*, for one-dimensional nanostructures (discussed below), their geometry is crucial for achieving high PEC-WS efficiency owing to the control over light absorption and orthogonal charge separation, as well as for establishing preferential percolation pathways.

### 1.2 Calculation of solar-to-hydrogen efficiency

To promote the advancement of PEC-WS technologies, the discovery and development of new materials are of primary importance. However, also the identification of protocols and standard parameters to cross-compare the materials (in terms of properties and efficiency) is highly necessary – reliable and reproducible protocols may accelerate knowledge transfer and process development.

Therefore, we discuss the most important parameters, that is we provide a description of the most useful figures of merit (see below) typically reported to characterize the performance of a photoelectrochemical device in terms of light to energy conversion efficiency and indicate the limits within which each of these benchmarks may be applied.

The photoconversion efficiency (η) depends on many factors such as the spectrum of the incident radiation, the band-gap of the semiconductor, the reflection of light, and the transport of $e^-/h^+$ through the semiconductor.

The comparison of photoconversion efficiencies obtained for different semiconductor materials requires that these efficiencies are presented for a standard solar spectrum, usually the Air Mass AM 1.5 global solar spectrum at a given power density (100 mW cm$^{-2}$).

Most measurements of the photoconversion efficiency are performed under illumination by artificial light sources. This is convenient mainly because artificial sources are stationary and their intensity is essentially constant with time (although the lamp spectrum varies in intensity with the lamp lifetime), while spectrum and intensity of solar radiation reaching the ground depend on the time of day and on the atmospheric conditions. A most suitable artificial light for this purpose is a Xe lamp which, adequately filtered and compared to other artificial lights, best replicates the solar spectrum.

Several measures are commonly used to determine the efficiency of materials for PEC-WS:

- solar-to-hydrogen conversion efficiency, STH
- applied bias photon-to-current efficiency, ABPE
- incident photon-to-current efficiency, IPCE
- absorbed photon-to-current efficiency, APCE.

STH describes the overall efficiency of a PEC-WS device exposed to a broadband solar AM 1.5G illumination, under zero bias conditions.[11] The working and the counter electrodes (WE and CE, respectively) should be short-circuited, that is, STH is measured in a 2-electrode system. In detail, STH represents the ratio of the chemical energy stored in the form of $H_2$ molecules to the solar energy input, and is expressed by (5):

$$STH = \left[\frac{r_{H_2} \times \Delta G}{P_{in} \times A}\right]_{AM\ 1.5\ G} \quad (5)$$

where $r_{H_2}$ is the rate of hydrogen production (mmol$_{H2}$ s$^{-1}$) measured with a gas chromatograph or mass spectrometer, $\Delta G$ is the change in Gibbs free energy per mole of $H_2$ produced (237 kJ/mol), $P_{in}$ is the incident illumination power density (mW cm$^{-2}$) and $A$ is the illuminated electrode area (cm$^2$). STH can also be calculated as a function of the short-circuit photocurrent density, $j_{SC}$ (6):[13]

$$STH = \left[\frac{j_{SC} \times 1.23\ V \times \eta_F}{P_{in}}\right]_{AM\ 1.5\ G} \quad (6)$$

where 1.23 V is the thermodynamic water splitting potential and $\eta_F$ the Faradaic efficiency for $H_2$ generation. It should be emphasized that (5) and (6) can be considered valid only if experiments are performed under AM 1.5 G illumination and in the absence of competing red-ox couples and sacrificial molecules in the electrolyte (that is, electron donors or acceptors).

A loss of faradaic efficiency for $H_2$ or $O_2$ evolution is observed not only when a sacrificial agent such as methanol, ethanol, *etc.*, is added to the electrolyte (this limits the $O_2$ evolution), but also when dissolved $O_2$ (from air) is present on the cathodic side of the PEC cell. That is, $O_2$ reduction to $O_2^{-\bullet}$ competes with the $H_2$ evolution reaction. $O_2^{-\bullet}$ radicals further react with water and form $H_2O_2$ which is a typical by-product detected in the cathodic compartment when the electrolyte is not purged with an inert gas prior to PEC-WS experiment, or if the anode material/electrolyte combination favors $H_2O_2$ production over $O_2$ production.

Therefore, a reliable evaluation of STH should imply (i) standard experimental conditions (such as purging the electrolyte to remove $O_2$, absence of sacrificial agents, checking for $H_2O_2$ in the cathodic compartment), and (ii) a comparison between the STH determined from measured $j_{SC}$ (6) and that calculated from the actual amount of evolved gases (5) – this is crucial to avoid an overestimation of the STH.

ABPE, IPCE and APCE are measured under applied bias in a 3-electrode PEC configuration.

Under electrical bias applied between working and counter electrodes, the current extracted from the device is higher compared to the bias-free condition of STH; therefore, it does not accurately reflect a solar-to-hydrogen conversion process but the so-called applied bias photon-to-current efficiency (ABPE) is determined (7):

$$ABPE = \left[\frac{j_{ph} \times (1.23\ V - |V_b|)}{P_{in}}\right]_{AM\ 1.5\ G} \qquad (7)$$

where $j_{ph}$ is the photocurrent density under the bias voltage $V_b$.

Noteworthy, $V_b$ is applied across the working and counter electrodes, that is, the two electrodes across which the photocurrent $j_{ph}$ flows. The product $(j_{ph} \times V_b)$ represents the electrical power loss that has to be subtracted in calculating the efficiency of the cell. By contrast, using the potential applied between working and reference electrodes ($V_{W-R}$) in place of $V_b$ typically leads to an overestimation of the cell efficiency (because $V_{W-R} < V_b$).[11]

One of the most adopted tools for the estimation of PEC properties of a semiconducting material is represented by IPCE. IPCE provides a measure of the efficiency of conversion of incident monochromatic photons to photocurrent flowing between WE and CE.

IPCE takes into account all the three fundamental steps of a light-driven process: (i) photon absorption (the absorption of one photon generates an $e^-/h^+$ pair); (ii) efficiency of $h^+$ transport to the solid-liquid interface and collection of $e^-$ at the back contact; (iii) efficiency of $h^+$ transfer at the semiconductor-electrolyte interface. It is expressed as:

$$IPCE = \frac{j_{ph}(\lambda)}{e\,I(\lambda)} \quad (8)$$

where $j_{ph}(\lambda)$ is the photocurrent measured at the specific wavelength λ, $I(\lambda)$ is the incident photon flux, and *e* is the electronic charge. Due to photon losses associated with the reflection of light or imperfect absorption, and due to the recombination of photogenerated e⁻/h⁺ pairs, typical IPCE values are below 100%,[11] except for rare cases where *e.g.* current doubling occurs.[14]

Compared to STH, the main advantage of IPCE is that it is measured as a function of the irradiation wavelength and normalized *vs.* the irradiation intensity (8), and it can be measured under irradiation of a common Xe lamp combined with a monochromator. A most reliable estimation of the semiconductor PEC efficiency can then be obtained simply by integrating the IPCE spectrum (measured at the cell bias voltage of maximum efficiency) over the AM 1.5 global solar spectrum (9):

$$\eta = e\,(1.23\,V - V_b) \int_0^\infty IPCE(\lambda)\,I_\lambda(\lambda)\,d\lambda / E_S \quad (9)$$

where $E_S$ is the incident irradiance that describes the spectrum of a light source.

Murphy et al. showed that his approach leads to efficiency results that are well in line with those measured under direct solar light irradiation, thus avoiding an under- or overestimation ascribed to spectral variations between different artificial sources.[11]

Finally, to understand the intrinsic PEC performance of a material, APCE is particularly useful to describe the collected current per incident photon *absorbed*. APCE takes into account also losses caused by reflection and/or transmission of photons. APCE can be obtained by combining (8) with (10), which provides the number of charge pairs generated per incident photon ($\eta_{e^-/h^+}$). Assuming that ($\eta_{e^-/h^+}$) is equal to the number of photons absorbed (that is, no photon losses occur):

$$\eta_{e^-/h^+} = \frac{I_0 - I}{I_0} = 1 - \frac{I}{I_0} \quad (10)$$

Then, (10) can be combined with the Lambert-Beer's law ($A = -log(I/I_0)$):

$$\eta_{e^-/h^+} = 1 - 10^{-A} \qquad (11)$$

APCE describes the collected current as a function of absorbed photons:

$$APCE = IPCE/\eta_{e^-/h^+} \qquad (12)$$

Therefore, (12) can be combined with (8) and (11):

$$APCE = \frac{j_{ph}(\lambda)}{e\,I(\lambda) \times (1 - 10^{-A})} \qquad (13)$$

However, regardless of the figure used to express the PEC efficiency of a certain device, an additional problem often encountered when cross-comparing the performances of different photoelectrodes is that PEC devices may also be operated under different electrochemical conditions such as applied potential, electrolyte composition, and reference electrode.

The lack of uniformity of electrochemical parameters can, in principle, be addressed by referring the applied bias to the Reversible Hydrogen Electrode (RHE) according to the Nernst equation:

$$E_{RHE} = E_{ref} + E^0_{ref} + 0.059\,pH \qquad \text{(eq. 13)}$$

where $E_{ref}$ is the measured potential referred to the used reference electrode (*e.g.*, Ag/AgCl, Saturated Calomel Electrode SCE, Hg/HgO, *etc.*), $E^0_{ref}$ is the potential of $E_{ref}$ with respect to the Standard Hydrogen Electrode (SHE) at 25 °C, and the pH is that of the electrolyte solution.

This approach, when comparing different pHs, however implies that no significant changes in the water splitting mechanism occur with pH. Additionally, the Nernst equation assumes the linear dependence of $E_{RHE}$ on the electrolyte pH. However, these assumptions are not always valid and, therefore, many material specific data are truly comparable only when different materials are tested in view of their PEC performance in the same electrolyte and at the same pH.

## 1.3 α-Fe$_2$O$_3$ and TiO$_2$ based photoanodes

Among all the investigated metal oxide semiconductors, α-Fe$_2$O$_3$ and TiO$_2$ have stimulated an immense research interest due to their natural abundance, chemical stability in liquid solutions, non-toxicity, high resistance to photocorrosion, and low production costs.[5,8,15–17]

TiO$_2$ has suitable conduction and valence band energy to drive water reduction (to H$_2$) and oxidation (to O$_2$) – E$_{CB}$(TiO$_2$) = *ca.* –0.2 V *vs.* RHE and E$_{VB}$(TiO$_2$) = *ca.* 3.0 V *vs.* RHE).[8] However, a severe limitation of TiO$_2$ is represented by its optical band gap (*i.e.*, 3.2 and 3.0 for anatase and rutile polymorphs, respectively). This means that only UV light (λ < 380 – 400 nm) is absorbed and thus only *ca.* 4% of the solar spectrum can promote charge carrier generation in the material, and the consequent photoelectrochemical reactions.

By contrast, α-Fe$_2$O$_3$ absorbs light in the visible range up to λ ~ 600 nm owing to its optical bandgap (E$_g$ = 2.0 – 2.2 eV). This corresponds to a light absorption of *ca.* 40% of the solar spectrum.[18] Nevertheless, while the valence band of hematite matches the thermodynamic requirements for oxygen evolution from water (E$_{CB}$(α-Fe$_2$O$_3$) = *ca.* 2.4 *vs.* RHE, that is, the VB lies at more positive potential than E°(O$_2$/H$_2$O)),[8] an external electrical bias (and thus a PEC configuration) is needed to drive hydrogen generation due to the more positive conduction band position with respect to the HER potential (E$_{CB}$(α-Fe$_2$O$_3$) = *ca.* 0.4 *vs.* RHE) – in other words, H$_2$ evolution does not occur on α-Fe$_2$O$_3$ under open circuit conditions.[8]

Furthermore, both TiO$_2$ and α-Fe$_2$O$_3$ suffer from inherently low electrical conductivity, owing to bulk electron-hole recombination. In addition, TiO$_2$ and α-Fe$_2$O$_3$ feature limited excited state lifetime (less than 10 ns and 10 ps, respectively) and a short hole diffusion distance (less than 20 nm and 5 nm, respectively), which significantly reduces the efficiency of holes transfer and collection at the SCLJ.[9,17] All these aspects are typically reported to be primary reasons for a limited PEC performance in H$_2$O splitting.[18,19]

While bulk materials have been extensively investigated (and reviewed) in classical photoelectrode configurations (deposited film layers, compacted particle layers or single crystals) – see *e.g.* 5,15,17,19–23, a main focus of this review is to highlight the use of nanostructuring techniques to fabricate efficient photoanodes. Particularly, we will provide an overview of recent breakthroughs relative to synthesis and use of one-dimensional (1D) $TiO_2$ and $\alpha\text{-}Fe_2O_3$ photoanodes. The most explored strategies for doping, surface sensitization and the coupling of Ti and Fe oxides with narrow-bandgap semiconductors will also be illustrated.

Moreover, we describe most recent results relative to $\alpha\text{-}Fe_2O_3\text{-}TiO_2$ heterojunction-based materials or combination compounds (*e.g.*, $Fe_2TiO_5$) that were shown to outreach the PEC-WS performance of the single counterparts. For these materials, modification and exploitation of the (as such detrimental) conduction band offset between $\alpha\text{-}Fe_2O_3$ and $TiO_2$ has led to various new designs that combine (i) the efficient charge carrier separation, provided by $TiO_2$ that suppresses electron back injection in hematite, with (ii) the strong visible light absorption promoted by an $\alpha\text{-}Fe_2O_3$ thin layer. A comprehensive description of this approach will be presented in the final section, along with studies on the charge carrier dynamic observed for these composites and useful tools for the rational design of potentially new platforms for efficient solar-driven water splitting.

## 2. Nanostructuring and nanostructures for PEC water splitting

Why micro- and nanostructuring? One of the reasons is that the properties of materials can be significantly different from the quantum-, to nano- and to bulk scale. Another reason is that defined structuring of electrode/electrolyte systems allows a much better control over light guiding, reactant diffusion, and electron pathways.

In regard to the first reason, nanostructured materials (NMs) can provide significantly modified physical, chemical, and biological properties and thus can exhibit unique functionalities.[24–27]

NMs are low dimensional materials, *i.e.*, in the nanoscopic size range, composed of assemblies of building units that have at least one dimension confined to the submicron- or nano-scale.[28] According to this definition, zero- (0D), one- (1D), two- (2D), and three-dimensional (3D) NMs can be distinguished (examples are sketched in Fig. 2.1).[29–34]

Typical features of NMs that are beneficial for PEC applications are: (i) the possibility of decoupling the direction of light absorption and charge-carrier collection; that is, instead of a diffusion through the bulk material, orthogonal separation of photogenerated charges occurs and the probability of charge recombination is reduced; (ii) improved photogenerated charge pair separation, promoted by the internal electric field (*e.g.*, the upward band bending in the walls of $TiO_2$ nanotubes facilitate $h^+$ transfer to the electrolyte and $e^-$ collection in the center of the material); (iii) the possibility of controlling crystal faceting that, in turn, influences the material band-bending, flat band potential, surface states, *etc*.

Moreover, compared to bulk materials, NMs exhibit remarkable *size effects* owing to the high surface-to-volume ratio. In particular, a larger specific surface area represents a great advantage when a material is intended for catalytic applications: a higher number of active surface sites facilitates the adsorption of reactants, offers extended contact with solutions, *etc.*, and typically leads to higher reaction rates.

Size effects imply that the nanomaterial dimensions (< 100 nm) are comparable to the critical length scale of physical phenomena, *e.g.*, charge transport distances and light absorption depth, the mean free path of electrons and phonons, or (typically for particles < 10 nm) the Bohr exciton radius.

On the one hand, this can be exploited for a more efficient charge carrier separation in the lattice of a nanomaterial in order to improve its PEC-WS performance. In other words, charge (*e.g.*, h$^+$) diffusion towards the environment occurs over minimized distances since the size of the semiconductor is within the length scale of solid state charge diffusion.

On the other hand, quantum size electronic effects, such as ballistic electron transport or optical gap widening, can occur due to quantum confinement. This was experimentally demonstrated, *e.g.* in the case of TiO$_2$, for nanoparticle suspensions,[35,36] and for ALD layers.[37] These works showed that a clear onset of quantum confinement for TiO$_2$ nanomaterials can be expected only if size-scales are < 5 nm. This can be observed for hydrothermal TiO$_2$ nanotubes,[38,39] which exhibit a measured band-gap of ~3.84 eV compared to $E_g \approx 3.2$ or 3.0 eV commonly reported for larger anatase or rutile TiO$_2$ crystals.[40] This difference is ascribed to the structure of hydrothermal tubes that is based on atomic sheets originating in quantum confinement effects.

Such size effects were already widely described in the 1960s,[41] but the first classification attempts in materials science dates back to 2000,[42] and only more recent reports consider also more complex architectures such as mesoscale assemblies particularly as nanocones, nanohorns, nanoegg-yolk, and nanoflowers (lilies, roses, tulips, *etc.*).[28]

In the following sections we will provide a description of the different nanostructures (0D to 3D) and hierarchical assemblies, in particular with respect to their application for the photoelectrochemical water splitting reaction.

## 2.1 From 0D to 3D

Zero-dimensional nanostructures feature all the three dimensions in the nanoscale regime (*i.e.*, no dimension is larger than 100 nm). Typical examples are nanoparticles (NPs), such as nanospheres, quantum dots (QDs), core-shell structures, hollow spheres, etc. An attractive feature is that 0D-NMs typically smaller than 5 – 10 nm (the Bohr exciton radius) show tunable band-gap. The smaller the size of a crystal the more affected its electronic structure: when a crystal is composed of a limited number of atoms, bulk electronic bands are discretized to distinct energy levels with a widening of the gap for smaller dot sizes. Subtraction or addition of just a few atoms alters the energy level distribution and thus leads to bandgap shifts in the QDs.[43]

Clearly, photoelectrochemical water splitting using a water dispersion of 0D nanoparticles is not possible, that is NPs should first be immobilized on an electrically conductive substrate (*e.g.*, ITO, FTO, etc.) to form various nanoparticulate assemblies or thin films that serve as electrodes in a photoelectrochemical cell arrangement.[44]

Therefore, despite some fundamental early works on the electrochemical and photoelectrochemical investigation of $TiO_2$ colloidal solutions that allowed the proposal of a detailed mechanism of redox reaction on hydroxylated titania surfaces,[45,46] electrochemical studies of $TiO_2$ solutions remained scarce in comparison to investigations of solid $TiO_2$ as electrode material.

However, films composed of stacked NPs often suffer from carrier recombination at grain boundaries (owing to the presence of trapping states) and long carrier diffusion paths (random walk) through the NP network (Fig. 2.2(a)). In this regard, replacing a nanoparticulate electrode with a one dimensional (1D) nanostructured electrode offers several advantages.

In terms of charge collection and transfer efficiencies, one-dimensional (1D) nanostructures exhibit superior performances in a wide range of applications (*e.g.*, nanoelectronics, nanodevices, energy-related fields) with respect to the other nanoarchitecture arrangements. In many cases, the size effects of a 0D material can be maintained but harvesting of charge carriers is facilitated.[17,47–51]

Charge collection efficiency depends on both the recombination lifetime of photogenerated charge carriers and on their collection lifetime, that is, the time needed by the $e^-$ ($h^+$) to reach the semiconductor-electric contact (semiconductor/electrolyte interface).[52] 1D nanostructures such as nanotubes (NTs), nanorods (NRs), and nanofibers (NFs) provide a preferential percolation pathway for charge carriers (Fig. 2.2(a)) that significantly enhances electron collection at the back-contact (typically, a metal foil or a conductive glass such as FTO and ITO). This leads to a remarkable improvement in the PEC efficiency.[48]

Also, compared to 0D nanoparticles, the recombination time of $e^-/h^+$ pairs is longer in 1D-NMs (*e.g.*, in $TiO_2$ nanotubes recombination times are 10 times higher than in a $TiO_2$ nanoparticle layer) and this leads to larger charge collection efficiency.[53–58]

We illustrate in detail one-dimensional nanoarchitectures and their application in PEC water splitting in Section 3.

Two-dimensional nanostructures have only one dimension in the nanoscale. Typical examples of 2D-NMs are nanoflakes, nanoplatelets, etc. (Fig. 2.2(b)), but also thin films (*i.e.*, with a thickness < 100 nm). 2D-NMs exhibit relatively large lateral size and a ultrathin thickness confined to the atomic scale regime that collectively lead to surface areas typically larger than those of 1D-NMs. Moreover, the ultrathin thickness of two-dimensional nanomaterials grants a high mechanical flexibility and optical transparency. These characteristics make 2D-NMs promising candidates particularly for high-performance flexible electronic and optoelectronic devices.[59,60]

However, 2D-nanostructures are also advantageous when used in PEC devices particularly when perpendicularly oriented to the electrode (Fig. 2.2(b)).[61–64] This arrangement simultaneously offers (i) enhanced charge transport properties, owing to a directional transport of charges to the back-contact, (ii) a facilitated access for electrolyte impregnation into the structure, owing to the high surface area, and (iii) a limited diffusion distance of the photogenerated holes to the electrode/electrolyte interface that reduces the charge recombination rate.

Typical methods for the fabrication of 2D-NMs are *e.g.*, sputtering techniques,[65] hydrothermal and chemical bath deposition methods,[61,62] thermal oxidation,[66] and spray pyrolysis techniques.[67]

According to the definition of NMs, 3D-nanostructures should be more appropriately considered bulk materials as none of their dimensions is confined to the nanoscale. However, 3D-NMs feature properties that are locally confined to the nanoscale, when composed of nanosized crystals. These 3D-nanomaterials may include layers of *e.g.* nanoparticles, nanowires, and nanoplatelets where the 0D-, 1D- and 2D-nanostructured elements are in close contact.[59,68,69]

A combination of their mesoporous matrix along with their nanoscopic structure (formed *e.g.* of nanoparticle aggregates) provides an ideal morphology for PEC-WS as the permeability of the scaffold is maximized and charge diffusion towards the environment has to occur over minimized distances (within each single nanoparticle). Moreover, 3D NMs have higher surface area and supply enough absorption sites for all involved molecules in a small space. In more details these structures are discussed in the following section and Section 4.

## 2.2 3D Hierarchical nanostructures

Hierarchical nanostructures (HNs) are considered 3D-NMs and, owing to their intrinsic properties, are gaining increasingly wider attention as scaffolds suitable for PEC-WS devices.

In general, a hierarchical nanostructure is composed of a distinct backbone (typically with a 1D-arrangement) onto which nanodimensional building blocks are grown – these usually include nanoparticles (0D), nanowires/rods/tubes (1D) and nanosheets (2D).

Therefore, HNs often resemble nanotrees that simultaneously combine the properties of the one-dimensional backbone (*e.g.*, enhanced carrier separation and directional charge transport) with a significantly increased surface area and a much higher amount of active sites, owing to the highly porous arrangement provided by the branches.[59,68,69]

In most cases, hierarchical structures can be produced by combining or modifying conventional processes for nanostructure synthesis: *e.g.*, hydrothermal growth, solution-phase chemical synthesis, chemical vapor deposition techniques or branch formation by in situ catalyst generation.[70–72]

Regardless of the synthetic procedure, hierarchical nanostructures can be composed either of a single material (Fig. 2.3(a,b)) or of a combination of two or more (Fig. 2.3(c–e)).[60]

In the first case, a hierarchical arrangement is mainly exploited in view of its significantly higher surface area and a major application is often as photoanode in a dye-sensitized solar cell (Fig. 2.3(a,b)). Higher dye loading, enhanced light harvesting, as well as the reduced charge recombination owing to preferential electron transport along the branches to the backbone, increase the power conversion efficiency of a $TiO_2$-HN photoanode compared to a more conventional 1D-$TiO_2$ arrangement.[72]

Hierarchical nanostructures composed of two or more materials exhibit an additional advantage, that is, the possibility of combining materials with different (complementary) properties and band structures. This in turns can either be exploited to enhance electron collection from the branched-material to the back-bone nanostructure (*i.e.*, host-guest approach) and/or to extend the light absorption (and, hence, its conversion) over a broader wavelength range.

The so-called "host–guest approach" typically provides a support material (the "host") for majority carrier conduction, into which a photoactive "guest" material can inject photogenerated electrons while, at the same time, providing proximity to the semiconductor/liquid junction (Fig. 2.3(c,d)). Therefore, the performance of such host–guest assemblies strictly depends on the availability of (i) a host material with good electronic transport properties, (ii) a guest material with optimized light absorption, and (iii) a suitable band alignment between the two counterparts that enable e$^-$ injection from the branches to the backbone, with holes accumulating at the (branched) semiconductor/electrolyte junction.[73] Examples of efficient host-guest HNs have been for instance reported for $WO_3$-$Fe_2O_3$,[74] $SnO_2$-$TiO_2$,[75] and $SnO_2$-$Fe_2O_3$ combinations.[73,76] In particular, the conducting properties of tin oxide as host-scaffold can be further improved by F- or by Sb-doping as a consequence of growing (and annealing) $SnO_2$ 1D nanostructures on a FTO glass.[73,77]

Recently, also more advanced hierarchical systems composed of n- and p-type semiconductors have been shown to provide improved solar-driven PEC-WS ability. In this context, an assembly composed of Si nanowires branched with $TiO_2$ nanowires has been reported by Liu an co-workers.[78] Here, the different intrinsic electronic properties of the two semiconductors allow for the production of $O_2$ and $H_2$ at the two separate sites (n-type ($TiO_2$) and p-type (Si) semiconductors, respectively): the photogenerated minority carriers (*i.e.*, h$^+$ in $TiO_2$ and e$^-$ in Si) promote the corresponding oxidation/reduction half reactions, while the majority carriers (*i.e.*, e$^-$ in $TiO_2$ and h$^+$ in Si) recombine at the ohmic contact. PEC *J-V* curves (Fig. 2.3(e)) showed that the system can promote solar water splitting under open circuit conditions, in contrast with the state-of-the-art materials typically used for photoelectrochemical applications.

## 3. α-$Fe_2O_3$ and $TiO_2$ nanostructures for PEC water splitting

In the following sections we will deal with the different one-dimensional arrangements of $TiO_2$ and α-$Fe_2O_3$ typically used for photoelectrochemical water splitting experiments.

We will mainly report on 1D-nanostructures such as nanorods (NRs), nanowires (NWs), and nanotubes (NTs).

In particular, we will show how the use of 1D-morphology is strategic to realize efficient electron transport and charge carrier separation and to overcome the short diffusion length of holes in both $TiO_2$ and $\alpha$-$Fe_2O_3$.

We will mainly focus on 1D-nanotubes, fabricated by various techniques such as hydrothermal methods and self-organizing electrochemical anodization.

Emphasis will be then on a number of modification strategies that (coupled to 1D-nanostructuring) have been adopted to address other critical aspects of hematite- and titania-based photoelectrodes, *e.g.*, the sluggish kinetic of hole injection into the electrolyte (that mainly affect $\alpha$-$Fe_2O_3$) and a limited use of solar light irradiation for promoting the PEC-WS reaction (for $TiO_2$).

## 3.1 Typical synthesis approaches

Solution-based growth techniques offer several major advantages including low-cost, simple processing, and good scalability. Many advanced nanomaterials that are currently commercially available are made via solution-based approaches, including colloidal NPs and QDs.[79]

Despite the large number of morphologies that can be grown via solution-based methods (nanoparticles of different size and shapes, nanorods, nanowires, nanoflowers, nanotubes, *etc.*), in recent years other techniques, such as template-assisted synthesis and self-organizing anodization, have also been developed. These techniques allow a higher degree of control on morphology and physical properties, both crucial aspects when considering 1D nanostructured materials.

1D nanostructure arrays can be typically fabricated by either bottom-up or top-down approaches. Bottom-up methods are solution- and vapor-based methods but, despite the several advantages offered by these approaches (see above), optimization and control over morphology and physical properties sometimes (still) represent a challenging aspect for practical applications.

On the other hand, top-down strategies (*e.g.,* (photo)lithography and focused-ion beam methods), in spite of a better control on the nanostructure morphology, are typically complex and expensive.

Here, we will give a larger attention to the most versatile methods (in our view), *i.e.*, hydrothermal synthesis and self-organizing electrochemical anodization.

A remarkable aspect is the possibility of combining, *e.g.* hydrothermal and self-organizing anodization methods to other synthesis procedures (such as solution-based and/or sputtering techniques) to grow hierarchical composite nanomaterials.

### 3.1.1 Sol-methods for nanoparticles

Sol-gel methods certainly represent the most widely used approach for the synthesis of nanoparticles. It is based on the slow hydrolysis and polymerization reaction of a colloidal suspension of the metal oxide precursor (*i.e.*, alkoxide, halide, nitrate salt, *etc.*) that, at controlled rate and under specific conditions, evolves toward the formation of a solid gel phase.[80–82]

Typical precursors for $TiO_2$ nanoparticles include titanium(IV) iso-propoxide (TTIP) or butoxide, and $TiCl_4$.[79] Crucial for the nucleation of nanoparticles is the formation of Ti-O-Ti chains that is favored with a low content of water (that is, low $Ti/H_2O$ ratio), low hydrolysis rate (that is, relatively low temperature), and in the presence of excess titanium alkoxide in the reaction mixture. As-synthesized nanoparticles are amorphous and annealing (typically, ~ 400–700 °C in air) is required to attain a crystalline solid.

Inorganic precursors, such as iron nitrate, are commonly used for α-Fe$_2$O$_3$ nanoparticles. In a typical synthesis procedure, Fe(NO$_3$)$_3$ is dispersed in an aqueous solution also containing EDTA as capping agent. Upon the formation of a solid gel, the evaporation of the liquid phase leads to iron oxide/oxy-hydroxide nanopowders. Annealing at 450–900 °C is typically reported to crystallize the amorphous powders into single phase hematite nanoparticles.[80,82,83]

Despite simple processing and large versatility (*e.g.*, doping can be achieved by simply adding the dopant source into the colloidal solution), sol-gel methods lack of a precise control on nanoparticle size and shape distribution. Therefore, alternative approaches have been developed that include micelle and inverse micelle, hydrothermal, and solvothermal methods.[81]

In particular, hydrothermal methods for TiO$_2$ lead to small particles (~ 5–25 nm), the size of which can be controlled by adjusting the concentration of Ti precursor and the solvent composition[84] – *e.g.*, the presence of additives such as peptizers largely influenced the nanoparticle morphology.[79] Typical precursors are Ti alkoxide (*e.g.*, TTIP and Ti butoxide) colloidal solutions that, under acidic conditions, mainly form TiO$_2$ anatase NPs, with no secondary phase formation.[79,84]

Colloidal synthesis of α-Fe$_2$O$_3$ nanoparticles through hydrothermal route has been first introduced in the early 1980s.[85] Nanosphere, disk, and plate morphologies were reported depending on the precursor concentration,[85] and follow-up works demonstrated that 60–100 nm α-Fe$_2$O$_3$ NPs synthesized through similar methods are ideal catalysts for the photocatalytic generation of hydrogen from water solutions, containing sacrificial agents (*e.g.*, methanol) or electron donors (*e.g.*, methyl viologen).[86–88]

More recently, TiO$_2$ and α-Fe$_2$O$_3$ nanoparticles and hollow spheres have been reported by solvothermal method,[89,90] that is, a procedure almost identical to the hydrothermal method except for the use of non-aqueous solvents – this enables the use of higher temperature, due to a large number of organic solvents with a high boiling point.[81]

However, as mentioned before, the use of nanoparticle water dispersion is not possible in view of photoelectrochemical applications, and immobilization of NPs on a conductive substrate in the form of assemblies or thin films has several drawbacks (*e.g.*, long (random) carrier diffusion pathways, enhanced carrier recombination at grain boundaries, lower surface area compared to NP suspensions, *etc.*).

Thus, a most elegant approach is one-dimensional nanostructuring of photoanode materials.

### 3.1.2 Hydrothermal methods

Hydrothermal synthesis certainly belongs to the most commonly used methods for the synthesis of nanorods/wires/tubes of various metal oxides.

As for sol-gel methods, titanium isopropoxide,[91] tetrabutyl titanate,[92] or titanium isobutoxide are typically used as $TiO_2$ precursors,[93–95] while iron(III) chloride is mostly used as $\alpha$-$Fe_2O_3$ precursor.[96–100]

Hydrothermal synthesis is carried out in a pressure vessel under controlled temperature and/or pressure, in aqueous or organic based-solutions (as anticipated, in the latter case, the term solvothermal is more appropriate) and, in some cases, in the presence of surfactant agents.[81] Temperature, solution volume and nature of capping agent (*e.g.*, $F^-$ vs. $SO_4^{2-}$) largely influence the internal pressure and, hence, the aspect ratio (*i.e.*, the diameter/length ratio) of NRs, NWs and NTs.[49]

$\alpha$-$Fe_2O_3$ hydrothermal nanotubes are mostly prepared from a $FeCl_3$ solution in the presence $NH_4H_2PO_4$ at 220 °C for several (2–3) days.[101–103]

Under these conditions, the formation of $\alpha$-$Fe_2O_3$ hydrothermal nanotubes has been reported to occur *via* a coordination-assisted dissolution process.[101] In particular, it was demonstrated that crucial factor to induce the formation of a tubular structure is the presence of phosphate ions (from $NH_4H_2PO_4$) that adsorb on the surface of hematite aggregates and coordinate with ferric ions.[101–103] In detail, nanotube formation proceeds through controlled "dissolution" along the long axis of the spindle-like precursors formed at the early stage of the reaction; this generates rod-like nanocrystals and semi-nanotubes, and for sufficiently long times hollow tubes – remarkably, the inner part of tubes is only partially dissolved, and a non-uniform dissolution from spindle to spindle has also been reported.

More recently also alternative reactants have been proposed (*e.g.*, $[Fe(CN)_6]^{4-}$ precursor in $H_2O_2$ based solution) that significantly lowered both the temperature and the reaction time needed (160 °C for < 2h).[104]

Regardless of the used precursors (and hence conditions), α-$Fe_2O_3$ tubes are single-crystalline (*i.e.*, hematite is the only crystalline phase) and are typically 200 nm – 1 μm long, with a diameter in the 100–150 nm range and a wall thickness of 25–30 nm.[101–104]

Hydrothermal $TiO_2$ NTs were first reported by Kasuga *et al.*,[38] by alkaline treatment of anatase $TiO_2$ powders in a NaOH solution at 20–110 °C. Approximately 100 nm thick nanotubes with a ~8 nm diameter were obtained upon acidic washing of the reaction suspension. Several follow-up works demonstrated that amorphous and crystalline (that is, also rutile and commercial powders) $TiO_2$ as well as metallic Ti are suitable precursors,[39,105] and that NaOH can be replaced by KOH – under these conditions, the reaction temperature can be increased and $TiO_2$ NWs can also be formed.[106] In general, large NaOH concentrations and high operating temperatures facilitate the formation of tubes – and in particular the higher the T, the longer the tubes.

Most interesting feature of $TiO_2$ hydrothermal tubes is their multi-walled morphology, featuring inter-wall distance of ~ 0.7 nm[107] and (in contrast to α-$Fe_2O_3$ tubes) a wall thickness in the range of atomic sheets.[38] That is, only $TiO_2$ hydrothermal tubes should be considered

"real" nanotubes, with quantum confinement effects such as bandgap widening being practically observable (only) in tube walls prepared through such a procedure.

Another remarkable aspect of hydrothermal methods is the possibility of growing 1D-nanostructures such as NRs and NWs anchored on a conductive substrate (usually a FTO layer immersed in the reaction vessel), for a direct use as photoelectrode – this, in contrast to nanotubes that grow in bundles/agglomerates, dispersed in a reaction media.[48]

Normally, the F-SnO$_2$ side of FTO layers has to face the bottom of the reactor, so to enable the growth of NRs/NWs on the conductive substrate, while the metal oxo-species (*e.g.*, TiO$_x$ and FeOOH), which form on the glass side due to gravimetric precipitation from the precursor solution, can be easily washed away.[108]

Growing thin films directly from a substrate not only does considerably improve the adherence and mechanical stability of the film compared to standard deposition techniques (such as spin coating, dip coating, screen printing, or doctor blading), but also grants better charge transfer and collection to the back-contact with no extra binder layers to be tunneled.[33,96]

Under these conditions, the formation mechanism relies on the constant supply of H$_2$O molecules at the hydrophilic surface of FTO (*i.e.*, the F-SnO$_2$ side) through the hydrolysis of the metal oxide precursor. Nuclei will appear on the entire substrate and if their formation rate is controlled and maintained limited by the precipitation conditions, epitaxial crystal growth will take place from these nuclei. If the concentration of precursors is high, a condensed phase of vertically aligned arrays perpendicular to the substrate will be generated.[96]

Concerning TiO$_2$, it is known that acidic media set the conditions for rutile TiO$_2$ NR formation; key aspect to such a growth is the presence of Cl$^-$ ions in solution, *i.e.*, Cl$^-$ ions preferentially adsorb on the rutile (110) plane and, by retarding the growth along this direction, promote NR formation.[109] Accordingly, TiO$_2$ NRs were not formed when HCl was replaced by HNO$_3$ or H$_2$SO$_4$.

However, a major limitation of Cl$^-$ ions is their blocking effect towards the TiO$_2$ surface sites that are active for the water splitting reaction.[109] Annealing at T > 200–250 °C is a most common approach to remove the Cl$^-$ termination.[110]

By contrast, alkaline media and an excess of OH$^-$ ions favor the crystallization of anatase TiO$_2$[111] – however, an alternative way to anatase is also the presence of an anatase-based seed layer on FTO.[112] Moreover, it is reported that the reaction temperature (not time) has a critical effect on rod/wire morphology and crystal structure.[113]

Inorganic iron salts such as FeCl$_3$[96,97,114] and FeSO$_4$[98,100] are most typical precursors for the hydrothermal synthesis of hematite rods/wires (~ 100 nm – 1 μm long and with a ~ 5–10 nm diameter, reaction temperature varies within 100–120 °C and reaction time up to 24 h).

In particular, FeCl$_3$ is usually dissolved in a NaNO$_3$ based solution, also containing HCl to establish an acidic environment.[96,97,114] On the other hand, FeSO$_4$ is used under milder conditions and in the presence of CH$_3$COONa.[98]

Common aspect to all these procedures is that, contrary to the TiO$_2$ case, the formation of hematite rods/wires proceeds through the initial nucleation and aggregation of β-FeOOH 1D nanoarrays and hence annealing at T > 400 °C[96] is required to convert the Fe oxide-hydroxide into hematite.

### 3.1.3. Template-assisted techniques

We mentioned that hydrothermal approaches for nanotube fabrication results in single tubes or loose agglomerates of tubes dispersed in a solution and that non-homogeneous tube lengths are typically obtained.

Most critical aspect to a use of these structures in electrically contacted devices (*e.g.*, photoelectrochemical cell) is that tubes need to be compacted to layers (similar to powders) on an electrode surface. This leads to an arbitrary orientation of the nanotubes on the electrode and, in turn, eliminates many advantages of their one-dimensional nature (*e.g.*, providing a 1D direct electron path to the electrode).[48]

However, using aligned templates (or self-organizing electrochemical anodization, see below) not only leads to an array of oxide nanotubes oriented perpendicular to the substrate surface, but also tubes in the template can relatively easily be contacted by metal deposition (before removal of the template by selective dissolution).

Numerous different nanotube morphologies can be obtained by simply tuning the morphology of the template.[40,115–117]

Most classic template for $TiO_2$ NTs is porous anodic aluminum oxide (AAO),[118–120] which can be produced with a hexagonal pattern of nanopores in a virtually perfect (long-range) order, and with interpore distances between 10 and 500 nm.[121]

Historically, the first attempt to produce $TiO_2$ NTs through a template-assisted method was seemingly that reported by Hoyer[118] who electrodeposited titanium oxide from a $TiCl_3$ solution. The oxide first is in the form of polymer-like hydrous titania that, after annealing, crystallizes into anatase $TiO_2$ (Fig. 3.1(a)). More recently, also other filling approaches have been developed that include sol-gel techniques[115,122,123] and atomic layer deposition (ALD), see Fig. 3.1(b).[124–126]

As already discussed for the synthesis of NPs (section 3.1.1), sol-gel methods are based on the hydrolysis reaction of Ti-alkoxide, $TiCl_4$, and $TiF_4$, followed by condensation into a solid gel phase.[80] After appropriate heat treatment, the alumina template can be easily dissolved. ALD techniques offer an even larger control on tube morphology as conformal coating of the template with one atomic layer after the other can be achieved, also by alternating different titania precursors.[124,125]

Another effective template technique is the use of a ZnO nanorod array as sacrificial template (Fig. 3.1(c)).[33,127,128] A thin ZnO seed layer is first deposited (*e.g.*, sputtered) on a FTO glass; hydrothermal growth of ZnO nanorods follows. As-synthesized ZnO NRs on FTO are then immersed into the Ti (or Fe) precursor solutions that also contain boric acid ($H_3BO_3$). Hydrolysis reaction of the metal oxide precursor to $TiO_x$ (or FeOOH) on the individual ZnO nanorods results in the formation of $TiO_2$ (or α-$Fe_2O_3$) based nanotubes with the simultaneous dissolution of the ZnO template in the acidic environment.[33] Remarkably, in a one-step "NT growth/template dissolution" synthetic approach, no extra reactions are needed that may damage the NT array.

### 3.1.4 Self-ordering anodization

Electrochemical anodization is often considered a most straightforward synthesis path to fabricate one-dimensional vertically oriented layers for photoelectrochemical applications since it is scalable (it allows one to coat virtually any shape of various metal surfaces), offers an extended control over nanoscale geometry, produces directly back-contacted photoanodes, and is a versatile approach as it can be used to grow nanostructures of various metal oxides (Fig. 3.2(a)).[47,48,129]

Classically, electrochemical anodization is carried out in a 2-electrode electrochemical arrangement where a metal substrate (M) is immersed in a suitable electrolyte (most commonly $H_2SO_4$) and subjected to a positive electrical bias (U).

In this configuration, with M as working electrode (anode) and platinum or carbon as counter electrode (cathode), oxidation of the metal substrate occurs and leads to the formation of a metal oxide ($MO_{z/2}$). Oxide growth is sustained by field-assisted ion-migration. Briefly, as long as high field conditions hold, metal cations are subjected to outward migration from the metal substrate towards the electrolyte, while at the same time $O^{2-}$ ions migrate (from water) towards the anode (Fig. 3.2(b)).

This process is self-limited since the formed oxide is stable in the anodizing electrolyte: with increasing the oxide thickness (d), the electric field (*i.e.*, $\Delta E = \Delta U/d$) gradually drops and so does the inward/outward ion migration. This results in the formation of an oxide layer with a finite thickness. Such an oxide film is compact or, more generally, does not exhibit specific morphological features.

However, when using a suitable electrolyte and under specific electrochemical conditions (that largely depend on the metal of interest), a steady-state equilibrium can be established between the electrochemical oxide formation and its dissolution (Fig. 3.2(b)).

In particular, electrochemical parameters can be adjusted to establish controlled oxide growth and etching to form porous oxide layers or, even more, to enable the growth of one-dimensional self-organized metal oxide structures. Typically, the use of electrolytes containing $ClO_4^-$, $NO_3^-$ or $F^-$ ions under self-organizing electrochemical conditions is key for the anodic growth of Ti and Fe oxide layers in the form of arrays of nanopores or nanotubes.[47,48,130,131]

Key advantage of anodization is that the self-ordering degree of the formed one-dimensional nanostructures, their morphology and physicochemical properties can be adjusted by simple tuning of electrochemical parameters (electrolyte composition and temperature, applied voltage, anodization time, *etc.*).

As the effects of such parameters on the growth of $\alpha$-$Fe_2O_3$ tubes are qualitatively comparable to what is observed for anodic $TiO_2$ NTs, general guidelines can be summarized as follows:

- electrolyte composition: tube growth in $H_2O$-based electrolytes is limited to a maximum length of ~ 2 – 3 μm. For $TiO_2$, significantly thicker NT layers (up to some hundreds nanometers) can be grown in organic-based electrolytes (ethylene glycol, glycerol, *etc.*). Also the fluoride concentration and the $H_2O$ content play a crucial role, that is, the growth of thicker layers requires higher $F^-$ concentration[130,132] while limited $H_2O$ contents lead to highly ordered tubes.[133,134]

- electrolyte temperature: the higher the temperature, the faster the tube growth.[135,136] Rapid tube growth can also be achieved in the presence of complexing agents (lactic acid) in the electrolyte.[137]

- anodization voltage: the larger the anodization voltage, the longer and wider the tubes. Long tubes are also grown by extended anodization time.[134]

A final aspect common to both α-$Fe_2O_3$ and $TiO_2$ is that as-formed tubes are amorphous and adequate thermal annealing (mostly in air at T ~ 400–700 °C) is required to convert the non-crystalline arrays into hematite and anatase (or rutile, or anatase/rutile mixed phases), respectively.[48,131]

### 3.1.5 Other methods

A range of other approaches to form α-$Fe_2O_3$ and $TiO_2$ nanotubes/nanofibers have been reported.[44,48] Among these, for instance, electrospinning represents a simple and versatile method for generating ultrathin fibers made of various materials (*i.e.*, functional polymeric materials such as hierarchical nanofibers, inorganic-doped hybrid nanofibers, core–shell nanofibers, *etc.*).

Moreover, if combined to other techniques, in particular sol-gel methods and annealing, electrospinning can be used to fabricate a large number of inorganic nanomaterials with controlled morphologies and properties.[138]

In this process, a strong electric field is used to pull a thin jet out of a drop of polymer solution or melt through a nozzle. The jet then is deposited in the form of a nanofiber. Ti and/or Fe precursors can be used to coat the fibers (most elegant is a simultaneous coating of the fibers while spinning, through a coaxial two-capillary spinneret nozzle system) and generate single hollow-nanofibers (nanotubes) that are well separated and can be uniformly distributed over a several centimeters range. Finally, annealing is carried out to induce thermal degradation of the organic fiber template and crystallize the amorphous metal oxide nanotubes (Fig. 3.3(a,b)).[139]

Typically, NTs by electrospinning exhibit extremely high aspect ratios and a diameter that ranges from a few tens of nanometers to a few tens of micrometers.

$TiO_2$ nanotubes are fabricated, for example, by using titanate as well as Ti-alkoxide (Ti(IV) iso-propoxide) precursors,[139–141] while iron inorganic salts (*e.g.*, iron nitrate) or organic precursors (*e.g.*, Fe(III) acetylacetonate) are used for α-$Fe_2O_3$[142,143] – examples of such nanotubes are shown in Fig. 3.3(c).

Key to the fabrication of these hollow nanostructures is a use of two immiscible liquids, that is, the solvent to dissolve the Ti or Fe precursor (*e.g.*, an ethanol-based solution) and the polymer to fabricate the nanofibrous array (*e.g.*, a heavy mineral oil) – Fig. 3.3(c).[139]

In addition, it is possible to use this method to obtain nanofibers/nanotubes, with specific surface topologies. Since electrospinning to a first approximation involves fast evaporation of the solvent, use of a solvent that evaporates rapidly (*e.g.*, dicholoromethane) induces rapid phase separation and the subsequent rapid solidification generates fibers as those reported in Fig. 3.3(d), with pores of regular shape and narrow size distribution (diameter ~ 100–200 nm).[144]

## 3.2 Typical modification approaches

Experimental conditions and parameters for the modification of α-Fe$_2$O$_3$ and TiO$_2$ may be significantly different. However, for both oxides two main strategies can be distinguished, namely methods that modify the intrinsic electronic properties of the oxide (charge mobility, donor density, bandgap, *etc*.), and methods that provide a co-catalytic effect for chemical reactions (*e.g.*, deposition of co-catalyst to improve the kinetic of water oxidation). In addition, also other methods are reported that enable for instance visible light PEC-WS and/or generate suitable surface states for enhanced charge transport/transfer properties (*e.g.*, surface decoration with plasmonic metals and quantum dots, or the formation of a heterojunction).

### 3.2.1 Doping

Semiconductors for practical applications are often modified (doped) by introducing a secondary active species into the lattice. In particular, doping is carried out to (i) improve the material electronic properties, by adjusting a desired conductivity (typically, at low dopant concentrations), or to (ii) adjust (that is, in the case of TiO$_2$, to narrow) its bandgap and extend the light absorption properties (usually at high dopant concentrations).

The incorporation of dopants into TiO$_2$ and α-Fe$_2$O$_3$ can be achieved through a range of different methods, *e.g.*, hydrothermal procedures, thermal diffusion of ions from the substrate (FTO), drop casting, wet chemical treatments, thermal treatments under controlled atmospheres (containing the dopant), high-energy ion implantation.[44,48,49,51]

For TiO$_2$, bandgap narrowing is necessary in view of enhancing its visible-light absorption and promoting photoelectrochemical and photocatalytic reactions under solar irradiation.

Since the report by Asahi on N-doped TiO$_2$ and the photocatalytic degradation of methylene blue and acetaldehyde under visible light,[145] there have been (and still there are) a large number of attempts to produce differently doped $TiO_2$. This is typically achieved by introducing a non-metal element into its lattice, such as C, N, S, P, *etc*.[146–152]

The presence of non-metallic dopants forms electronic intra-gap states and this influences the oxide light absorption characteristics by leading to a redshift in the light absorption threshold.

However, more recently, some studies underlined the difficulty in definitively assigning a photoresponse under visible light to the influence of anion dopants. In some cases, the anion dopant may be confined to the surface and promote the formation of surface trap states that typically lead to detrimental charge recombination, which in turn limits the photoactivity of the material.[153]

Also some metal dopants such as Cr, V and Fe were reported to be successful in activating a $TiO_2$ response under visible light.[154] However, the photocurrents for, *e.g.*, Fe- or Cr-doped $TiO_2$ are typically low, and a use of cation doping to improve the visible light absorption of wide-bandgap semiconductors is only partially considered in photoelectrochemistry.[44]

Finally, an aspect that is often overlooked in the literature is that to an extended light absorption to visible light wavelengths (that is, determined by reflectivity measurements) does not always correspond the generation of mobile charge carriers. This was deeply investigated by Murphy, who pointed out that the effects on vis light induced PEC activity of *e.g.*, $TiO_2$ should be anyway considered valid only if IPCE spectra that demonstrate this ability are provided. Even more, a most reliable estimation of photoresponse is provided if photocurrent spectra are associated to photocurrent transients measured under monochromatic $\lambda$.[11,155]

Another promising way to improve the photoelectrochemical properties of $TiO_2$ is to improve its charge transport abilities, therefore reducing charge carrier recombination. This can be achieved, *e.g.*, by introducing a metal dopant (electron donor) such as Ru, Nb or Ta into $TiO_2$ lattice.[156–158] The amount of dopants is usually limited to 1–2 at.% that corresponds to a concentration of ~$10^{21}$ cm$^{-3}$. Higher concentrations are not likely to be effective, and may even lead to segregation of the dopant phase.

Finally, it is well established that, regardless of their non-metallic/metallic nature, some dopants such as F and Nb also inhibit the transition from anatase to rutile when annealing TiO$_2$ at relatively high temperature (> 400 °C) – rutile is detrimental in photoelectrochemical applications due to its poor electronic properties compared with anatase.[159–161] The presence of such dopants retards the seeding of the rutile phase and, therefore, allows annealing at relatively high temperature. This in turn increases the crystallinity degree of TiO$_2$ and enhances the charge transport properties as electron transport becomes considerably faster.[162,163]

On the other hand, α-Fe$_2$O$_3$ exhibits absorption properties that already fit to a solar spectrum. That is, light absorption of hematite extends up to λ ~ 600 nm and any doping attempt is clearly performed with the only attempt of increasing its intrinsically limited conduction properties that are inadequate for PEC applications.[17]

This is commonly achieved by substitutional doping using dopants with oxidation state +4 to +6, such as Ti$^{4+}$, Sn$^{4+}$, Zr$^{4+}$, Nb$^{5+}$.[164,165] By substituting at sufficient levels, high carrier conductivities can be attained.

In general, titanium is less effective compared to other +4 ions as it may form Ti$_{surf}$ sites that may also act as electron–hole recombination centers. However, it has also been recently reported that Ti$^{4+}$ surface species might effectively capture and store photogenerated holes and facilitate their transfer to the electrolyte for O$_2$ evolution by water oxidation.[166]

Zr$^{4+}$ doped into α-Fe$_2$O$_3$ single crystals led to one order of magnitude higher donor densities (in the order of $10^{19}$ cm$^{-3}$), and increased conductivity and electron mobility compared to undoped hematite.[167]

Also for hematite, optimum impurity concentrations are *ca.* 1 at.% or less and, while introducing inter-bandgap energy states that are beneficial for α-Fe$_2$O$_3$ conductivity, doping does not significantly alter the oxide bandgap or its absorption properties.[17]

### 3.2.2 Heterojunction engineering

The recombination of photopromoted charge carriers in solid-state semiconductors typically takes place within μs, while the time for electrons and holes to react with acceptor and donor species, respectively, is in the ns–μs.

The time scale for charge transfer to a donor or acceptor molecule depends very much on its specific nature, in particular, on its red-ox potential and chemical adsorption to the surface. It is, therefore, apparent that charge recombination and charge transfer to the electrolyte are competing processes.[50,168–171]

One of the most common strategies to tackle the rapid charge recombination and thus to increase solar conversion efficiencies is increasing the charge spatial separation by creation of a heterojunction.

Heterojunctions are typically designed to combine the functionality of the two (or more) constituent phases, with the aim of addressing (at least one of) the primary requirements for an efficient PEC-WS using solar energy: visible light activity, chemical stability, appropriate band-edge characteristics, and potential for low-cost fabrication.[172]

Heterojunctions may be formed by combining, *e.g.* (i) two semiconductors (S–S), or (ii) a semiconductor and a metal (S–M).

Enhancement of charge separation can be achieved through various mechanisms that mainly depend on the electronic properties of the partner materials:

(i) via the formation of an internal electric field at the heterojunction interface, in S–S type heterostructure. Upon favorable band bending, e$^-$ will tend to move from the higher to the lower lying CB, while h$^+$ will follow the opposite direction (*i.e.*, from the lower to the higher lying VB).

(ii) driven by the Schottky barrier that typically forms when interfacing a semiconductor with a metal (in S–M type heterojunction). This promotes electron flow from the material with the higher Fermi level to that with a lower Fermi level, until steady-state equilibrium is reached.

In details, different configurations may apply to the S–S type heterojunction: both semiconductors are photoactive, that is, they generate e$^-$/h$^+$ pairs (*e.g.*, Fe$_2$O$_3$-WO$_3$, TiO$_2$-WO$_3$, TiO$_2$-BiVO$_4$, Fe$_2$O$_3$-SrTiO$_3$, TiO$_2$-ZnO, *etc.*)[74,173–177] or only one component absorbs light and generate energetic charge carriers (*e.g.*, metal oxide/QDs, metal oxide/OEC). It is also remarkable that narrow bandgap semiconductors can be combined to the larger bandgap semiconductors and complementary light absorption properties can be exploited (*e.g.*, NiO/TiO$_2$ systems). Therefore, an additional advantage is the photosensitization of the large bandgap material, as observed for instance in metal oxide/QDs configurations (see below).

A detailed description of heterojunction formation and working principles is out of the scope of this review and additional information can be found elsewhere.[1,178,179] In the following sections we will provide examples and describe strategies that apply to α-Fe$_2$O$_3$ and/or TiO$_2$ for photoelectrochemical applications.

### 3.2.3 Decoration with a co-catalyst

A widely investigated strategy towards higher PEC ability is the surface decoration of photoanodes with an oxygen evolution catalyst (OEC) and/or the decoration of cathodes with a hydrogen evolution catalyst (HER, such as MoS$_2$ and WS$_2$[180,181]).

As discussed in section 1.1, the oxygen evolution reaction (OER) kinetics are sluggish: this reaction requires a four-electron oxidation of two water molecules (that is, the concerted transfer of four holes from the photoanode surface to the electrolyte) coupled to the removal of four protons to form a relatively weak oxygen-oxygen bond.

If $O_2$ evolution from water splitting represents only a minor challenge for $TiO_2$-based photoelectrode, the poor match between the orbitals of hematite valence band (that partially exhibits Fe 3d character) and those of $H_2O$ (that is, O 2p states – see section 3.3 for discussion) significantly limits the efficiency of hematite in water photooxidation.

To overcome the limitation of poor OER kinetics, various catalysts have been coupled to hematite photoanodes to assist hole transport from the hematite surface to the electrolyte.

Typical OEC are noble metals such as $IrO_2$ and $RuO_2$. Though efficient and stable under working conditions, Ir and Ru based catalysts are however preferentially replaced with cheap and earth-abundant elements for a significant reduction of the costs.

In this context, Co oxide-based catalyst[182] and the more recent amorphous cobalt-phosphate (Co–Pi)[183] catalysts represent valuable candidates. While the mechanism behind the improvement obtained with Co–Pi is still not completely understood, the electrocatalytic activity of cobalt oxides/hydroxides for water oxidation is well established and involves the $Co^{II}/Co^{III}$ and $Co^{III}/Co^{IV}$ couples and the formation of high-valent Co(IV)–O intermediates that can "store" up to four holes from the valence band of hematite. Holes injection into the electrolyte for the generation of $O_2$ then regenerates again $Co^{II}$ centers.[182]

Finally, recently also more complex structures (*e.g.*, NiFe double-layered hydroxide (LDH) and NiFe oxide)[184,185] have been introduced. NiFe based co-catalysts have been shown to possess excellent electrocatalytic activity and durability for OER catalysis; even more, although the active site of NiFe still remains unclear, its transparency towards visible light irradiation (that can therefore be absorbed by hematite) represents an intrinsic advantage, particularly with respect to more common cobalt-based co-catalysts.[184,185]

### 3.2.4 Other strategies

Aside from doping and surface-decoration with a co-catalyst, a number of various other approaches have been developed to improve the photoelectrochemical response of α-$Fe_2O_3$ and $TiO_2$. Some of these methods are here described, but a more complete overview can be found in 3.3.3 and in 3.4.3 for hematite and titania, respectively.

In view of the "too" large optical bandgap of $TiO_2$ for efficient sunlight absorption, the recent finding of "black" $TiO_2$ by Chen and Mao[186] is often mentioned as a promising approach to (partially) address this issue.[93,187–189]

Black $TiO_2$ is typically obtained by exposing $TiO_2$ to a reduction treatment (*e.g.*, high pressure/high temperature treatment in $H_2$, vacuum annealing, electrochemical reduction, Ar/$H_2$ annealing under atmospheric conditions) that produces the typical dark coloration ascribed to the formation of oxygen vacancies or $Ti^{3+}$ defect states.[190] This in turn generates a strong light absorption in the visible range.

In particular, Chen and Mao reported that compared to conventional "white" $TiO_2$, black $TiO_2$ loaded with a Pt co-catalyst could reach a significantly higher open-circuit water-splitting activity, with the photoactivity enhancement ascribed to a thin amorphous $TiO_2$ hydrogenated layer encapsulating the $TiO_2$ anatase core.

However, the findings of Mao and many of the follow-up studies seem to limit the outstanding properties of black $TiO_2$ to its visible-light extended photoabsorption.

Only more recently, a unique co-catalytic effect towards efficient water splitting could be observed for "black" $TiO_2$ anatase nanoparticles and nanotubes treated with $H_2$ under high pressure[191,192] or high-energy proton implantation.[193] Such a co-catalyst effect enabled strongly enhanced photocatalytic activity for hydrogen production in the absence of any noble metal co-catalyst. A combination of different characterization techniques pointed out that hydrogenation generates voids in the anatase nanoparticles[191,192] and nanotubes (an additional discretization of walls on the length scale of around 10−20 nm was observed for the implanted tubes)[193] that are responsible for the observed outstanding photoactivity.

Therefore, "black" titania not only should be regarded as a tool to extend the absorption properties of $TiO_2$ into the visible light region, but hydrogenation should also be considered a powerful tool for the engineering of suitable structural defects that can activate $TiO_2$ for noble metal-free photocatalytic water splitting.

Thermal reducing treatments have also been explored for hematite. However, the reduction of α-$Fe_2O_3$ crystals is typically reported to generate a $Fe_3O_4$ surface layer which typically acts as a recombination center.[194] For instance, the exposure of α-$Fe_2O_3$ pellets to an inert He-atmosphere produced conductive pellets that are also visibly darker than untreated hematite, but a photocurrent drop due to the enhanced surface recombination promoted by $Fe_3O_4$ is observed.

More recent studies suggested that a low amount of oxygen vacancies in α-$Fe_2O_3$ may be beneficial for solar water splitting, as long as no lattice distortions are formed and no incorporation of $Fe_3O_4$ inclusions occurs.[195]

Another popular approach for the adjustment of α-$Fe_2O_3$ and $TiO_2$ optical properties, namely for enhancing their light absorption, is the incorporation of a plasmonic metal nanostructure (*e.g.*, nanoparticles, nanorods, nanopillars, nanohole arrays) on the semiconductor surface.[196–203]

Briefly, a plasmonic metal (such as Ag, Au and Cu) can harvest visible light due to localized surface plasmon resonance (LSPR), *i.e.*, the oscillation of free electrons with the incident radiation that promotes either electron injection (hot $e^-$) or energy transfer (PRET) into the adjacent semiconductor. In other words, the plasmonic metal can be thought as an antenna that absorbs light, and the semiconductor as the reaction center that promotes the photoreaction.

Since the resonance frequency of plasmonic metals depends (primarily) on the nature of the metal and can be tuned from visible to near infra-red wavelengths by adjusting the nanostructure shape and size, LSPR can be envisaged as a strategy for the sensitization of semiconductors below their bandgap and/or to increase the range of solar light absorption.[202–204]

For solar water splitting applications, studies have been limited to Ag and Au (LSPR at λ = 400–450 nm and 530–560 nm, respectively). This is mainly due to the relatively large optical cross sections of the two metals and, in particular for Au, due to its stability under the harsh conditions (pH, potential) that are needed for water splitting.[200]

Early studies on water splitting with plasmonic-metal/semiconductor photoelectrodes report Au/TiO$_2$ and Ag/TiO$_2$ nanocomposites, synthesized by immersion of mesoporous TiO$_2$ films in solutions containing nanoparticles of either Au or Ag.[205,206]

Photocurrent and photovoltage action spectra of Au/TiO$_2$ and Ag/TiO$_2$ showed an increased photoresponse compared to a pure TiO$_2$ photoelectrode, that also matched the LSPR of the metal nanoparticles (*i.e.*, λ = 450 and 550 nm for Ag and Au, respectively). Although the increase in the absolute value of IPCE was modest (~ 1–2 %), these works paved the way for follow-up studies.

In particular, more recently, plasmonic-metal/semiconductor arrangements other than the more conventional "plasmonic-nanostructure on metal oxide-array" have been developed, *e.g.*, by using a thin Ag mirror underneath a α-Fe$_2$O$_3$ film, the photocurrent of hematite could be significantly enhanced due to the metallic film promoting photon re-trapping schemes that increase resonant light trapping into the oxide film.[207]

Further improvements in terms of water splitting efficiency have also been reported for a plasmonic/semiconductor/mirror photoanode arrangement, with TiO$_2$ "sandwiched" between an Au/Cr-based metallic mirror and plasmonic gold nanodisks.[208] The optimized architecture exhibited multiple resonance peaks due to hybridized LSPR in the nanodisks and multiple internal photoemission between the mirror and the plasmonic nanostructures that enabled an 8–10 times photocurrent enhancement under resonant conditions.

Besides sensitization by plasmonic resonance, also quantum dots of semiconductors such as CdS, CdSe, PbS, PbSe, InAs, $MoS_2$, *etc.* can be used as light harvesting functionalities and combined to a semiconductor photoelectrode for solar water splitting.

Quantum dots (QDs) are semiconductor nanoparticles (with $d \sim 5$–10 nm) subjected to quantum-confinement effect, *i.e.*, their dimension is smaller than the exciton Bohr radius. Above mentioned semiconductor QDs can absorb visible light ($E_g \sim 2.2$–2.4 eV) and generate photoexcited states (electron–hole pairs) that once separated can be used to enable redox processes with suitable acceptor and donor species.

In our context, the electron acceptor is the conduction band of a (large) bandgap semiconductor (*e.g.*, $TiO_2$) – clearly, for an efficient charge injection, a suitable alignment of the CB of the QD semiconductor with that of *e.g.* $TiO_2$ is a prerequisite.[209–211]

Moreover, some QDs (*e.g.*, PbSe and CdS-CdSe dual QDs) allow charge recombination that leads to the emission of photons that are then absorbed by the semiconductor where they generate reactive $e^-/h^+$ pairs, according to the so-called up-conversion photoluminescence process, where the emitted λ is lower (more energetic) than absorbed λ.[212,213]

Numerous studies of QDs show that their light absorption/emission features can be controlled adjusting their particle size.[43]

However, in spite of the large number of studies on QD-sensitized semiconductor scaffolds for light conversion reactions and of the significant IPCE enhancement (*e.g.*, for $CdS/TiO_2$ an IPCE as high as 20% has been determined at $\lambda \sim 450$ nm)[210,214], two main limitations still remain to be addressed: (i) the slower interfacial hole transfer rate (to a donor species, *e.g.* in the electrolyte) compared to electron injection into the semiconductor; that is, for the efficient

II–VI materials, QDs typically suffer of photochemical instability due to hole-induced corrosion; (ii) a low internal quantum efficiency in the visible region, ascribed to a slow hole transfer kinetics compared to the rate of surface-mediated electron–hole recombination.

Recently, carbon quantum dots (CDs) have been introduced and regarded as promising materials due to their lower toxicity and eco-friendly properties compared with traditional metal sulphide and selenide-based QDs.[215] In addition, in view of PEC-WS applications, two other characteristics of CDs are beneficial for enhancing solar energy conversion efficiency: first, the electrons in the excited energetic level of CDs can be efficiently injected into the conduction band of *e.g.* $TiO_2$ and α-$Fe_2O_3$, and also the holes left in the ground state have a redox potential to oxidize a suitable donor species in the environment. Thus, a co-catalytic effect towards the oxygen evolution reaction can be exploited. Also, the relatively weak intensity of fluorescence from "naked" carbon nanoparticles can be easily addressed by surface passivation with organic molecules or polymeric species – in this case, a main concern remains however the photostability of the surface ligands.[216,217]

## 3.3 α-Fe$_2$O$_3$ for photoelectrochemical applications

In its fully oxidized form, Fe$_2$O$_3$ exists as crystalline as well as amorphous material. In particular, crystalline Fe$_2$O$_3$ may assume one of the four α, β, γ, ε possible forms.

α-Fe$_2$O$_3$ (hematite) is the most thermodynamically stable polymorph of iron oxide under ambient conditions. It crystallizes in a rhombohedral lattice system, with lattice parameters $a$ = 5.0356 Å and $c$ = 13.7489 Å, and exhibits the same crystal structure as corundum (Al$_2$O$_3$). That is, O$^{2-}$ atoms arrange in a hexagonally close-packed lattice along the [001] direction and Fe$^{3+}$ atoms occupy octahedral interstices in the (001) basal planes.[218]

A combination of hexagonally packed oxygen atoms with interstitially positioned iron atoms results in a highly dense structure ($d$ = 5.26 g cm$^{-3}$), with high polarizability and, hence, high refractive index. This in turn may lead to complex behavior while interacting with photons and electrons.

In particular, the electronic nature of the bandgap in hematite is of great interest to understand its performance as a material for solar energy conversion, and Tauc analysis of the bandgap absorption onset has shown an indirect (phonon-assisted) bandgap nature of the material.[219] Ab initio calculations by the Hartree-Fock approach [220] and density functional theory approximation,[221] also supported by soft X-ray absorption and emission spectroscopies at the O K-edge,[222] showed that the valence band of α-Fe$_2$O$_3$ primarily exhibits a O 2p character, with a significant contribution from the Fe 3d orbitals (much higher compared to other metal oxide semiconductor such as TiO$_2$).[223]

Photon absorption of hematite starts in the visible-near infrared region at λ > 650 nm (with low absorption coefficient α ~ 10$^3$ cm$^{-1}$), that is well below its optical bandgap (*i.e.*, E$_g$ ~ 2.1 eV that corresponds to λ ~ 550–580 nm). This is most likely due to spin-forbidden d–d transitions associated to Fe$^{3+}$ ions ($Fe^{3+} + Fe^{3+} \xrightarrow{h\upsilon} Fe^{4+}(h^+) + Fe^{2+}(e^-)$).[224] However, although increasing its conductivity, photoexcitation of hematite at these wavelengths does not produce any significant photocurrent in PEC systems and confirms that, in view of a measurable photocurrent, efficient $e^-/h^+$ pairs are associated to a charge transfer process of the type $O^{2-} + Fe^{3+} \xrightarrow{h\nu} O^-(h^+) + Fe^{2+}(e^-)$, that is, an O (2p)–Fe (3d) indirect transition.[17,44]

The first report on the use of α-Fe$_2$O$_3$ as a photoanode material for splitting water dates back to 40 years ago, when Hardee and Bard produced hematite thin films on Ti or Pt foils by chemical vapor deposition and observed a photocurrent from λ = 550 nm, under 0.8 V vs. SCE in a KCl-based electrolyte.[15]

Later works explored crystal faceting of α-Fe$_2$O$_3$,[225] and n-type doping with elements from the IVa (*i.e.*, Si, Ge, Sn, and Pb)[226–228] and IVb (*e.g.*, Ti and Zr)[229] groups as a mean to ameliorate the performance of hematite by tackling its intrinsic low carrier mobility. In particular, the introduction of an extra $e^-$ in the conduction band of α-Fe$_2$O$_3$ by substitutional doping leads to higher photocurrent (under anodic bias).[228]

Particularly relevant is the work by Sanchez *et al.* who reported on Nb-doped (1.5 mol%) α-Fe$_2$O$_3$ single crystals with a quantum efficiency for water splitting as high as *ca.* 40% at 370 nm at +0.5 V vs. SCE.[230,231]

This value was only recently outperformed by Pt-doping single-crystalline α-Fe$_2$O$_3$ nanostructures. In particular, IPCE as high as 60–80 % in the 300–400 nm range was reported upon decoration of the Pt/α-Fe$_2$O$_3$ surface with a Co-Pi layer.[232]

Common to all these early studies is the conclusion that hematite exhibits a flat-band potential that cannot spontaneously reduce water to H$_2$ and, therefore, an external bias is needed to complete the water splitting reaction. Not only this, but also a few other critical points have to be addressed when considering α-Fe$_2$O$_3$ as photoanode material for PEC-WS applications, namely low water oxidation (OER) kinetics, low light absorption coefficient, and short diffusion length of holes.

Poor OER kinetics, which translates into a relatively high overpotential for the anodic reaction, is related to two different factors:[17,44] (i) the presence of oxygen vacancies and crystalline disorder on the surface of hematite that, by trapping and accumulating $h^+$ at the interface, may lead to Fermi level pinning, and (ii) the increased $Fe^{3+}$ character of the valence band that results in the energy mismatch between d orbitals (where the hole moves) and the p orbitals of *e.g.* the hydroxide donor species in solution (*i.e.*, the acceptor) resulting in a slow charge transfer.

One of the most commonly adopted strategies to overcome surface crystalline disorder is the decoration of α-$Fe_2O_3$ with thin surface passivation layers that reduce ($h^+$ trapping) surface defects and limit hole back injection (see below).[233–237]

In addition, the use of a co-catalyst (*e.g.*, Co-Pi and $IrO_2$, see below) is known to (i) increase the kinetic of hole transfer to the electrolyte, limiting their accumulation at the interface and therefore leading to an overall higher photocurrent, (ii) reduce the photocurrent-onset overpotential and enhance the overall energy conversion efficiency, and (iii) extend the long-term stability of the photoanode under operating conditions.

More recent efforts have been devoted to optimizing the hematite-based electrode morphology by nanostructuring to address the intrinsic low absorption coefficient and short hole diffusion length of α-$Fe_2O_3$. However, as described in the next paragraphs, a combination of all mentioned strategies (*i.e.*, doping, surface decoration and nanostructuring) is key to an ideal hematite photoanode that features an optimal material composition/structural configuration and therefore a maximized PEC ability.

In the next sections we will primarily focus on one-dimensional hematite-based nanostructures for improved PEC performance, along with examples of the most promising and rationally designed architectures obtained by a tailored combination of above mentioned strategies. Table 1 provides some examples of different synthetic/modification approaches to 1D α-Fe$_2$O$_3$ nanostructures and their PEC performances.

### 3.3.1 Doping to an enhanced conductivity

As briefly outlined in previous sections, α-Fe$_2$O$_3$ suffers from a very low electrical conductivity (~ 10$^{-14}$ ohm$^{-1}$), low carrier density (10$^{18}$ cm$^3$), and limited electron mobility (10$^{-2}$ cm$^2$ V$^{-1}$ s$^{-1}$).[238,239] These low values result all from a Fe$^{3+}$/Fe$^{2+}$ valence alternation on spatially localized 3$d$ orbitals.

Concerning conductivity, empirical data well match the theoretical predictions of a conduction mechanism based on small polarons' hopping for both electron and hole transports (Fig. 3.4(a)).[240] In particular, the extremely low mobility of electrons leads to severe energy loss due to electron–hole recombination.

Doping hematite with a metallic element, either n-type doping (*i.e.*, with ions with a net charge higher then 3+) or p-type doping (*i.e.*, with elements in their I or II oxidation state), is a practical solution to these limitations.[240–242] DFT calculations showed that transition metals (TMs, *e.g.*, Ti, Cr, Mn, Ni, *etc.*) can provide a dual functionality, namely not only TMs enhance the electron conductivity, but they can also influence other parameters, such as morphology, stoichiometry, crystallinity, lattice symmetry, grain size, *etc.*[243]

For instance, addition of titanium carbonitride (TiCN) as Ti precursor during the hydrothermal growth of α-Fe$_2$O$_3$ changes the morphology from porous α-Fe$_2$O$_3$ to Ti-doped 1D urchin-like nanostructures and leads to a photocurrent density of 1.9 mA cm$^{-2}$ (measured at 1.23 V vs. RHE under Am 1.5 and in 1 M NaOH).[244]

Combining Ti-doping with the formation of oxygen vacancies introduced in hematite by controlling the oxygen pressure during α-Fe$_2$O$_3$ nanorod growth has been reported to lead to Ti-doped α-Fe$_2$O$_3$ with a coral-like morphology that exhibit a significantly larger surface area with respect to the undoped α-Fe$_2$O$_3$ case. The high PEC performance obtained (2.25 mA cm$^{-2}$ and 4.56 mA cm$^{-2}$ at 1.23 V and 1.6 V vs. RHE, respectively) was attributed to the synergistic effect of the high specific area of the nanorod/nanocoral morphology and the increased donor density, due to oxygen vacancies and Ti doping (Fig. 3.4(b,c)).[245]

Clearly, one of the most critical aspects to take into account when introducing a dopant element into a "host" structure is the identification of the optimal dopant amount, in order to achieve the desired effects (*e.g.*, increased donor density and charge conductivity), with a minimum of side effects.

For α-$Fe_2O_3$, for instance, an excess amount of Ta has been reported to negatively affect nanorod growth along the [110] direction, with a consequent drop of the material overall PEC performance[246] – remarkably, the (110) plane of hematite with its surface termination dominated by Fe(III) ions features high conductivity.[182]

A similar effect was also observed for an excess amount of Nb dopant, as reported in Fig. 3.4(d).[247]

A significant lattice cell reduction, which in turn promotes larger probability for localized charge carrier (polaron) hopping and thus increases the conductivity, has been calculated for Mn-doped α-$Fe_2O_3$ (Fig. 3.4(e)).[248,249] Mn-doped hematite nanorods (with a dopant content in the 5–7 mol.% range) has been fabricated through hydrothermal methods and photocurrent density as high as 1.6 mA $cm^{-2}$ at 1.23 V vs. RHE has been reported.

Also Pt and Ru have been considered as suitable TM dopants for α-$Fe_2O_3$. Guo *et al.* recently developed a synthetic procedure to dope ultrathin α-$Fe_2O_3$ nanorods with Ru cations.[250] α-FeOOH nanorods coated with a ruthenium(III) acetylacetonate-based layer were deposited on FTO substrate via doctor blading technique; annealing in air 700 °C transformed the amorphous nanorods into crystalline hematite NRs and produced Ru nanoparticles with different valence states (both +4 and +3). Such photoanodes showed a photocurrent density as high as 5.7 mA $cm^{-2}$ that was attributed to (i) an enhanced carrier transport promoted by tetra- and lower-valent Ru cations and to (ii) the absence of intra-gap trapping states in α-$Fe_2O_3$.

A most straightforward strategy for doping α-Fe$_2$O$_3$ nanostructures grown on FTO glass is the thermally-induced diffusion of Sn from the substrate, while annealing at temperatures above 750 °C.

Among the first reports on "unintentional" Sn-doping of hematite layer on FTO is the study of Sivula *et al*.[16] The significant enhancement of photocurrent observed upon annealing at 800 °C was attributed to a two-fold enhancement of the optical absorption coefficient, due to the structural distortion of the hematite lattice induced by tin-doping.

A similar strategy was applied by Ling *et al*.[114] who reported a photocurrent density of 1.24 mA cm$^{-2}$ at 1.23 V vs. RHE and attributed the enhanced PEC performance to an increased donor density (*i.e.*, from $1.89 \times 10^{19}$ cm$^{-3}$ and $5.38 \times 10^{19}$ cm$^{-3}$ for undoped and Sn-doped hematite, respectively) and hence higher carrier conductivity.

From these early studies it is apparent that to achieve a desired functionality associated to Sn-doping, high temperature annealing is required; however, annealing FTO at 800 °C also leads to a significant change in the material stoichiometry, already after 20 min, and typically doubles the tin oxide resistivity – a 10 min annealing was reported to be optimal to establish a trade-off between a desired increase in conductivity and the drawbacks related with treatment at high T (Fig. 3.4(f)).[251] Moreover, another side effect associated with high T treatment is the collapse of α-Fe$_2$O$_3$ nanostructure morphology. Clearly, both effects have a negative impact on the photoanode overall PEC performance.

Therefore, different techniques including *e.g.* solution-based approaches, microwave hybrid annealing or chemical encapsulation have also been reported that overcome the requirement of a high temperature annealing.[114,252,253] For instance, efficient Sn doping of hematite was achieved starting from hydrated tin (IV) chloride as the metal precursor. As for Ti-doped α-Fe$_2$O$_3$,[244] a morphology change from nanowires for undoped hematite to nanocoral-like structures for Sn-hematite was observed, together with a surface area increase and an improved electrical conductivity that were identified as dominant factors leading to a higher photoactivity.

In situ doping of α-Fe$_2$O$_3$ nanotube layers was investigated by Lee *et al.* and Chitrada *et al.*[254,255] Self-organizing anodization on Fe-Si alloys with various Si content (1, 2, and 5 at.%) led to the simultaneous formation of iron oxide-based NT arrays and to the incorporation of Si. The introduction of Si decreased the layer growth rate and led to a more pronounced nanotubular morphology.

In particular, hematite layer with 5 at.% Si exhibited a 5-fold increase in the photocurrent, along with a negative shift of the onset potential compared to the undoped sample.[254]

Iron oxide based nanoporous layers were also synthesized by electrochemical anodization of Fe$_{14}$Nd$_2$B permanent magnet. Highly regular structure with a pore diameter ranging from 30 to 60 nm and a thickness of about 0.5 μm were produced, while incorporation of Nd$^{3+}$ and B$^{3+}$ resulted in a direct bandgap of 2.05 eV for doped iron oxide, and shifted the flatband potentials to –0.8 V vs. Ag/AgCl.[255]

Despite the abovementioned beneficial effects observed upon doping α-Fe$_2$O$_3$, the introduction of a single dopant into the oxide lattice may induce also some undesired side effects, such as the narrowing of space charge layer and the generation of impurity states within the oxide bandgap that act as recombination centers. A rational combination of two (or more) dopants can then be crucial for overcoming such drawbacks.

This has been *e.g.*, reported in the case of Sn$^{4+}$ and Be$^{2+}$ co-doped α-Fe$_2$O$_3$ nanorods. Sn$^{4+}$ alone, despite providing an enhanced charge conductivity, (i) introduces shallow n-type defect states beneath the conduction band minimum of hematite, thus leading to an undesired band edge shift, and (ii) increases micro-strain in the α-Fe$_2$O$_3$ lattice, decreasing the Fe–O bond ordering.

Nevertheless, the co-presence of $Be^{2+}$ ions attenuates these negative effects, by recovering α-$Fe_2O_3$ CBM and reducing lattice distortion in hematite – remarkably, the sequence in which the co-doping is carried out was also found to be very crucial. For the optimized Sn(4%)-Be(6%) containing α-$Fe_2O_3$ NR array, a maximum photocurrent density of 1.7 mA $cm^{-2}$ at 1.23 V vs. RHE has been reported.[256]

Similar effects have been demonstrated also for other co-dopant couples, including *e.g.* Ti-Sn,[257] and Ti-Si.[258]

Worth mentioning are also Zn-doped (*i.e.*, p-type doping) α-$Fe_2O_3$ NTs fabricated by means of electrodeposition.[259] Precursor ZnO nanorod array simultaneously served as sacrificial template and Zn source, while a $FeCl_2$ aqueous solution was used as the electrolyte for iron deposition. Under optimized conditions (1 $V_{SCE}$ anodic potential and 5–15 min reaction time), $Fe^{2+}$ to $Fe^{3+}$ oxidation occurs in the electrolyte, with trivalent-Fe ions that precipitate and form an amorphous r-FeOOH layer on the surface of ZnO nanorods. During this oxidation/precipitation process, ZnO is gradually dissolved due to the acidic environment and the applied anodic potential and some of the $Zn^{2+}$ ions are eventually incorporated as p-type doping species into the r-FeOOH layer. A positive flat band potential of 1.82 V vs. RHE and an acceptor density of $4.19 \times 10^{18}$ $cm^{-3}$ were determined for the optimally Zn doped α-$Fe_2O_3$ NTs array, together with a cathodic photocurrent of 40.4 μA $cm^{-2}$ measured at 0.5 V vs. RHE.

### 3.3.2 Co-catalyst decoration

A most effective approach to accelerate the oxygen evolution reaction (OER) kinetic in α-$Fe_2O_3$ by increasing $h^+$ transfer rate to the electrolyte is the modification of hematite with an oxygen evolution catalyst (OEC).

Tilley *et al.* demonstrated that $IrO_2$ nanoparticles (*ca.* 2 nm diameter) on hematite nanostructure photoanodes lead to a 200 mV cathodic shift in the onset potential and increase the photocurrent from 3.45 to 3.75 mA cm$^{-2}$,[260] while non-stoichiometric IrO$_x$ nanoparticles (with $x = ca.$ 2.1) on α-Fe$_2$O$_3$ were reported to decrease the overpotential for the water oxidation reaction to an even larger extent.[261] However, despite the fact that iridium oxide (IrOx) co-catalysts typically exhibit the highest performance in terms of OER activity,[260] only a few reports deal with IrO$_x$-decorated hematite for PEC applications, the main limitation being the high cost of Ir.

Other reported efficient co-catalyst materials for α-Fe$_2$O$_3$ photoelectrode include cobalt-phosphate (Co-Pi), Ni-Fe and Zn-Co based layered double hydroxides (LDH), FeOOH, NiOOH, *etc*.

In particular, Co-Pi on 1D α-Fe$_2$O$_3$ nanostructures improves the overall photoanode current density and shifts the current onset potential to more cathodic values, due to the bridging effect of phosphate ions that facilitate hole injection from α-Fe$_2$O$_3$ to Co ions.[233,234,262]

Several methods are reported to deposit Co-Pi on hematite nanostructure surface, such as chemical immersion deposition, electrodeposition and photoassisted electrodeposition.[233,235]

In particular, photoassisted electrodeposition of Co-Pi catalyst enables the formation of an optimized junction between hematite and the catalyst for a maximized interfacial charge transfer, with the amount of co-catalyst that can be easily controlled by deposition time and current density. For these reasons, such technique provides Co-Pi/α-Fe$_2$O$_3$ layers with superior photoelectrochemical water oxidation ability compared to both Co-Pi electrodeposition and Co$^{2+}$ wet impregnation methods.[263,264]

More recently, NiOOH, FeOOH, and Ni-Fe LDH and Zn-Co LDH layers have shown remarkable performance used in combination with α-Fe$_2$O$_3$ for photoelectrocatalytic water oxidation.[236,237,265] Several methods are known to deposit oxy-hydroxide and LDH layers onto hematite nanostructures, the most common being *e.g.* solvothermal techniques, electrodeposition or photoelectrodeposition and pulsed laser deposition (PLD).[266–269] For all these co-catalyst/α-Fe$_2$O$_3$ combination, a remarkable increase in photocurrent and decrease in water oxidation overpotential has been reported compared to naked hematite layers.

### 3.3.3 Other strategies

As previously mentioned (section 3.2.4), strategies that can be used to improve the PEC performance of 1D hematite photoanodes also include the formation of heterostructures, sensitization by quantum dots and plasmonic metal nanoparticles, the use of graphene derivatives for enhancing the photogenerated charge dynamics, *etc*.

Moreover, very often co-catalytic systems consisting of multiple components with different functions are coupled to enable a maximized PEC performance.

In the context of C-based co-catalytic materials, a multi-heterojunction assembly consisting of α-Fe$_2$O$_3$ (core)/reduced graphene oxide (rGO, interlayer)/BiV$_{1-x}$Mo$_x$O$_4$ (shell) has been investigated by Hou *et al.* as photoanode material for PEC water splitting.[270] According to the authors, the system provides multiple beneficial functionalities: the BiV$_{1-x}$Mo$_x$O$_4$ shell enhances light harvesting properties, while the rGO interlayer bridges the photoactive shell with the hematite core and enables shuttling of photogenerated electrons from BiV$_{1-x}$Mo$_x$O$_4$ to α-Fe$_2$O$_3$, significantly reducing charge recombination due to enhanced e$^-$/h$^+$ separation. For such a system, a negative shift of the onset potential has observed, along with an increased photocurrent density of 1.97 mA cm$^{-2}$ measured at 1.6 V vs. RHE.

A synergistic co-catalytic effect has been also described for carbon nanodots (CDots) and a Co$_3$O$_4$ layer, both deposited on α-Fe$_2$O$_3$.[271] In particular, the interaction between CDots and Co$_3$O$_4$ co-catalyst improves the PEC efficiency of pure α-Fe$_2$O$_3$ by accelerating the two-step-two-electron O$_2$ evolution form H$_2$O. That is, the kinetically slow process of water oxidation molecule to H$_2$O$_2$ that takes place at Co$_3$O$_4$ sites is significantly accelerated by the fast oxidation reaction of H$_2$O$_2$ to O$_2$, catalyzed by CDots.

Other efficient heterojunction partners for α-Fe$_2$O$_3$ include ZnFe$_2$O$_4$,[272] MgFe$_2$O$_4$[273] and Ag$_x$Fe$_{2-x}$O$_3$.[274]

The heterojunction based on Co-doped α-Fe$_2$O$_3$ nanorod array branched with a MgFe$_2$O$_4$ ultrathin layer has been reported by Hou *et al*.[273] A Co-doped α-Fe$_2$O$_3$ nanorod photoanode prepared by hydrothermal method is overcoated by an MgFe$_2$O$_4$ film using a wet impregnation method, in combination with thermal annealing. Due to the suitable band alignment between the two components, electrons photopromoted in the MgFe$_2$O$_4$ overlayer migrate to the conduction band of Co-Fe$_2$O$_3$ and along the Co-Fe$_2$O$_3$ nanorods to the back contact. At the same time, holes from the valence band of Co-Fe$_2$O$_3$ are transferred to that of MgFe$_2$O$_4$ where they react with water to form O$_2$.

A significant enhancement in PEC performance has also been reported for an inverse heterostructure consisting of ZnO nanowires (core)/ultrathin α-Fe$_2$O$_3$ (~ 5 nm, shell). This n/n heterojunction generates a negative shift in the flat-band potential and increases surface band bending, thus promoting charge carrier separation. Both these features resulted in a 2-fold enhancement of the PEC activity compared to that of a bare ZnO photoanode.[177]

Mao *et al.* reported on a template-assisted method for use of noble metals (such as Pt and Au) as conducting support to enhance charge conductivity in α-Fe$_2$O$_3$.[275] At first, Au nanorods are grown in an aluminum oxide template (AAO). Afterwards, Fe ions are electrodeposited onto the Au nanorods, followed by annealing in oxygen atmosphere. Upon removal of the AAO template, a photocurrent density of 8 mA cm$^{-2}$ was measured, with 10 μm long NRs. Following a similar procedure, Pt-supported α-Fe$_2$O$_3$ nanorods were also prepared that exhibited a photocurrent density of 10 mA cm$^{-2}$ at 1.5 V vs. RHE.

Clearly, for both Au- and Pt-supported heterostructures, a crucial contribution to such a high PEC performance is given by the high conductivity of the noble metals, although also their contribution towards an increased light absorption cannot be excluded.

Similar to the more commonly reported black $TiO_2$, also reduced α-$Fe_2O_3$ has been prepared by hydrogen plasma treatment of an α-$Fe_2O_3$ layer under ultra-pure hydrogen plasma.[276] "Black-like" hematite consists of mixed $Fe_3O_4$ and α-$Fe_2O_3$ phases, with the mixed layer Fe:O stoichiometry (due to the valence switching between $Fe^{3+}$ and $Fe^{2+}$) being crucial to increase the hematite donor density. This aspect, in combination with a reduced bandgap led to a considerable increase of photocurrent density and the most significant cathodic shift in photocurrent onset potentials (from 1.68 to 1.28 V vs. RHE upon hydrogen plasma treatment).

Finally, an example of a sequential-multifold approach applied to α-$Fe_2O_3$ nanorods to simultaneously reduce electron–hole recombination in the bulk oxide, promote electron transport to the back-contact, while simultaneously enabling fast hole injection into the electrolyte has been provided by Cho *et al* (Fig. 3.5).[277]

A flame-treatment in the presence of a Ti precursor can be used to produce Ti-doped α-$Fe_2O_3$ nanorods, with reduced bulk and surface charge recombination, without affecting the morphology of the hematite NRs. The addition of a dense-layer between the hematite array and the fluorine-doped $SnO_2$ substrate effectively reduced the interfacial recombination by suppressing electron back-injection. Finally, the deposition of a FeOOH on the surface of Ti-doped α-$Fe_2O_3$ nanorods passivated hematite surface states and provided a highly active interface for hole injection into the electrolyte. These sequential treatments led to an overall significant decrease of the photocurrent onset potential (*ca.* 400 mV).

## 3.4 TiO$_2$ for photoelectrochemical applications

Titanium dioxide crystallizes in three major different polymorphs, namely anatase (A), brookite (B), and rutile (R). In particular, anatase and rutile TiO$_2$ represent the crystalline forms more widely investigated for photoelectrochemical and photocatalytic applications.[190] It is commonly reported that, for bulk TiO$_2$, anatase into rutile transition occurs at annealing T > 600 °C, under ambient atmospheric conditions.[278]

While holding for TiO$_2$ nanopowders, this observation is however not always valid: as we will discuss in the following section, amorphous TiO$_2$ may even crystallize directly in the form of rutile or the anatase into rutile transformation may occur at lower temperature. These are the cases of *e.g.* hydrothermal TiO$_2$ nanorods that, grown at high pressure (~ 15–20 atm) and at T ~ 150–200 °C, typically nucleate with a rutile crystal structure, or of anodic TiO$_2$ nanotubes on Ti metal foil.[279,280]

Both anatase and rutile TiO$_2$ exhibit a tetragonal structure with lattice cell parameters $a = b = 3.782$ Å and $c = 9.502$ Å for anatase, and $a = b = 4.594$ Å and $c = 2.953$ Å, respectively.[190] In both structures, crystalline units consist of TiO$_6$ octahedra (*i.e.*, each Ti$^{4+}$ atom is surrounded by six O$^{2-}$ atoms), with a more or less pronounced distorted configuration: two long bonds between the (central) Ti atom and the O atoms at the apices and four shorter Ti–O equatorial bonds (*i.e.*, 1.966 Å *vs.* 1.937 Å in anatase, and 1.983 Å *vs.* 1.946 Å in rutile).[281] Despite having the same chemical composition, differences in the coordination environments and chemical bonding of anatase and rutile TiO$_2$ strongly affect the electronic properties of the material.

Early studies by Earle in 1942 demonstrated that TiO$_2$ is an n-type semiconductor, with a valence band that exhibits a strong O 2p character and a conduction band mainly composed of Ti 3d orbitals.[282]

In its amorphous form, $TiO_2$ has a mobility gap of 3.2–3.5 eV and a very low electron mobility. Upon crystallization to anatase or rutile an indirect optical bandgap of 3.2 eV and 3.0 eV is obtained, respectively, with both polymorphs showing a higher mobility than amorphous $TiO_2$.[283] In particular, a crystallization into anatase is often reported to be crucial to obtain a high electron mobility.[284]

Not only the crystal structure, but also oxygen vacancies and $Ti^{3+}$ states in the lattice largely influence the electronic and optical properties of $TiO_2$. Electrons may be trapped in $Ti^{3+}$ states (or holes in the oxygen vacancies) and injected in the nearby conduction (or valence) band by thermal excitation. Electron conductivity of $TiO_2$ is therefore dominated by $Ti^{3+}$ trap states and, in the polycrystalline materials, also by the grain size.[285,286]

Spectro-/photo-/electrochemical characterization[287,288] performed on anatase and rutile crystals showed that both polymorphs exhibit positive holes with adequate oxidative power to promote $H_2O$ oxidation to $O_2$ and mainly differ for the position of their conduction band edges; in particular, while the conduction band of rutile is slightly more positive than $E°(H^+/H_2)$, the flat-band potential of anatase is 0.2 V more negative than that of rutile. Therefore, splitting of water into $H_2$ and $O_2$ is in principle thermodynamically possible (that is, spontaneous) without external bias on anatase only. However, due to the slow kinetics of the cathodic HER, also a $TiO_2$ anatase-based photoanode is often assisted by an applied voltage that, inducing favorable band bending, increases charge separation.

The first study on the use of $TiO_2$ as electrode material in a photoelectrochemical arrangement dates back to 1968;[289] rutile (single crystal) reduced in $H_2$ atmosphere was shown to evolve $O_2$ from a phosphate buffered aqueous solution, under chopped UV illumination from a mercury arc lamp and 0–50 V vs. SCE applied bias. Only the anodic reaction was considered and tunneling of electrons from occupied levels at the surface directly into the empty conduction band was suggested to produce upward band bending near the surface, at the TiO$_2$ electrode.

However, only after the ground-breaking work of Fujishima and Honda in 1972 on the "complete" photoelectrochemical water splitting reaction (*i.e.*, into H$_2$ and O$_2$), TiO$_2$ has been regarded as the key semiconductor material for photoelectrochemical and photocatalytic applications.

This not only because of its intrinsic electronic and optical properties, but also because it is abundant, cheap, non-toxic and stable against (photo)corrosion – all relevant aspects when targeting the production of *e.g.* large-scale devices for energy conversion.

A large number of studies have been promoted to address one of the (still) biggest limitations of TiO$_2$, that is, its relatively wide bandgap (as mentioned, 3.0–3.2 eV) that rules out any possible use of λ > 400 nm for light-promoted reactions.

After various early findings [290,291] and, in particular, since the report by Asahi on the visible light activity of nitrogen-doped titania,[145] doping has been (one of) the *leitmotivs* for the fabrication of visible light active materials.

As discussed in the previous section (3.2.1), doping of TiO$_2$ is achieved by replacing the metal (titanium) or the nonmetal (oxygen) component with a nonmetal or a transition metal cation (*i.e.*, substitutional doping), or by filling in interstitial lattice spaces (*i.e.*, interstitial doping). Typical nonmetal dopants for TiO$_2$ nanomaterials are B, C, N, F, P, S, Cl, and Br;[81,145,152,153,292] on the other hand, Cr, V, W, Ta and Nb are commonly used as metal dopants.[293–297]

Along with the development of strategies for its sensitization, nanostructuring (in particular one-dimensional nanostructuring) of TiO$_2$ has also attracted tremendous interest in view of improving charge separation and collection in TiO$_2$-based (photo)electrodes.

In the following sections we will deal with one-dimensional TiO$_2$-based nanostructures,

that exhibit improved PEC-WS ability and that were obtained by a rational design of complementary strategies (doping, co-catalytic effects, and other (more recent) strategies) to tackle the specific (mentioned) limitations of the material – Table 2 reports some examples of different synthetic/modification approaches to 1D TiO$_2$ nanostructures and their PEC performances.

### 3.4.1　TiO$_2$ one-dimensional arrays

The beneficial characteristics of one-dimensional nanoarrays (*e.g.*, enhanced charge carrier separation, improved e$^-$ transport, large carrier collection efficiency, *etc.* – see Section 2.1) combined with the intrinsic electronic and optical properties of TiO$_2$ (see above) contributed to the development of 1D-TiO$_2$ based nanostructures for a large variety of (but not limited to) light-related applications: for photoelectrochemistry and photocatalysis, in Grätzel-type solar cells, for photovoltaic and electrochromic devices, *etc*.[47–49,298,299]

In many of these applications, the efficiency depends to a large extent on the available surface area, that is, on the surface area of the electrode that is accessible to both the electrolyte and the incident irradiation.

However, it has been observed that for nanoparticle and nanotube layers with the same thickness, although the effective surface area of the particles is higher than that of the nanotubes, the quantum efficiency for the nanotubes is higher (Fig. 3.6(a)).[300] Therefore, also other factors should be considered to play an important role in the photoelectrochemical characteristics of 1D-TiO$_2$ based nanostructures, *e.g.*, the lower number of surface states in nanotubes and the upward band bending.

Moreover, the photoelectrochemical performance of one-dimensional nanostructures is largely influenced by the geometrical aspect of the arrays, that is, it varies with the thickness and/or diameter of the nanostructures; overall, the different carrier mobility (in different structures and morphologies) seems to dominate the efficiency.

For TiO$_2$ nanotube arrays, the work of Lynch *et al.* outlined how the morphology of nanotubes affects their photoelectrochemical properties also in comparison to other TiO$_2$ nanostructures.[300] In particular, the IPCE efficiency of tubes was shown to increase with increasing the layer thickness up to a critical maximum length (*i.e.*, $d$ = *ca.* 5–6 μm for ethylene glycol-based nanotubes – Fig. 3.6(b)). When an e$^-$/h$^+$ pair is generated in the tube walls, holes are promptly transferred to the electrolyte – this as long as depletion conditions are established in the tube walls,[301,302] while electrons must be transported to the substrate (*e.g.*, metallic back-contact, FTO, ITO). The longer the tube, the larger the contribution of charge recombination.

Other morphological features were also shown to be crucial to enhance the PEC-WS ability of TiO$_2$ NTs, such as the smoothness of the tube walls,[300] and the geometry of the tube top.[303] In particular, tubes grown in organic electrolyte exhibit superior performance compared to tubes grown in aqueous electrolyte (*i.e.*, ripple-walled NTs); a possible reason could be the increased wall smoothness and the greater order of the formers that also lead to a significant increase in the charge carrier diffusion length compared to ripple-walled tubes.[300] On the other hand, the presence of "grass" on the tube tops was shown to be highly detrimental for photoelectrochemical solar light water splitting, more likely due to a trap filling mechanism predominant in the "grass" part of the tubes.[303]

Compared to classic TiO$_2$ thin films (with no specific 1D morphology), also TiO$_2$ NW arrays exhibit significantly higher PEC ability.[304] In particular, also for nanowires on conductive ITO substrate photocurrent measurements showed that the layer thickness plays a significant role and a maximum photocurrent can be achieved for ∼ 5 μm thick layers,[305] confirming that under specific conditions a critical thickness exists to enable efficient transport and collection of electrons in TiO$_2$-based one-dimensional arrays.[91,300,305]

Moreover, for TiO$_2$ in particular a crystallization into anatase is crucial for photoelectrochemical aplications.[284,306]

However, for titania nanotubes, a common problem of annealing tubes on Ti metal substrate is that such a thermal treatment also induces the formation of undesired rutile phase at the metal/oxide interface, due to the thermal oxidation of metallic Ti, and typically catalyzes the anatase to rutile transition in the NT walls.[279] Therefore, large efforts are dedicated to the identification of strategies for overcoming this detrimental crystal phase transition: *e.g.*, optimized annealing rate profile,[307] doping the TiO$_2$ NTs with low amount of Nb,[308] detachment of tubes from the metallic substrate,[280] *etc.*

Nevertheless, also TiO$_2$-based nanomaterials with a rutile crystal structure are reported for efficient photoelectrochemical water splitting.

This is particularly relevant for single-crystalline TiO$_2$ NRs grown on FTO glass by hydrothermal methods. Such nanorods exhibit a rutile crystal structure due to the significantly low lattice mismatch (~ 2%)[309] between tetragonal FTO ($a = b = 0.4687$ nm)[310] and rutile TiO$_2$ ($a = b = 0.4594$ nm),[190] and typically grow vertically oriented from the FTO substrate, along the (110) crystal plane with a preferred (001) orientation.[113,309,311,312] The single-crystalline nature of these NRs plays a critical role for the observed PEC efficiency. Clearly, the absence of grain boundaries, which contribute to poor electron mobility in conventional polycrystalline TiO$_2$ films, suppresses charge carrier losses due to recombination and leads to relatively high performance (~ 0.5–1.0 mA cm$^{-2}$ under AM 1.5 illumination).[109,311,312]

More recently, also the synthesis of branched single crystal rutile TiO$_2$ nanorods has been reported for PEC hydrogen production.[313] Due to their branched structure (Fig. 3.6(c–e)), these NRs exhibit larger surface area and excellent light-trapping characteristics due to enhanced light scattering phenomena that collectively contribute to the *ca.* four-fold enhancement in photocurrent density compared to the more conventional NRs counterpart.

Although in view of single crystal materials, rutile crystals are much easier to synthesize, produce defined surfaces, and are much better characterized on an atomic level, also the fabrication of single crystal anatase TiO$_2$ 1D nanostructures has been recently developed for light-induced applications. In this case, the use of an alkaline reaction media and, particularly, the presence of a seed layer on the FTO substrate are essential prerequisites (see Section 3.1). Due to the higher electron mobility that characterizes anatase with respect to rutile TiO$_2$,[284,306] single crystal anatase TiO$_2$ nanorods/wires mostly find applications as photoanode for dye-sensitized solar cells.[111,112]

### 3.4.2 Doping and defect engineering

To consider TiO$_2$ for solar PEC water splitting, rational design strategies are needed that address its intrinsic limitations, namely (i) a bandgap too large for an efficient use of solar light and (ii) a conduction band minimum (more appropriately, a Fermi level E$_F$) only slightly negative on an electrochemical potential scale to the hydrogen evolution potential (see section 3.2.1).[153,314] Typically, (i) is overcome by nonmetal doping TiO$_2$, while (ii) is addressed with the introduction of a metal dopant into the oxide lattice.

For instance, the highest PEC-WS current density reported for anodic TiO$_2$ tubes is under most optimized conditions 0.25–0.3 mA cm$^{-2}$ measured in 1 M KOH under AM 1.5 G illumination.[315] This is ascribed to the slow transport of electrons along the nanotube length, *i.e.*, in the order of seconds,[300] which cannot be overcome by simply adjusting the tube morphology. Nevertheless, clearly better results are obtained when the tubular morphology is combined to the modification of the TiO$_2$ physicochemical properties (see below).

In general, since the valence band minimum of TiO$_2$ is much lower than the H$_2$O/O$_2$ redox potential, bandgap reduction should be approached by shifting the VBM upwards, *i.e.*, without affecting the more "critical" position of the conduction band. This is typically accomplished by introducing a high density of states (or a band) above the original valence band of TiO₂, that is, by incorporating into the TiO$_2$ lattice anions with *p* orbitals that are more energetic than those of O 2p. To this regard, most appropriate dopants are substitutional nitrogen and carbon that, beside introducing suitable acceptor levels, also exhibit ionic radii similar to those of oxygen; hence, their incorporation causes minimal distortion in the oxide lattice.[153]

Clearly, depending on the synthetic approach to 1D-nanostructured TiO$_2$ materials and also on the dopant element, different doping strategies are available.

For instance, nitrogen-doping and visible light driven PEC water splitting have been reported upon annealing single-crystal rutile TiO$_2$ NWs in NH$_3$ atmosphere under optimized conditions (*i.e.*, annealing temperature, time, and NH$_3$ pressure).[91] The low-energy threshold of IPCE spectra for N–doped TiO$_2$ was red-shifted to λ ~ 520 nm (Fig. 3.7(a)). A more significant shift of photocurrent threshold to λ ~ 570 nm was also reported upon nitridation (in NH$_3$ atmosphere) and hydrogenation (in H$_2$/Ar atmosphere) of TiO$_2$ NW arrays. The interaction between substitutional N- and the Ti$^{3+}$-centers induced by hydrogenation (*i.e.*, between midgap levels close to the valence and conduction bands, respectively – Fig. 3.7(b,c)) has been proposed as key for spectrum extension and visible light photoactivity.[188]

However, the effects induced by substitutional N and by midgap state formation on the N-TiO$_2$ PEC activity are strongly debated. Despite several reports showing an improved PEC performance for N-doped TiO$_2$ nanostructures,[91,188,316] other studies demonstrated that to a promoted vis-light activity of N-TiO$_2$ also corresponds an increase of carrier recombination rate under UV irradiation, particularly when the N amount exceeds the optimum concentration.[146]

It should be noted that, with respect to various methods used to achieve nitrogen doping, different states of nitrogen are observed, and the active species may be present in the bulk TiO$_2$ or on the surface. To determine the nature of doping-induced states, typically non-destructive surface spectroscopic techniques such as XPS are used. Low-dose N ion-implantation into $TiO_2$ tube layers shows an XPS peak at ~396 eV,[147,148] which is in accordance with the peak position of Ti–N bonds in titanium nitride.[149]

On the other hand, wet treatments in amine-based solutions typically lead to peaks at ~400 eV (or higher) and also yield visible light response. This is the case, for instance, of nitrogen-doping from $NH_4F$ contained in the electrolyte for the anodic growth of tubes,[317] or of N-doping by hydrothermal treatment of $TiO_2$ nanotubes in the presence of trimethylamine.[318] However, such a peak in many cases corresponds to adsorbed molecular nitrogen on $TiO_2$ and often leads to neither (or negligible) visible photocurrent nor photocatalytic activity.[150,319] That is, although absorption spectra of such N-$TiO_2$ materials provide evidence of extended sub-bandgap absorption, corresponding photocurrent spectra (if provided) may not show significant response.

Visible light photocurrent response and IPCE spectra of $TiO_2$ anodic layers modified with urea,[320,321] and the photocurrent transients of aminated $TiO_2$ nanotubes under monochromatic laser ($\lambda$ = 474 nm)[322] provided evidence of a significantly improved solar PEC activity compared to undoped $TiO_2$ (Fig. 3.7(d,e)). Remarkably, the decoration of the tubes with amino group (*i.e.*, formation of surface states), along with interstitial nitrogen doping, not only leads to an increase in the photocurrents for both UV and visible regions of the light spectrum, but also significantly improves the electrode conductivity due to a lower charge transfer resistance and hence a faster carrier transport in the nanotubes after amination (Fig. 3.7(f)).[322]

In the case of carbon-doped $TiO_2$, reported findings are to a certain extent ambiguous and particularly limited amount of IPCE data on C-doped $TiO_2$ questions the actual role of C-doping on the visible light PEC-WS ability of these materials.[155]

According to DFT calculations, carbon doping $TiO_2$ leads to a modest variation of the oxide bandgap, while at the same time induces several midgap occupied states which may account for the observed red shift in the light absorption threshold and for the formation of oxygen vacancies, but may also promote charge carrier trapping.[323]

From the experimental point of view, little to no photocurrent in the visible range is usually reported for C-doped $TiO_2$ nanostructures.[150–152] For instance, C-doping by annealing titania nanotube arrays under controlled CO gas flow showed a slight enhancement in the PEC-WS performance measured under visible light illumination with respect to undoped $TiO_2$ NTs.[152] Interestingly, C-containing tubes (that is, tubes with a double-walled morphology where carbon uptake is due to the decomposition of the anodizing organic electrolyte under applied voltage) also exhibit a change in the $TiO_2$ optical properties and sub-bandgap response has been reported for carbon contaminations as low as ~ 3 at.% (Fig 3.8(a)).[300]

By contrast, in line with theory predicting that the incorporation of carbon and/or related defects strongly affects the electronic structure of $TiO_2$,[323] a drastic change to a semimetallic $TiO_xC_y$ compound,[324] also with improved mechanical stability,[325] has been reported by thermal treatment of $TiO_2$ nanotubes in an acetylene atmosphere (Fig 3.8(b,c)).

Finally, although theoretical calculations do not consider F as a suitable substitutional dopant for $TiO_2$ because it is in principle a donor at O sites,[153] Fang *et al.* reported on porous single-crystal F-doped $TiO_2$ NRs where fluorine-doping is shown not only to improve the electron transport properties of single crystal rutile $TiO_2$, but also to enhance the PEC-activity under both UV and visible irradiation (Fig 3.8(d–f)).[326,327]

Doping $TiO_2$ with a metal (that is, a donor species) is reported to promote faster electron transport and, hence, to reduce charge recombination. As mentioned, it is of primary importance that any doping attempt in $TiO_2$ should maintain the conduction band minimum (or, even better, slightly upshift the CBM) to satisfy the correct band-edge position relative to the $H^+/H_2$ redox potential and provide sufficient driving force for hydrogen production. For this, transition metals (TMs) are suitable dopant candidates, since they have $d$ orbitals with energies higher than those of Ti and are similar in size. DFT calculations, in particular, showed that most suitable metal dopants are Ta, Nb, W, and Mo.[153] However, experimental evidence has been provided that also other transition metals are suitable for $TiO_2$ doping and achieving better charge carrier transport.

In the case of self-organized NTs, a most unique approach to X-doped $TiO_2$ is offered by the possibility of anodizing Ti–X alloys (with *e.g.*, X = Ru, Ta, W, Nb, Cr). In this case, simple tailoring of the anodization parameters is sufficient to fabricate highly aligned and variously doped $TiO_2$-based nanotube arrays.[294,297,308,328–335]

For instance, $RuO_2$-modified $TiO_2$ nanotubes grown from Ru–Ti alloys exhibit PEC characteristics that depend on the Ru content. In particular, strong and stable enhancement of the PEC activity has been obtained for an Ru concentration as low as 0.5 at.% and for annealing at 450 °C (Fig. 3.9(a)). Under these conditions, a 6-fold increase in photocurrent density was measured compared to pristine $TiO_2$ NTs, and the effect was also shown to be far superior compared to anodic $TiO_2$ NTs modified with $RuO_2$ nanoparticles.[315] Here $RuO_2$ is considered to be dopant and co-catalyst for $O_2$ evolution.

In optimally doped and annealed $TiO_2$ tubes, both Nb and Ta act as electron donors and increase the electron conductivity of the NT layers. This was observed for nanotube arrays doped with small amounts (up to 0.1 at.%) of Nb[330,332,333] or Ta[294,334] for which a significant enhancement of photoelectrochemical water splitting and dye-sensitized solar cell performances was observed (Fig. 3.9(b,c)). In particular for Nb-doped $TiO_2$ tubes, a more recent study shows that Nb doping is also beneficial in view of retarding rutile formation at the metal/oxide interface, while annealing the tube layers (Fig. 3.9(d–f)).[308]

Similar effects on the charge carrier transport of TM-doped $TiO_2$ arrays have also been observed for layers synthesized by wet chemical methods.

For instance, Xu *et al.* reported on the preparation and PEC activity of Sn-doped single-crystalline rutile $TiO_2$ nanowires.[92] Due to the low lattice mismatch between rutile and $SnO_2$,[309] Sn dopants could be easily incorporated into $TiO_2$ NWs by hydrothermal synthesis. For low Sn doping levels (up to $Sn/TiO_2 \sim 12\%$), a photocurrent increased under simulated sunlight illumination was observed compared to pure $TiO_2$ NWs. In particular, photocurrent spectra revealed that this effect is mainly ascribed to the enhancement of photoactivity in the UV region, while electrochemical impedance measurement confirmed the higher density of n-type charge carriers within the NWs produced by Sn doping.

Interestingly for some TM-doped $TiO_2$ combinations, beside the effect on charge carrier transport, also a light edge absorption shift from UV to visible has been reported and, correspondingly, a visible-light activated photocurrent has been observed.

For instance, $TiO_2$ NRs doped with Fe, Mn or Co and synthesized through a hydrothermal method (Fig. 3.10(a)) produced higher photocurrent than the undoped $TiO_2$ NRs, both under solar and visible light irradiation (Fig. 3.10(b,c)).[295] In particular, Mott-Schottky analysis showed that a significant enhancement in the charge carrier density $N_d$ can be obtained by Fe-doping; it was concluded that not only does iron significantly increase the concentration of electrons in the conduction band, with an overall beneficial effect on charge carrier separation and transport, but it also generates impurity states near the conduction band of $TiO_2$ improving the visible light absorption (by apparently reducing $TiO_2$ bandgap) and enabling visible light PEC activity.

For single crystal rutile $TiO_2$ nanowires doped with V, Cr, Mn, Fe, Co, Nb, Mo, and Rh by means of a molten-salt synthesis, a dual effect due to TM doping was reported:[296] not only the doped layers exhibited optical absorption extended to the visible light (though not supported by the corresponding photocurrent spectra), but most interestingly a decrease in the overpotential for the OER was measured with respect to pure $TiO_2$ layers (Fig. 3.10(d)). Liu *et al.* attributed this result to the homogeneous distribution of dopants achieved by means of the molten-salt procedure. In particular, EXAFS spectra (Fig. 3.10(e)) showed that the introduction of TM atoms occurs uniformly at subsurface Ti sites with no significant alteration in the rutile lattice (that is, neither crystal defects nor other phases could be observed). This alters the adsorption energy of water splitting reaction intermediates (O, OH, and OOH radicals) on the surface of the material, and accounts for the substantial decrease in the overpotential for OER.

Crystal defect engineering can be instead achieved by reduction of $TiO_2$ to black titania, typically carried out by thermal treatment in a $H_2$-containing atmosphere – more recently also alternative methods such as high-energy proton implantation have been investigated. As anticipated in 3.2.4, beside the extension of the absorption properties of $TiO_2$ to visible light wavelengths, hydrogenation introduces voids and defects in the material (mainly in the outer shell) that exhibit a co-catalytic effect towards a stable and noble metal-free photocatalytic $H_2$ generation.[191–193]

Hydrogenation of $TiO_2$ has been reported also for photoelectrochemical water splitting. For instance, $H_2$-treated single crystal $TiO_2$ NWs have been shown to provide stable photocurrent and higher PEC efficiency compared to $TiO_2$ NWs annealed in air. Mott-Schottky plots demonstrated an enhanced donor density in H-$TiO_2$, due to the increased concentration of oxygen vacancies. This in turn positively affects the electron conductivity and enables a more efficient charge transport in the reduced array.[93]

Similar results were also reported for $TiO_2$ NWs that underwent a rapid flame-reductive treatment in a $CO/H_2$ atmosphere (Fig. 3.11). Compared to other reduction methods, flame reduction can be operated at ambient conditions and provides an adjustable concentration of oxygen vacancy by controlling the reduction time. That is, reduction treatment as short as 5 s has been reported to double the photocurrent density of $TiO_2$ NWs, compared to the untreated counterparts. Clearly, the significantly short exposure time of the nanoarray to high temperature is of great advantage also in view of preserving the electrode morphology and crystallinity.[94]

### 3.4.3  Surface decoration and heterojunction formation

Decoration of $TiO_2$ with metals, semiconductors, and polymers is frequently used to influence the electronic and physicochemical properties of the oxide and achieve a desired increment in PEC activity.

In particular, metal nanoparticles can introduce local variations in the band bending of the adjacent semiconductor (*i.e.*, pinning of the Fermi level of an energy that depends on the work function of the metal), inducing an effect similar to that observed by applying an external bias[336] – remarkably, a great advantage is that such a Fermi level pinning is obtained under "open circuit conditions", and thus this approach is often preferred for open circuit photocatalytic applications.

As for hematite, also the surface properties of $TiO_2$ can significantly limit the overall PEC reaction due to a limited kinetics of water oxidation as well as the charge recombination loss at the electrode surface. Photogenerated holes can be trapped by surface states and recombine with electrons before they can diffuse to the $TiO_2$/electrolyte interface for interfacial charge transfer and water oxidation.[336]

Therefore, techniques for passivating surface trap states and decreasing trap state-mediated charge recombination often apply also to $TiO_2$ photoanodes. Typically, atomic layer deposition (ALD)[312,337] or $TiO_2$ precursor (*e.g.*, $TiCl_4$, $Ti(OiP)_4$, $Ti(OBu)_4$) based treatments[95] are used to decorate the surface of $TiO_2$ nanoarrays (usually rods or wires) with a $TiO_2$ thin shell layer (Fig. 3.12(a)).

A combination of photoelectrochemical and photovoltage measurements and impedance spectroscopy demonstrated that the presence of a thin $TiO_2$ (anatase or rutile) shell passivates the surface trap states and significantly increases the PEC activity, compared to the naked $TiO_2$ arrays, by promoting interfacial charge transfer at the $TiO_2$/electrolyte interface as well as a better carrier utilization (Fig. 3.12(b–d)).[95,312,337]

Also the more classical Co-based oxygen evolution catalyst is used to passivate $TiO_2$ surface states and significantly higher photocurrent density has been reported with respect to unmodified $TiO_2$.[91]

Metal nanoparticles with plasmonic effect (mainly Ag and Au) can be used to enable visible light photoelectrochemical water splitting on $TiO_2$ photoanodes (Fig. 3.13(a,b)).

Overall, this mechanism is particularly effective as the photogenerated holes remain on the surface of the photoanode, overcoming the relatively short exciton diffusion length of $h^+$, while electrons percolate through the one-dimensional array to the back-contact.

A range of different methods is available for the decoration of $TiO_2$ surfaces with plasmonic metals: (i) photoreduction of the precursor (*e.g.*, $HAuCl_4$ to Au NPs),[338] (ii) anodization of noble metal alloys of Ti such as Au-Ti (or Pt-Ti); in contrast to previous examples on Ti–X alloy (with X being a non-noble metal – section 3.4.2), this approach leads to noble metal nanoparticle-decorated tubes as the noble metals are not oxidized during the anodization process (Fig. 3.13(c,d));[339–341] (ii) templated dewetting and dewetting/dealloying of sputtered metal thin films;[197,342–344] when applied to $TiO_2$ nanotube arrays, this approach is entirely based on self-ordering principles, that is, self-organized nanotubes offer an ideal topological substrate for a controlled split up of a metal thin film into nanoparticles whose size, density and placement are influenced by the nanotubes geometry itself (Fig. 3.13(e–g)).[345]

Remarkably, for both (ii) and (iii), the decoration density of noble metal co-catalytic nanoparticles can be easily controlled by the amount of the noble metal in the alloy or sputtered-deposited, respectively.

Visible light sensitization due to charge (electron) injection into $TiO_2$ is instead achieved by decoration with semiconductive quantum dots (CdS, CdSe, ZnS, *etc.* – see Fig. 3.14(a) and section 3.2.3 for discussion on the mechanism).

Sun *et al.* reported that the synergy between the unique morphology of nanotubes and the visible light sensitization promoted by surface deposited QDs led to a 35-time higher photocurrent with respect to that of plain $TiO_2$ nanotubes.[210] Due to their small band gap of 2.4 and 1.7 eV, respectively, CdS and CdSe are the most common sensitizers or co-sensitizers for $TiO_2$.[210,346–348]

QDs are usually deposited on the nanotube/wire walls by different methods: (i) in situ growth of QDs by sequential chemical bath deposition (CBD), (ii) deposition of pre-synthesized colloidal QDs by direct adsorption, and (iii) by linker-assisted adsorption.

The CBD method is based on repetitive cycles of immersion/rinsing of the oxide array into the quantum dot precursor solution. QDs form according to a deposition/aggregation process and uniformly cover the surface of the $TiO_2$ nanoarrays (Fig. 3.14(b));[210,346,347] clearly, the control on the number of cycles is crucial to avoid excessive particle growth and aggregation that would in turn increase the chance of carrier recombination.

The adsorption of pre-synthesized QDs certainly offers a higher control on particle aggregation; in particular, the solvent used to disperse the QDs plays a critical role as the establishment of a favorable interaction of the QDs with both the solvent and oxide surface leads to a high degree of coverage with a low degree of aggregation.[349]

However, the adsorption of pre-synthesized colloidal QDs often leads to a low surface coverage (*i.e.*, ~14%)[349] that limits the efficiency of the QDs/$TiO_2$ assemblies compared to the CBD method.[347] Also, a direct contact between the oxide matrix and the sensitizers is beneficial for a higher PEC efficiency – that is, QDs deposition by direct adsorption is preferred over the linker-assisted method.[349]

Regardless of the formation/deposition method, sensitization of $TiO_2$ with *e.g.* CdS QDs leads to a broadband light absorption that extends up to λ ~ 600 nm.[346,349] Remarkably, light absorption up to λ ~ 750 nm has been reported for $TiO_2$ NT arrays decorated with CdS/CdSe/ZnS QDs,[347] that is, as for plasmonic metals, also for semiconductive QDs the extent of extended light absorption depends on the material (but also on the size and shape of the nanoparticles). Typically, an enhancement in the IPCE and photocurrent is correspondingly observed (Fig. 3.14(c,d)).[346]

In particular, comparative studies showed that $CdSe/TiO_2$ assemblies exhibit an initial higher photocurrent with respect to $CdS/TiO_2$, due to the smaller bandgap of the selenide-based QDs; however, $CdSe/TiO_2$ efficiency decreases quite rapidly as a consequence of the QDs low photostability. On the other hand, despite the lower PEC performance, $CdS/TiO_2$ exhibits higher e⁻ injection from the QDs and a lower charge recombination rate.

In line with this, a maximized photoelectrochemical performance was therefore observed for CdS/CdSe co-sensitized $TiO_2$ nanorod arrays that could exploit the advantages of both single cadmium-based materials.[347,348]

Suitable electronic heterojunctions can be "built" also by surface decoration of $TiO_2$ with metal oxides including $CuO_2$,[350,351] NiO,[352] ZnO[353,354] and $Bi_2O_3$.[355] Slow hydrolysis of precursors, electrochemical techniques, and CVD or PVD deposition are the typically adopted methods.

One of the most followed up schemes is the combination of $TiO_2$ with a p-type material (*e.g.*, $Cu_2O$ or NiO) to establishing useful p–n heterojunctions. This configuration enables solar light water splitting by combining a narrow bandgap semiconductor (p-type) to a larger bandgap semiconductor (n-type $TiO_2$) and utilizes both electron and hole majority charge carriers (from the n-type and p-type counterparts, respectively) for the PEC-WS reaction.[351,352] Also, for $TiO_2$/NiO assemblies, a higher visible light photoresponse in comparison to other classical approaches (*e.g.,* N-doping) has been observed and the beneficial effect attributed to charge injection from the NiO excited states to the conduction band of titanium oxide.[352]

Nevertheless, it should be noted that for many of these compounds (namely, all II–VI and red-ox active materials) the long-time stability in photoelectrochemical applications still needs to be improved, not only in terms of resistance to corrosion or photocorrosion, but also due to the instability of some of the co-catalysts under applied voltage.

More complex systems such as hierarchical $CuInS_2/TiO_2$[356,357] and $TiO_2/In_2S_3/AgInS_2$[358] core-shell structures and "umbrella" hybrid $(Bi_2S_3/rGO)_5/TiO_2$ NR heterostructure[359] have been recently introduced (Fig. 3.15).

Common to all these hierarchical assemblies is a beneficial band misalignment that enables charge carrier separation, with the preferential collection of electrons in $TiO_2$ (and, thus, to the back-contact) through a cascade down-hill process, with holes accumulating in the shell material (*i.e.*, in the valence band with the least positive redox potential – see Fig. 3.15(a,b)).

This clearly suppresses charge carrier recombination and overall contributes to the enhancement of the multi-junction electrode PEC performance.

## 4. $Fe_2O_3$-$TiO_2$ heterojunction-based materials – the $Fe_2TiO_5$ case

We have already pointed out that the relative misalignment of $Fe_2O_3$ and $TiO_2$ band edges can be exploited to design a suitable heterojunction for enhanced charge carrier separation, that is, holes generated in $TiO_2$ will be transferred to $Fe_2O_3$ with electrons preferentially accumulating in $TiO_2$ at the same time.[33,360,361]

Under specific conditions, hybrid composites, *i.e.*, iron titanates (*e.g.*, $Fe_2TiO_5$, $Fe_3TiO_4$, and $FeTiO_3$) form at the interface between hematite and titania that, despite some controversial results, exhibit remarkable performances as photoanode materials in PEC devices.

Typically, iron titanates are formed via solid-state reaction between $TiO_2$ and α-$Fe_2O_3$. Due to the similar ion sizes of Ti and Fe, thermal annealing of the two oxides generates the composite – the various stoichiometric forms of titanates are given by the different valence configurations of the metal ions (such as $Fe^{2+}Ti^{4+}$, $Fe^{3+}Ti^{3+}$, *etc.*).

Among all the possible titanates, $Fe_2TiO_5$ has been recently the most widely explored Fe-Ti combination compound as photoanode material.[362–366]

$Fe_2TiO_5$ has a pseudobrookite structure and, as $TiO_2$, has a valence and a conduction band that are (in principle) suitable to drive spontaneous water oxidation and reduction reactions, respectively. Moreover, it has a band gap of *ca.* 2.0 eV and thus can generate $e^-/h^+$ pairs under visible light illumination (Fig. 4.1). Clearly, both aspects are highly desired for materials intended for PEC applications.

Key to efficient $Fe_2TiO_5$ for PEC-WS is a layer thickness of only a few nm and a use of $Fe_2TiO_5$ in combination with $TiO_2$ or α-$Fe_2O_3$.

In fact, early studies on iron titanates based photoanodes for direct solar water oxidation,[367] showed that for a "pure" $Fe_2TiO_5$ layer (not supported on $TiO_2$ or α-$Fe_2O_3$) an applied bias higher than that required for α-Fe$_2$O$_3$ is needed to promote efficient PEC water splitting, and that composites such as FeTiO$_3$ and Fe$_2$TiO$_4$ suffer from significant photocorrosion.

Typically two different approaches are reported that instead lead to improved PEC efficiency in the presence of Fe$_2$TiO$_5$: (i) a use of nanostructured TiO$_2$ array as core material, where a thin α-Fe$_2$O$_3$ overlayer is deposited, or (ii) nanotubes/nanorods of α-Fe$_2$O$_3$ on FTO as supporting scaffold modified with a thin TiO$_2$ layer. In both cases, thermal annealing at 500–700 °C in air converts the thin overlayers into Fe$_2$TiO$_5$.[362–366]

For the Fe$_2$TiO$_5$(overlayer)/TiO$_2$(core) configuration, two aspects contribute to the enhancement of the PEC activity: the titanate layer significantly extends the absorption spectrum of the material to visible wavelengths, and contributes to the generation of e$^-$/h$^+$ pairs up to λ ~ 600 nm. In addition, while h$^+$ are preferentially transferred to the electrolyte, e$^-$ are injected into the TiO$_2$ CB.

Although, in principle, the CBs of TiO$_2$ and Fe$_2$TiO$_5$ are not suitably located to promote e$^-$ transfer from Fe$_2$TiO$_5$ into TiO$_2$ – see Fig. 4.1, enhanced charge separation has been reported and suggested as the origin for the higher photoelectrochemical activity compared to both TiO$_2$ and Fe$_2$O$_3$ photoelectrode counterparts.[362]

Different synthesis methods have been explored for the deposition of α-Fe$_2$O$_3$ thin layers on one-dimensional TiO$_2$ scaffolds: (i) electrodeposition (*e.g.*, from a Fe(NO$_3$)$_3$ solution – Fig. 4.2(a)), (ii) thermal pyrolysis of Fe-based organic solution (*e.g.*, FeCl$_3$ ethanol solution – Fig. 4.2(c)), (iii) chemical bath deposition.[363,364]

In general, PEC performance is strictly related to the thickness of the Fe$_2$TiO$_5$ layer, that is, there is a threshold thickness (~ 100 nm) that grants both significant light absorption and carrier separation. Above this limit, h$^+$ and e$^-$ tend to recombine due to a high charge transport resistivity, with a detrimental effect on the overall PEC performance of the composite (Fig. 4.2(d)).[364]

To overcome this issue and further promote carrier utilization, Liu *et al.* reported on the use of a Co-based layer as a (more) efficient co-catalyst for the anodic reaction.[363] A current density of 4.1 mA cm$^{-2}$ at 1.23 V *versus* RHE was measured along with a lower photocurrent onset potential (*i.e.*, 0.18 V *versus* RHE) with respect to pure TiO$_2$. The authors proposed that the natural band bending at the interface between the three oxides (Fe$_2$TiO$_5$, TiO$_2$, and CoO$_x$) enables holes produced in Fe$_2$TiO$_5$ to accumulate in CoO$_x$, while electrons from Fe$_2$TiO$_5$ are transferred to TiO$_2$ (Fig. 4.2(b)).

For the second core/shell configuration, that is, Fe$_2$TiO$_5$(shell)/Fe$_2$O$_3$(core) composite, enhanced PEC efficiency is mainly attributed to the beneficial role of Fe$_2$TiO$_5$ in passivating hematite surface trapping states. This clearly reduces surface e$^-$/h$^+$ recombination. Moreover, due to the favorable band offset between the two oxides, e$^-$ photopromoted in Fe$_2$TiO$_5$ are injected into the CB of α-Fe$_2$O$_3$ that supplies a direct transport path to the back-contact; similarly to the previous case, holes migrate to the Fe$_2$TiO$_5$/electrolyte interface and the increased charge separation results in an overall increase of the photoefficiency.[365]

Experimental evidence for this mechanism was provided by transient absorption spectra (TAS) measured with Fe$_2$TiO$_5$/Fe$_2$O$_3$ and pure α-Fe$_2$O$_3$ (Fig. 4.3).[368]

With respect to pure α-Fe$_2$O$_3$, Fe$_2$TiO$_5$ features an increased density of energy states near the CB and improved hole dynamics. This creates competitive trapping pathways for photopromoted electrons, decreases e$^-$/h$^+$ recombination probabilities and, clearly, increases the lifetime of photogenerated holes.

Therefore, the pseudobrookite/hematite heterojunction generates a large amount of long-lived surface holes and, at the same time, acts as a block layer for the photogenerated electrons transferred to α-Fe$_2$O$_3$. By preventing e$^-$ back injection (and therefore e$^-$/h$^+$ recombination), Fe$_2$TiO$_5$ leads to the overall increase of the photoanode efficiency.

Two possible approaches have been proposed to grow a $Fe_2TiO_5$ ultra-thin film over vertically aligned hematite α-$Fe_2O_3$ nanorods. In the first case, a FeOOH photoanode is immersed in a HF solution containing the Ti precursor; the second procedure is based on the evaporation of $TiCl_4$ on the surface of FeOOH. In both cases, adequate thermal treatment produced a $Fe_2TiO_5$ film.[366]

$Fe_2TiO_5$ was identified by synchrotron-based soft X-ray absorption spectroscopy (Fig. 4.4(a,b)) and TEM mapping (Fig. 4.4(c)), and a suitable lattice match between pseudobrookite $Fe_2TiO_5$ and α-$Fe_2O_3$ was suggested to favor a specific crystallographic growth direction, *i.e.*, the (110) plane for hematite and the (101) plane for pseudobrookite.

As for the $Fe_2TiO_5$/$TiO_2$ composite system, also $Fe_2TiO_5$/$Fe_2O_3$ exhibits enhanced solar water oxidation activity, also supported by IPCE spectra; particularly, the HF-assisted synthesis of $Fe_2TiO_5$/$Fe_2O_3$ resulted in a photocurrent density of 2.0 mA $cm^{-2}$ at 1.23 V *versus* RHE that was significantly higher than that measured with pure hematite nanorods, and that could be further improved by coupling with Co-Pi catalyst (Fig. 4.4(d)).

The beneficial effect of an oxygen evolution co-catalyst on the overall PEC activity of $Fe_2TiO_5$/$TiO_2$ assemblies has been also pointed out with the use of less "conventional" materials (other than Co-Pi[366] and Co-based[369] co-catalysts).

Passivation of surface states, reduction of electron–hole recombination, and, therefore, a photocurrent improvement have been reported by decorating $Fe_2TiO_5$/$TiO_2$ photoanodes with an $SnO_x$[370,371] (Fig. 4.5) or with a $FeNiO_x$[372] overlayer (Fig. 4.6). This overall enabled enhanced photoanode stability for up to 5 h under operating conditions.[372]

## 5. Conclusion and Perspective

Decades of experimental research on $TiO_2$ and $\alpha\text{-}Fe_2O_3$ have pointed out that the two semiconductor metal oxides are promising candidates as anodes for photoelectrochemical water splitting. The reason for this is that, compared to other efficient materials such as CdS, CdSe, GaAs, *etc.*, $TiO_2$ and $\alpha\text{-}Fe_2O_3$ are chemically stable against (photo)corrosion in a wide range of (electro)chemical conditions. Additionally, they are cheap, abundant and nontoxic.

In spite of these positive features, still some specific limitations hamper the development of highly efficient PEC devices based on the use of $TiO_2$ and $\alpha\text{-}Fe_2O_3$ that could meet the requirements for high-scale applications. A main issue is certainly represented by the limited charge transfer kinetics intrinsic to both oxides. In this review, we have shown that key to overcoming such a limitation in PEC arrangement is the use of one-dimensional oxide nanostructures (*i.e.*, nanotubes, nanorods, nanowires).

To this regard, we provided a range of available techniques and illustrate their possibilities and potentials. Particularly, we focused on the most promising routes that, in our view, are electrochemical anodization and hydrothermal methods: by adjusting simple experimental parameters, they offer a fine morphology control over the fabricated materials. In this context, we outlined how the resulting 1D geometry of the oxides is crucial towards a photoelectrochemical enhancement. Besides, these structures represent suitable platforms for further modifications, namely to induce doping effects or by the deposition of co-catalysts, light absorbers, charge transfer mediators, *etc.*

In addition to the (surface or bulk) modification of these two oxides, we illustrated achievements and outlooks regarding the use of $TiO_2$–$Fe_2O_3$ systems that combine the complementary features and functionalities of the two counterparts in terms of light absorption, charge transport and, for open circuit photocatalysis, in terms of band energy positions. Although only limited attention has been given to such composites and the mechanism underlying their working principle is still under debate, promising results have been reported in the recent literature. Further investigations on these $TiO_2$–$Fe_2O_3$ heterostructures may generate a new promising research trend in the frame of energy conversion processes.

## Acknowledgements

The authors gratefully acknowledge the support by Project No. LO1305 and Project No. 8E15B009 of the Ministry of Education, Youth and Sports of the Czech Republic, Project No. 15-19705S of the Grant Agency of the Czech Republic, and the Research Infrastructure NanoEnviCz, supported by the Ministry of Education, Youth and Sports of the Czech Republic under Project No. LM2015073. ERC, DFG, and the DFG cluster of excellence "Engineering of Advanced Materials", as well as DFG "funCOS" are also gratefully acknowledged for financial support.

# Figures and captions

## Table 1

Selected examples of 1D α-Fe$_2$O$_3$ based photoanodes and their PEC performance.

| Structure | Preparation | SEM | Electrolyte | $j$ / mA cm$^{-2}$ | Ref. |
|---|---|---|---|---|---|
| Fe$_2$O$_3$ NTs | Anodization (H$_2$ annealing) | 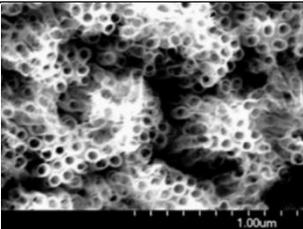 | 1 M KOH | 1.41 mA/cm$^2$ at 1.5 V$_{RHE}$ | 373 |
| Ni-doped Fe$_2$O$_3$ NTs | Anodization and electrochemical deposition | 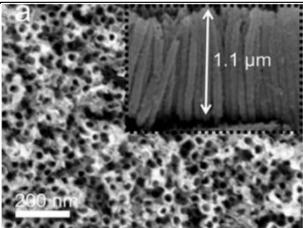 | 1 M KOH | ~3.3 mA/cm$^2$ at 1.45 V$_{RHE}$ | 374 |
| Fe$_2$O$_3$ NR | Hydrothermal synthesis | 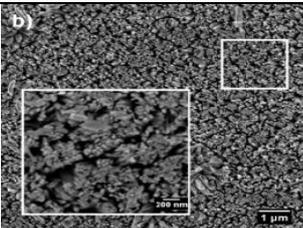 | 1 M NaOH | 3 mA/cm$^2$ at 1.6 V vs RHE | 375 |
| Fe$_2$O$_3$ NR | Template synthesis | 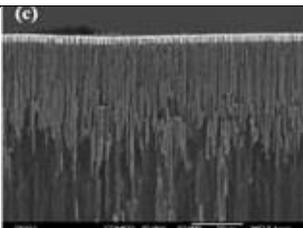 | 1 M NaOH | 8 mA/cm$^2$ at 1.6 V vs RHE | 275 |
| Fe$_2$O$_3$ NR Pt-doped | Template synthesis | 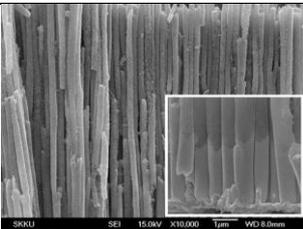 | 1 M NaOH | 10 mA/cm$^2$ at 1.5 V vs RHE | 376 |

| Fe$_2$O$_3$ NR, Pt-doped with Co-Pi co-catalysts | Solution-based methods and two-step annealing | 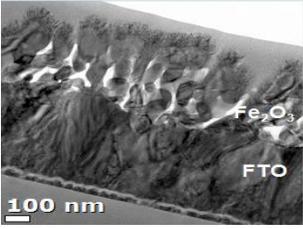 | 1 M NaOH | 4.32 mA/cm$^2$ at 1.23 V vs RHE | 232 |

SEM images (from top to bottom): reproduced from Ref. 373 with permission from the American Chemical Society; reproduced from Ref. 374 with permission from the American Chemical Society; reproduced from Ref. 375 with permission from Elsevier; reproduced from Ref. 275 with permission from the Royal Society of Chemistry; reproduced from Ref. 376 with permission from IOP Publishing; reproduced from Ref. 232 with permission from Springer Nature.

## Table 2

Selected examples of 1D TiO$_2$ based photoanodes and their PEC performance.

| Structure | Preparation | SEM | Electrolyte | $j$ / mA cm$^{-2}$ | Ref. |
|---|---|---|---|---|---|
| TiO$_2$ NTs | Anodization | 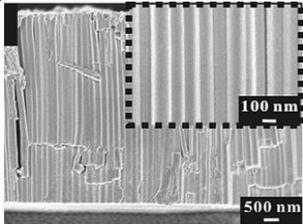 | 1 M KOH | 0.90 at 1.2 V$_{RHE}$ | 377 |
| Ru-doped TiO$_2$ NTs | Anodization | 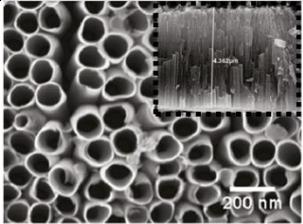 | 1 M KOH | 1.8 at 1.2 V$_{RHE}$ | 315 |
| TiO$_2$ NW | Hydrothermal | 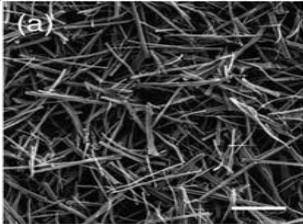 | 1 M NaOH | 2.8 at 1.24 V$_{RHE}$ | 378 |
| Fe-doped TiO$_2$ NRs | Hydrothermal method | 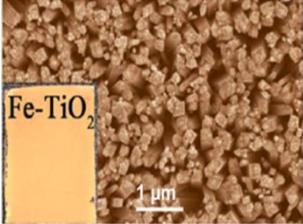 | 1 M KOH | 2.92 at 0.25 V$_{Ag/AgCl}$ | 295 |
| CdS-Au-TiO$_2$ sandwich NRs | Hydrothermal method and chemical bath deposition | 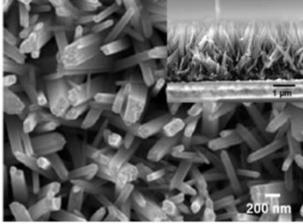 | 0.25 M Na$_2$S and 0.35 M Na$_2$SO$_3$ | 4.07 at 0 V$_{Ag/AgCl}$ | 379 |
| TiO$_2$/ln$_2$S$_3$/AgInS$_2$ core–shell branched NRs | Two-step hydrothermal method and SILAR | 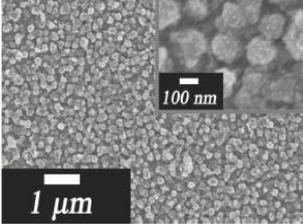 | 1 M KOH | 22.13 at 0.56 V$_{Ag/AgCl}$ | 358 |

* In all cases solar light AM 1.5 G (100 mW cm$^{-2}$) was used for irradiation.

SEM images (from top to bottom): reproduced from Ref. 377 with permission from Elsevier; reproduced from Ref. 315 with permission from the American Chemical Society; reproduced from Ref. 378 with permission from the Royal Society of Chemistry; reproduced from Ref. 295 with permission from the Royal Society of Chemistry; reproduced from Ref. 379 with permission from the American Chemical Society; reproduced from Ref. 358 with permission from John Wiley & Sons.

**Figure 1.1**

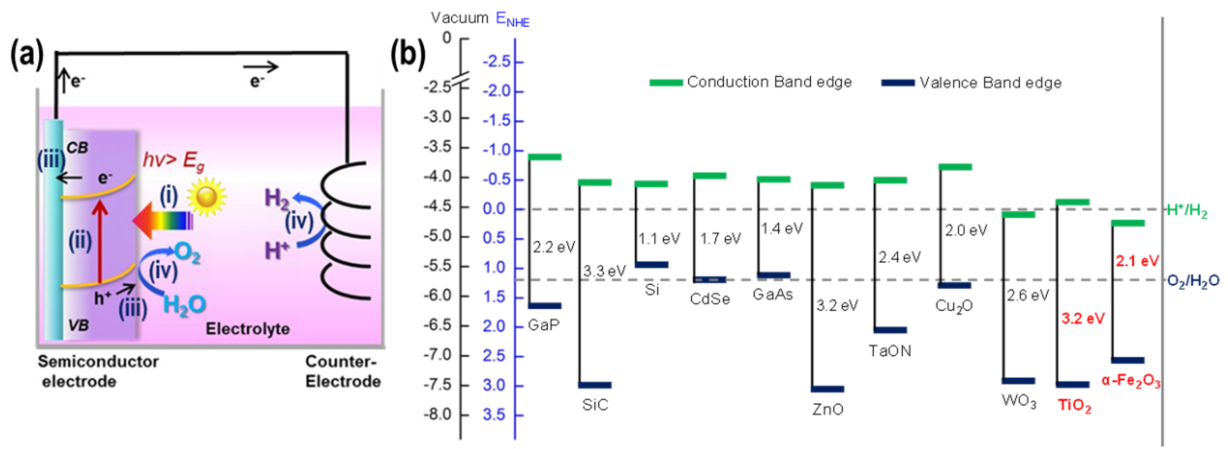

**Fig. 1.1** – (a) Principle of operation of a photoelectrochemical cell based on n-type semiconductor; the water splitting reaction steps (i–iv) are described in the text (Section 1.1). (b) Band edge positions of semiconductors in contact with an aqueous electrolyte at pH = 0, relative to NHE and to the vacuum level. For comparison HER and OER redox potentials are also reported.

**Figure 2.1**

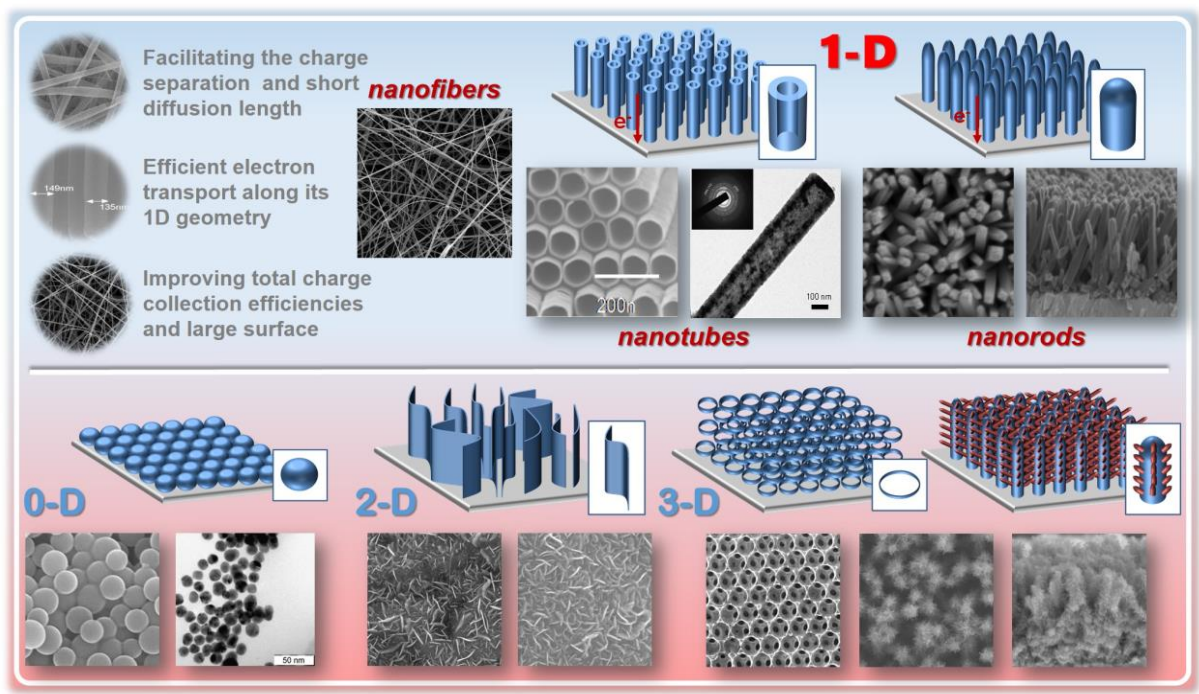

**Fig. 2.1** – Classification of nanostructures based on nanoscale dimensions – their structural and functional features are discussed in Section 3. 1D nanostructures (upper panel) include *e.g.* i) carbon nanofibers fabricated by the spin-electrodeposition (reproduced from Ref. 29 with permission from Elsevier), ii) $TiO_2$ nanotubes by self-organizing electrochemical anodization (reproduced from Ref. 30 with permission from the American Chemical Society) and iii) $TiO_2$ nanorods grown on FTO substrate by a hydrothermal procedure (reproduced from Ref 31 with permission from the American Chemical Society). The lower panel of the picture reports SEM and TEM images of i) 0D nanoparticles ($TiO_2$ and Au nanoparticles, respectively – unpublished results); ii) 2D sheaths of $TiO_2$ grown perpendicularly on FTO substrate, and of iii) 3D nanostructures – left: inverse opal $TiO_2$ (reproduced from Ref. 32 with permission from the National Academy of Science); right: $TiO_2$ nanorods branched with $Fe_2O_3$ nanosheets (reproduced from Ref. 33 with permission from the Royal Society of Chemistry).

**Figure 2.2**

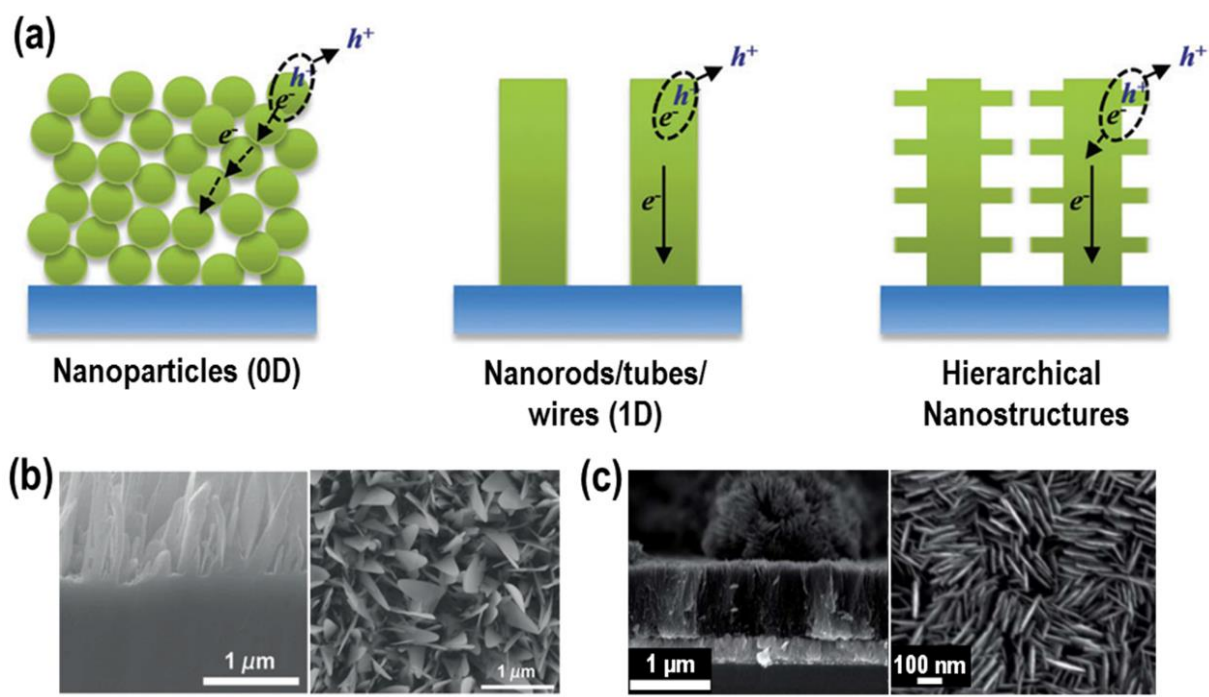

**Fig. 2.2** – a) Schematic illustrating three typical photoelectrodes and their corresponding charge separation/transport behaviors. b) Cross-sectional (left) and top (right) scanning electron microscopy (SEM) images of $Fe_2O_3$ nanoflake array, grown by thermal oxidation of Fe foil. c) Cross-sectional (left) and top view (right) SEM images of single crystalline rutile $TiO_2$ nanoplatelet arrays, grown by chemical bath deposition on FTO glass. Fig. (a) is reproduced from Ref. 69 with permission from the royal Society of Chemistry. Fig. (b) is reproduced from Ref. 66 with permission from John Wiley & Sons. Fig. (c) is reproduced from Ref. 62 with permission from the Royal Society of Chemistry.

**Figure 2.3**

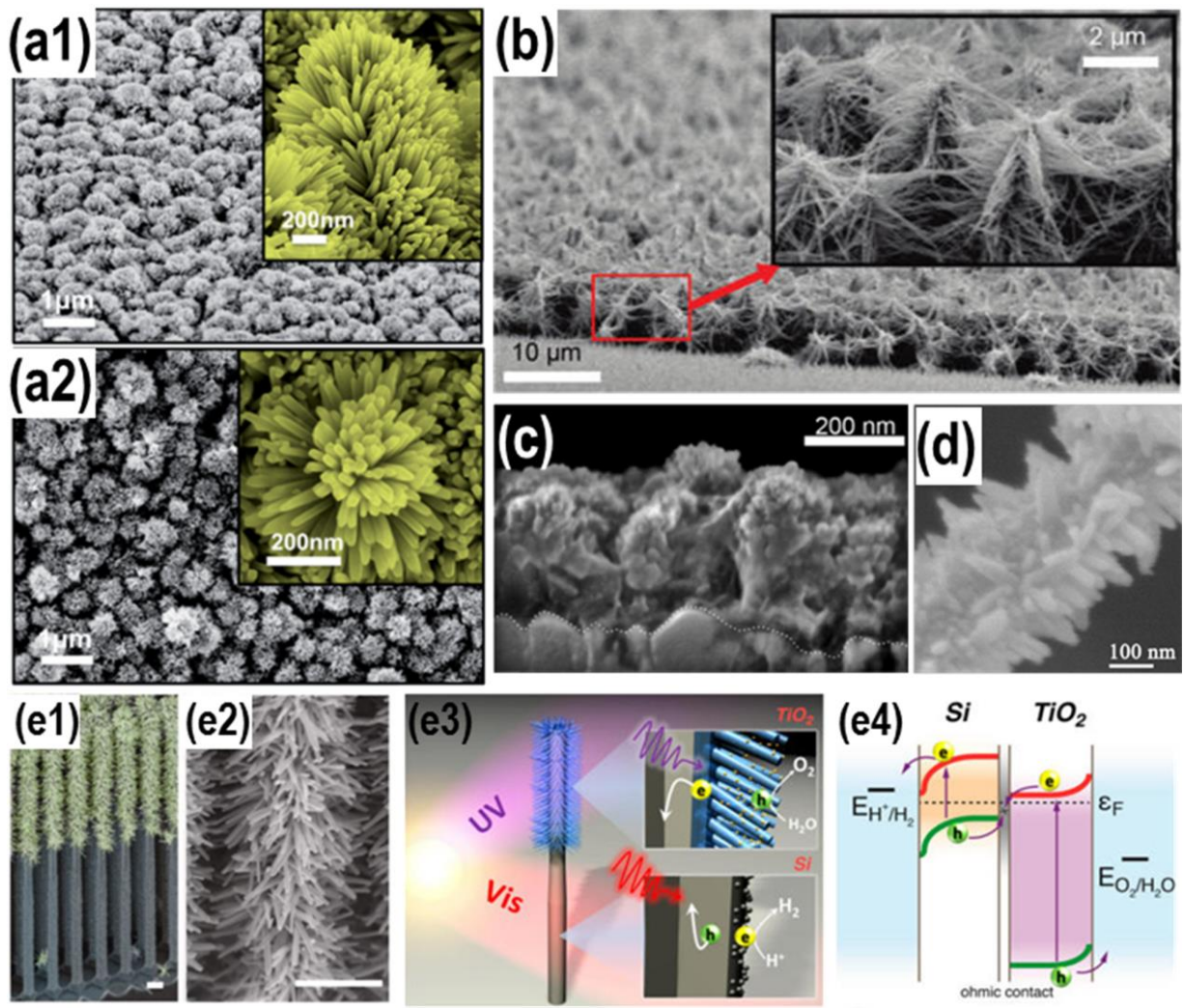

**Fig. 2.3 –** a) SEM images of 3D hierarchically branched TiO$_2$ NWs: (a1) tilted view, (a2) top view. The insets show high-magnification images of an individual hierarchically branched array. b) Tilted view SEM picture of ZnO NW nanoforest. c) Cross-sectional view SEM image of WO$_3$-Fe$_2$O$_3$ host-guest film. The broken white line indicates the FTO/WO$_3$ interface. d) FESEM image of SnO$_2$ nanowire covered with TiO$_2$ nano needle leaf-like structures via TiCl$_4$ treatment. e) SEM image of a nanotree heterostructure; e2) magnified SEM image showing the large surface area of the TiO$_2$ segment used for water oxidation – scale bars are 1 μm. Structural schematics of the nanotree heterostructure: small diameter TiO$_2$ nanowires (blue) and Si nanowires (gray). The two insets report the separation of photoexcited

electron−hole pairs. Energy band diagram of the nanotree heterostructure for solar-driven water splitting. Fig. (a1,2) are reproduced from Ref. 72 with permission from the Royal Society of Chemistry. Fig. (b) is reproduced from Ref. 71 with permission from the American Chemical Society. Fig. (c) is reproduced from Ref. 74 with permission from the American Chemical Society. Fig. (d) is reproduced from Ref. 75 with permission from Elsevier. Fig. (e1–4) are reproduced from Ref. 78 with permission from the American Chemical Society.

**Figure 3.1**

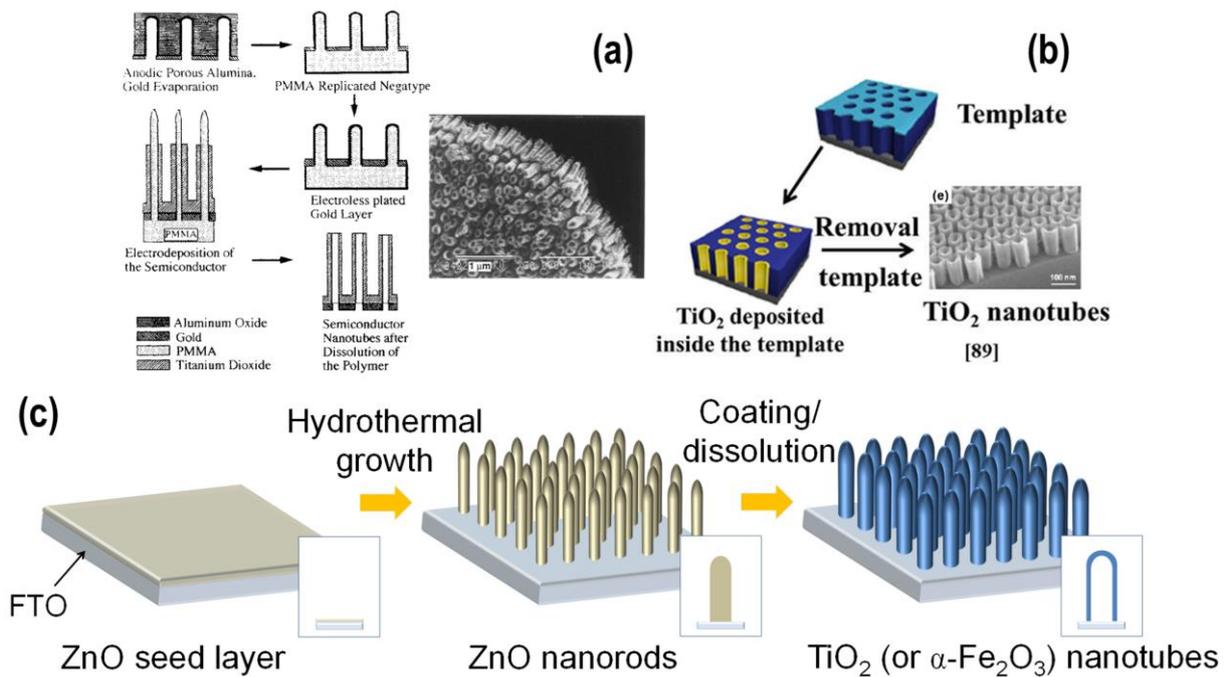

**Fig. 3.1** – a) Schematic view of the replication process and SEM image of the corresponding TiO$_2$ NTs. b) Schematics of the process to fabricate highly ordered freestanding TiO$_2$ nanotube arrays using Si-containing block copolymer lithography and atomic layer deposition, and cross-sectional SEM images. c) Schematic representation of TiO$_2$ (or α-Fe$_2$O$_3$) nanotube formation process: first a ZnO seed layer is deposited on FTO, and ZnO nanorods are grown through a hydrothermal method; afterwards, TiO$_2$ (or α-Fe$_2$O$_3$) nanotubes are grown and the ZnO sacrificial template is simultaneously dissolved. Fig. (a) reproduced from Ref. 118 with permission from the American Chemical Society. Fig. (b) reproduced from Ref. 126 with permission from IOP Publishing. Fig. (c) reproduced from Ref. 33 with permission from the Royal Society of Chemistry.

**Figure 3.2**

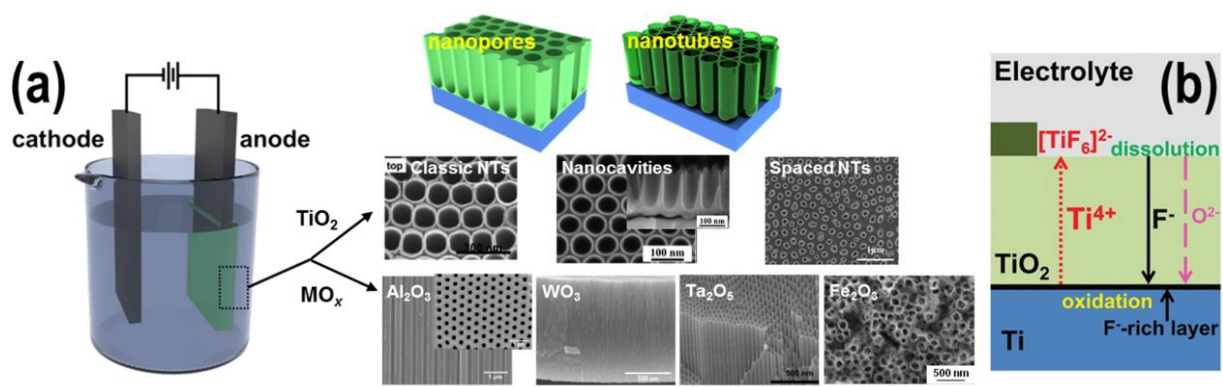

**Fig. 3.2** – a) Typical 2-electrode electrochemical anodization set-up and possible anodic morphologies for different metal oxides: $TiO_2$ (upper row), and $Al_2O_3$, $WO_3$, $Ta_2O_5$, and $Fe_2O_3$ (lower row). b) High field oxide ($TiO_2$) formation in the presence of fluoride ions: a steady-state is established between the oxide formation at the inner interface and its dissolution at the outer interface (due to dissolution/complexation of $Ti^{4+}$ as $TiF_6^{2-}$). Rapid fluoride migration leads to the formation of a fluoride-rich layer at the $Ti/TiO_2$ interface. Reproduced from Ref. 129 with permission from the Royal Society of Chemistry.

**Figure 3.3**

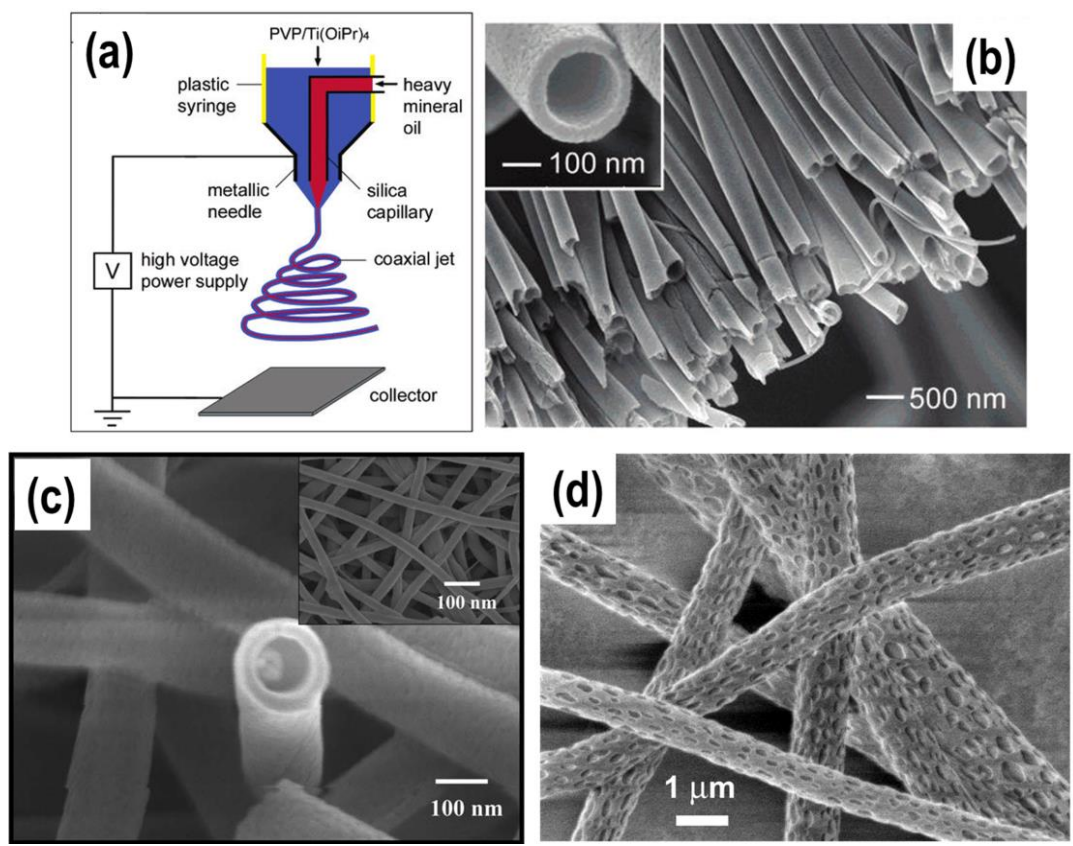

**Fig. 3.3** – a) Schematic illustration of the setup for electrospinning nanofibers. Spinneret is fabricated by means of a two-capillary coaxial jet, through which heavy mineral oil and an ethanol solution of Ti precursor are simultaneously ejected. b) SEM image of a uniaxially aligned array of anatase hollow-nanofibers. c) FESEM images of α-$Fe_2O_3$ hollow fibers by electrospinning an iron acetylacetonate (Fe(acac)$_3$) and polyvinylpyrrolidone (PVP) solution. d) SEM image of porous fibers obtained via electrospinning of a polymer solution in dichloromethane. Fig. (a,b) reproduced from Ref. 140 with permission from the American Chemical Society. Fig. (c) reproduced from Ref. [142] with permission from the Royal Society of Chemistry. Fig. (d) reproduced from Ref. 144 with permission from John Wiley & Sons.

**Figure 3.4**

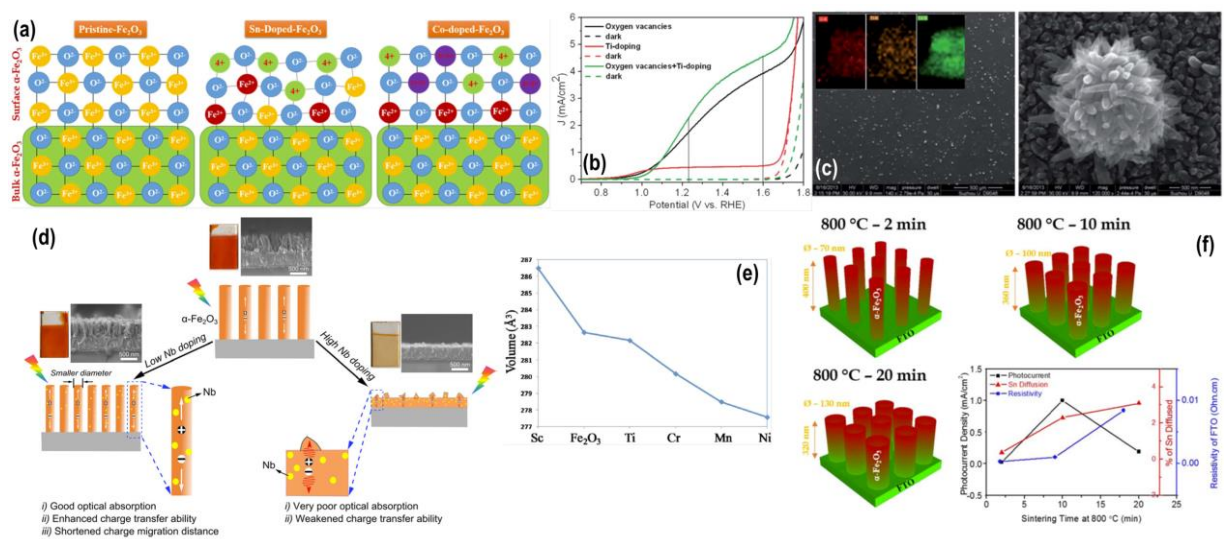

**Fig. 3.4** – (a) Schematic representation of the destabilizing effect induced in the α-Fe$_2$O$_3$ lattice by Sn$^{4+}$ dopants, and the recovery of a more ordered structure promoted by Be$^{2+}$ dopants. (b) Linear sweep voltammetry scans of reduced and Ti-doped α-Fe$_2$O$_3$ nanoarrays, showing the synergistic effect induced by Ti and oxygen vacancies on the photoanode PEC performance (measurements were performed under simulated solar light AM 1.5 irradiation, 100 mW cm$^{-2}$). (c) SEM images and TEM elemental mapping of Ti-doped α-Fe$_2$O$_3$ nanorods. (d) Effects on α-Fe$_2$O$_3$ morphology and properties induced by low (left) and high (right) Nb doping amounts. (e) Calculated volumes of defective α-Fe$_2$O$_3$ supercells containing 3$d$ TM substitutions. (f) Schematic representation of thermally induced Sn-doping of α-Fe$_2$O$_3$ nanorod arrays and dependence of photoanode performance, extent of Sn diffusion and FTO resistivity on the annealing time. Fig. (a) reproduced from Ref. 277 with permission from John Wiley & Sons. Fig. (b,c) reproduced from Ref. 245 with permission from the Royal Society of Chemistry. Fig. (d) reproduced from Ref. 247 with permission from the Royal Society of Chemistry. Fig. (e) reproduced from Ref. 243 with permission from the American Institute of Physics. Fig. (f) reproduced from Ref. 251 with permission from the American Chemical Society.

**Figure 3.5**

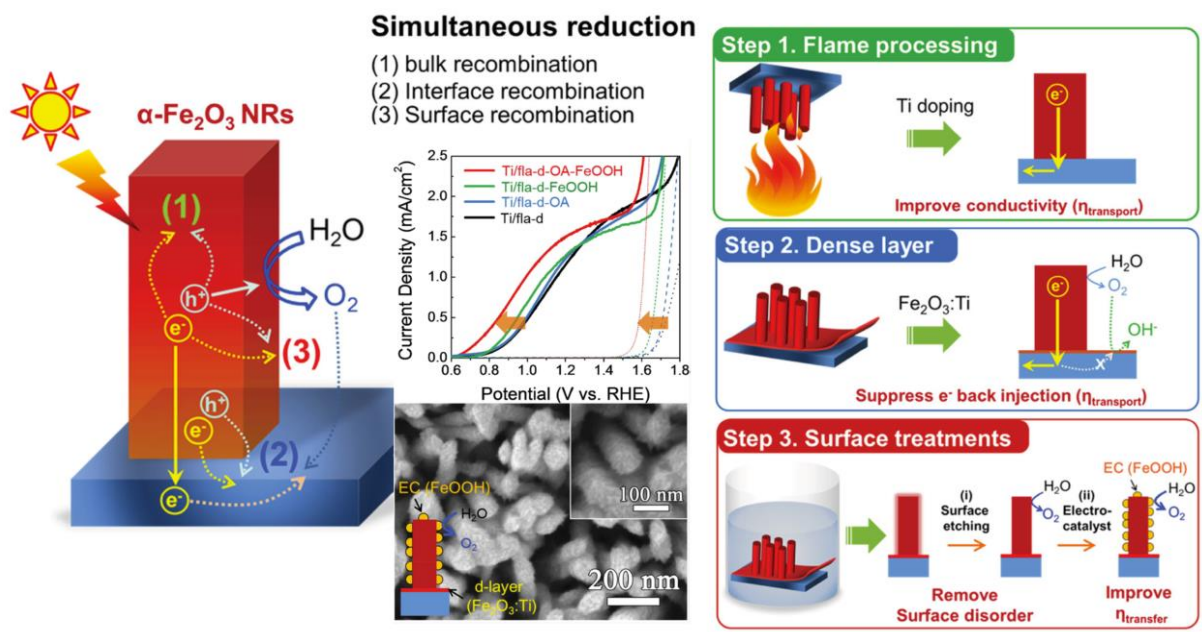

**Fig. 3.5** – Schematic representation of (i) the main electron–hole recombination pathways in hematite NRs photoanodes for solar water splitting, and of (ii) the three-step approach to reduce bulk, interface, and surface recombination. SEM image of the Ti/flame hematite NRs after sequential oxalic acid etching and FeOOH decoration. J–V curves measured at 0.8 $V_{RHE}$ of Ti/flame hematite NRs, also modified according to different surface treatments: (i) oxalic acid (OA) etching, (ii) FeOOH deposition, and (iii) combined OA etching and FeOOH deposition. Reproduced from Ref. 277 with permission from John Wiley & Sons.

**Figure 3.6**

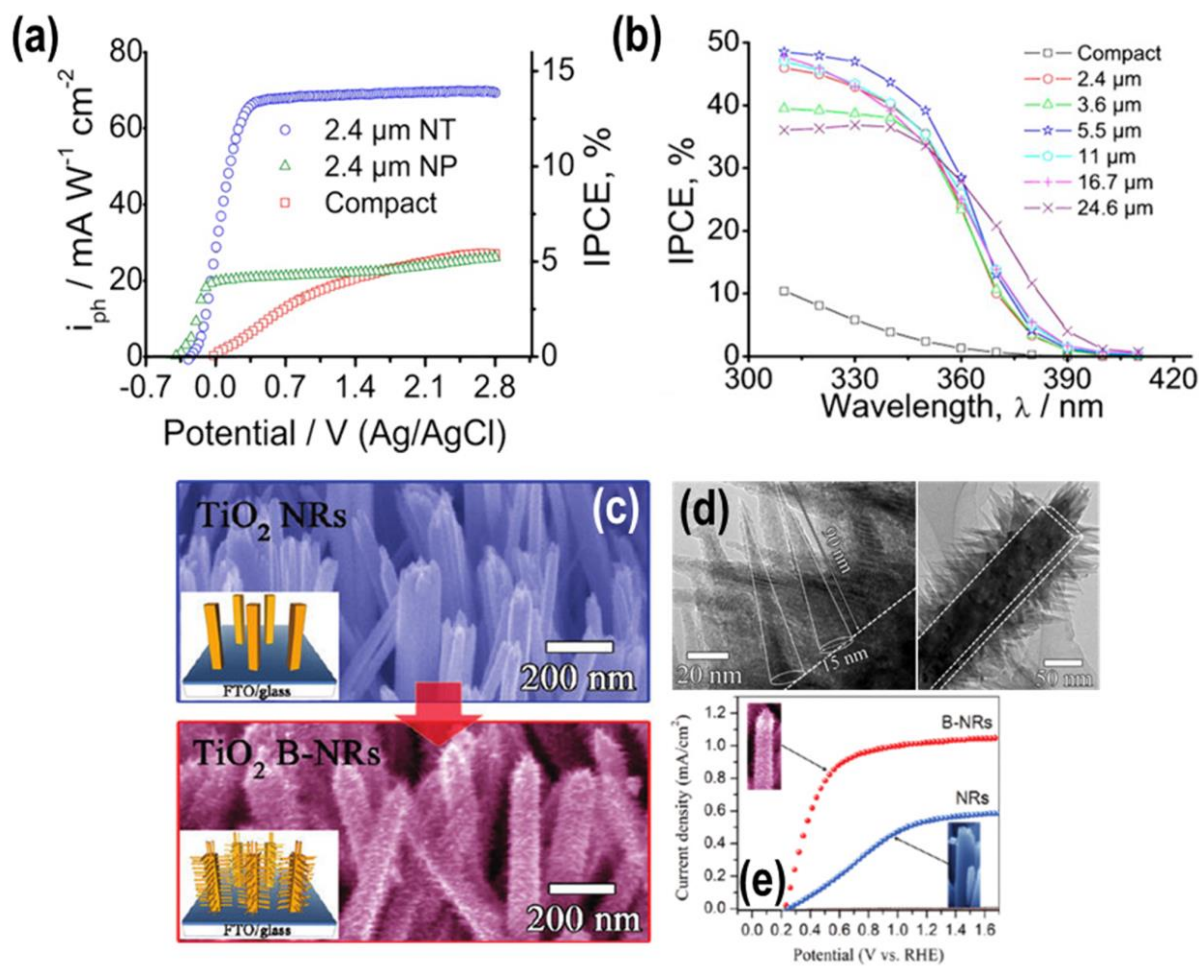

**Fig. 3.6 –** a) Photocurrent and IPCE vs. applied potential measured for flat compact oxide (60 nm thick), 2.4 μm thick nanoparticle, and 2.4 μm thick nanotube $TiO_2$ layers under incident λ = 350 nm. b) IPCE vs. wavelength spectra measured for different thicknesses of nanotube layers anodized in an ethylene glycol-based electrolyte. c) SEM images of $TiO_2$ NRs and branched $TiO_2$ NRs (B-NRs). d) TEM images of a branched nanorod. e) Photocurrent vs. applied potential measured for $TiO_2$ NRs and B-NRs. Fig. (a,b) reprinted with permission from *J. Electrochem. Soc.*, 2010, **157**(3), G76. Copyright 2010, The Electrochemical Society. Fig.(c,d) reproduced from Ref. 313 with permission from the American Chemical Society.

**Figure 3.7**

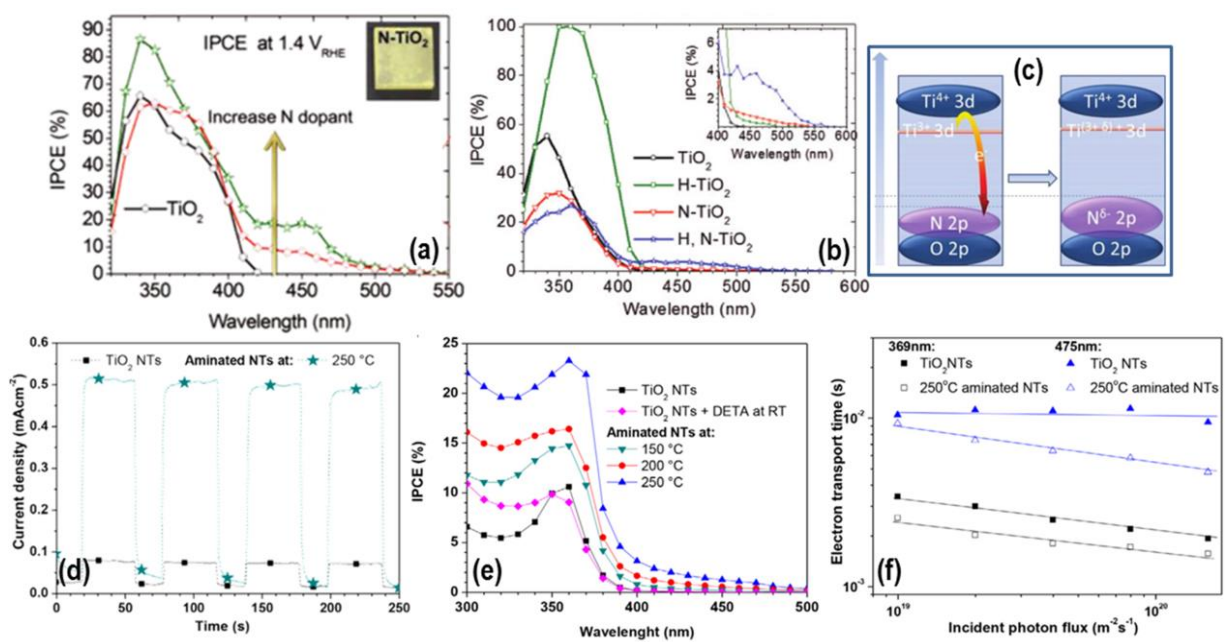

**Fig. 3.7** – a) IPCE spectra of N-modified TiO$_2$ films at 1.4 V$_{RHE}$, in 1 M KOH. b) IPCE spectra measured at 1.23 V$_{RHE}$, 1 M KOH electrolyte and under AM 1.5 irradiation. c) Proposed mechanism for the interaction between Ti$^{3+}$ (formed upon hydrogenation) and substitutional N. d) Photocurrent transient under monochromatic laser ($\lambda$ = 474 nm) for aminated nanotubes; e) IPCE spectra of the TiO$_2$ nanotube before and after hydrothermal treatment for amination; at different temperature in DETA; f) comparison of transport time constants for bare and aminated TiO$_2$ nanotube as a function of the incident photon flux, for monochromatic 369 and 475 nm light illumination. Fig. (a) reproduced from Ref. 91 with permission from the American Chemical Society. Fig. (b,c) reproduced from Ref. 188 with permission from the American Chemical Society. Fig. (d–f) reproduced from Ref. 322 with permission from Elsevier.

**Figure 3.8**

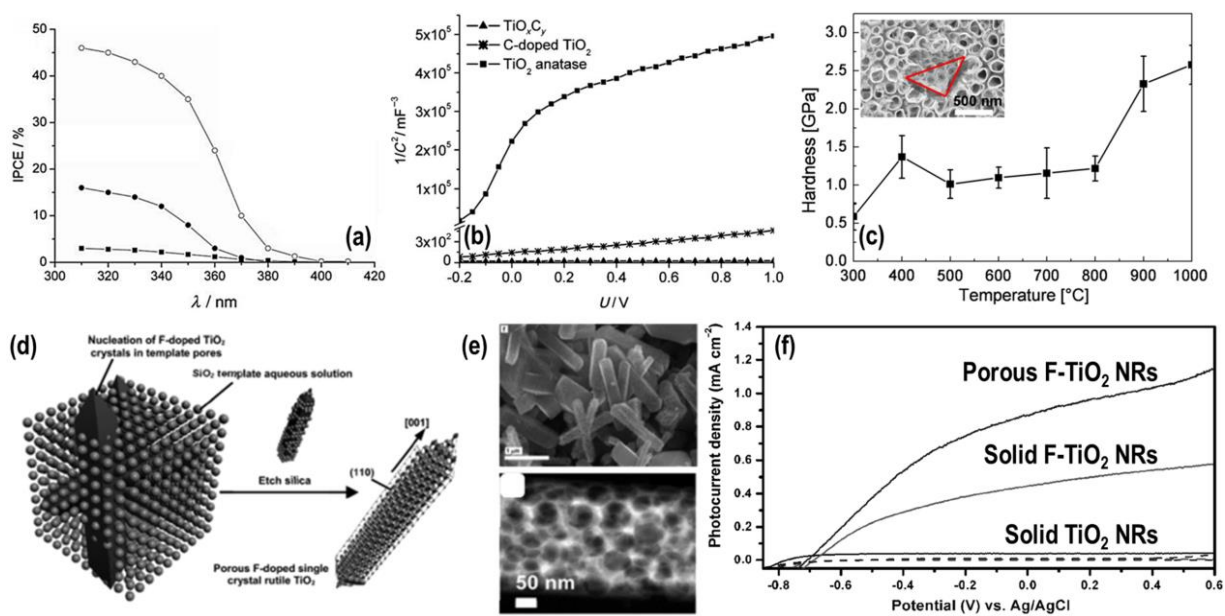

**Fig. 3.8** – a) IPCE spectra of different C-contaminated TiO$_2$ nanotube layers that show sub-band gap response: (■) amorphous NTs, (●) NTs annealed in air, (○) NTs annealed in CH$_3$OH-containing atmosphere. b) Potential-dependent capacity measurements of the different nanotube layers in 0.1 M Na$_2$SO$_4$ solution at 1 kHz and plotted in a Mott–Schottky graph. c) Hardness vs. carbonization temperature plot for various treated TiO$_2$ nanotube layers, obtained from nanoindentation tests (the inset reports the SEM image of a nanoindentation test spot). d) Schematic representation of the growth pathway of porous single-crystal rutile TiO$_2$ nanorods; e) and f) SEM image and dark-field TEM image of F-doped porous single-crystal rutile TiO$_2$ nanorods; g) linear sweep voltammograms of undoped solid TiO$_2$ NRs, and F-doped solid and porous TiO$_2$ NRs. Fig. (a) reprinted with permission from *J. Electrochem. Soc.*, 2010, **157**(3), G76. Copyright 2010, The Electrochemical Society. Fig. (b) reproduced from Ref. 324 with permission from John Wiley & Sons. Fig. (c) reproduced from Ref. [325] with permission from Elsevier. Fig. (d–g) reproduced from Ref. 327 with permission from John Wiley & Sons.

**Figure 3.9**

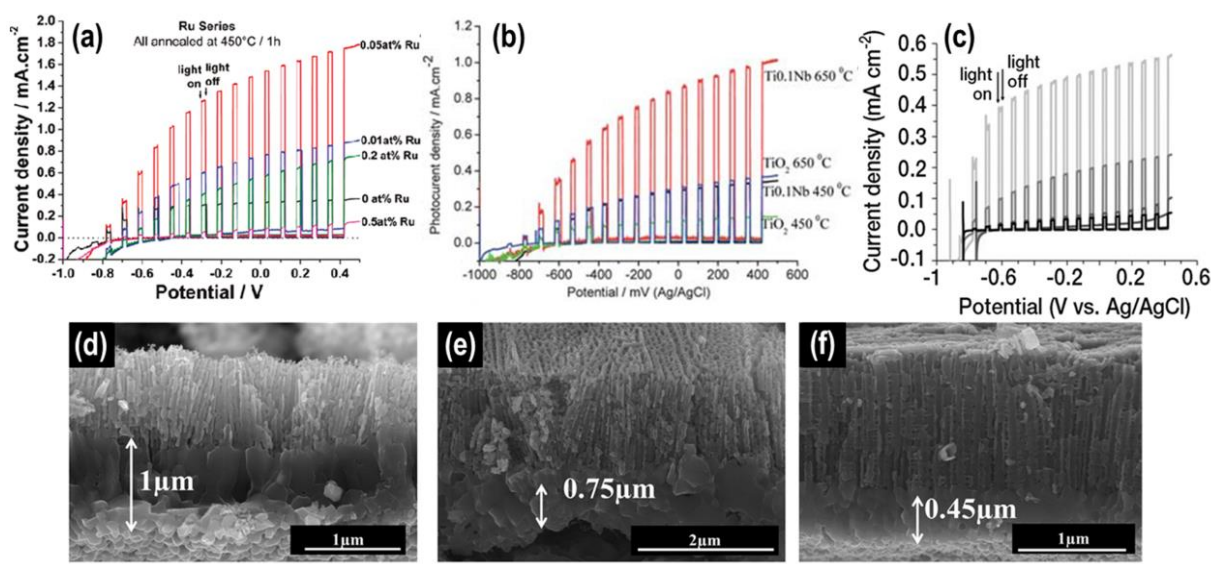

**Fig. 3.9** – a) Photocurrent transient vs. potential curves of TiO$_2$ nanotube layers with different Ru contents (0–0.5 at.%) after annealing at 450 °C for 1 h. Measurements were carried out in 1 M KOH electrolyte under AM 1.5 illumination. b) Photocurrent transients vs. potential curves for TiO$_2$ and Ti0.1Nb nanotube layers after annealing at 450 or 650 °C. Measurements were carried out in 1 M KOH electrolyte under AM 1.5 illumination. c) Photocurrent transient vs. potential curves of Ta-doped TiO$_2$ nanotube layers after annealing at 450 °C for 1 h, measured in 1 M KOH solution under AM 1.5 illumination – Ta contents (at.%): light grey, 0.1% Ta; medium grey, 0.03% Ta; dark grey, 0.4% Ta; black, pure TiO$_2$. d–f) Cross-section SEM images of 2 μm TiO$_2$ nanotubes annealed at 650 °C in air: (d) pure TiO$_2$ NTs, (e) 0.1 at.% Nb TiO$_2$ NTs, (f) 0.2 at.% Nb TiO$_2$ NTs. A clear decrease of the rutile layer thickness underneath the nanotubular layer can be observed, from ~ 1 μm for undoped TiO$_2$, to ~ 0.75 μm and ~ 0.45 μm for 0.1 at.% and 0.2 at.% Nb doped TiO$_2$, respectively. Fig. (a) reproduced from Ref. 315 with permission from the American Chemical Society. Fig. (b) reproduced from Ref. 333 with permission from the Royal Society of Chemistry. Fig. (c) reproduced from Ref. 334 with permission from John Wiley & Sons. Fig. (d–f) reproduced from Ref. 308 with permission from Elsevier.

**Figure 3.10**

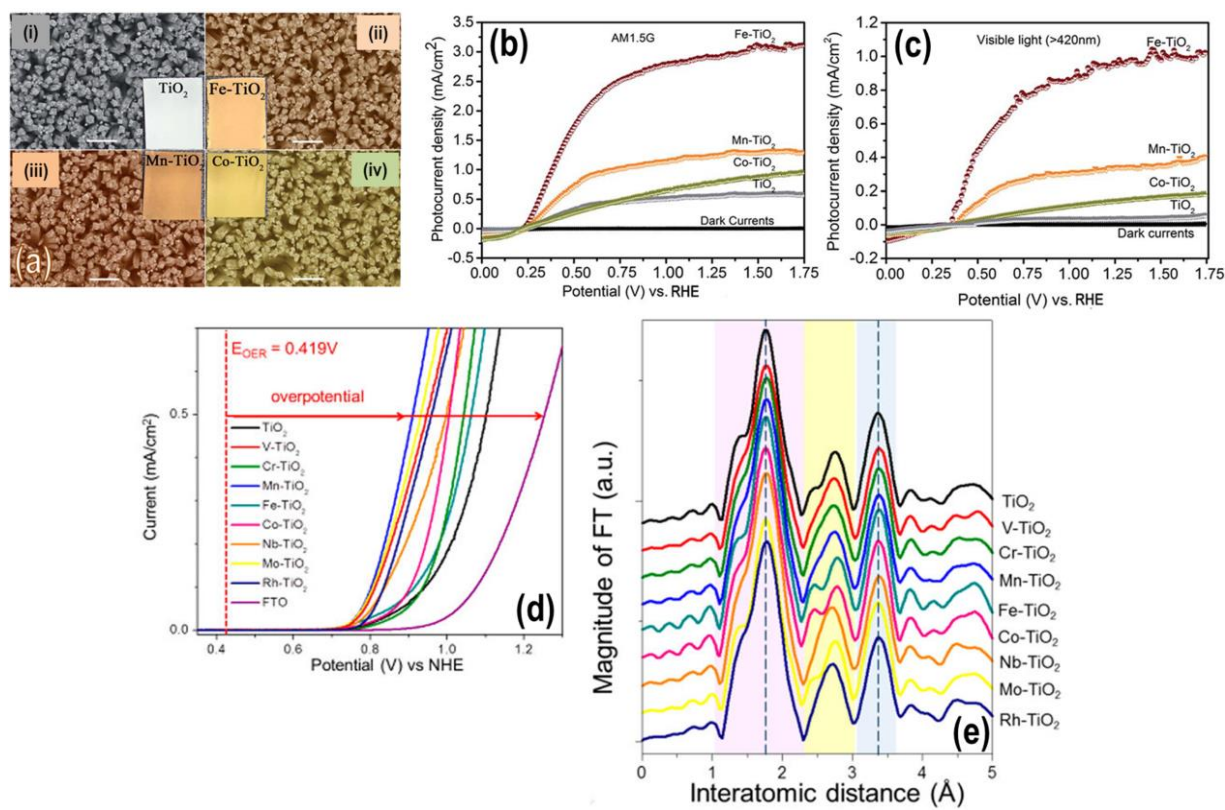

**Fig. 3.10** – a) SEM images of (i) TiO$_2$, (ii) Fe–TiO$_2$, (iii) Mn–TiO$_2$, and (iv) Co–TiO$_2$ nanorod arrays on FTO substrates. Insets represent the corresponding digital pictures. All scale bars are 1 μm. b) Photocurrent transient vs. potential curves of undoped and variously doped TiO$_2$ nanorod layers, under AM 1.5 G solar illumination. c) Photocurrent transient vs. potential curves of undoped and variously doped TiO$_2$ nanorod layers, under visible light illumination (λ > 420 nm). d) Overpotential measurement for OER using FTO and TiO$_2$ nanowire with/without transition metal doping of electrodes in basic solutions. The dashed vertical line represents the thermodynamic redox potential for water oxidation at pH ~13.6. e) Fourier transforms of EXAFS spectra of various transition metal-doped samples. Fig. (a–c) reproduced from Ref. 295 with permission from the Royal Society of Chemistry. Fig. (d,e) reproduced from Ref. 296 with permission from the American Chemical Society.

**Figure 3.11**

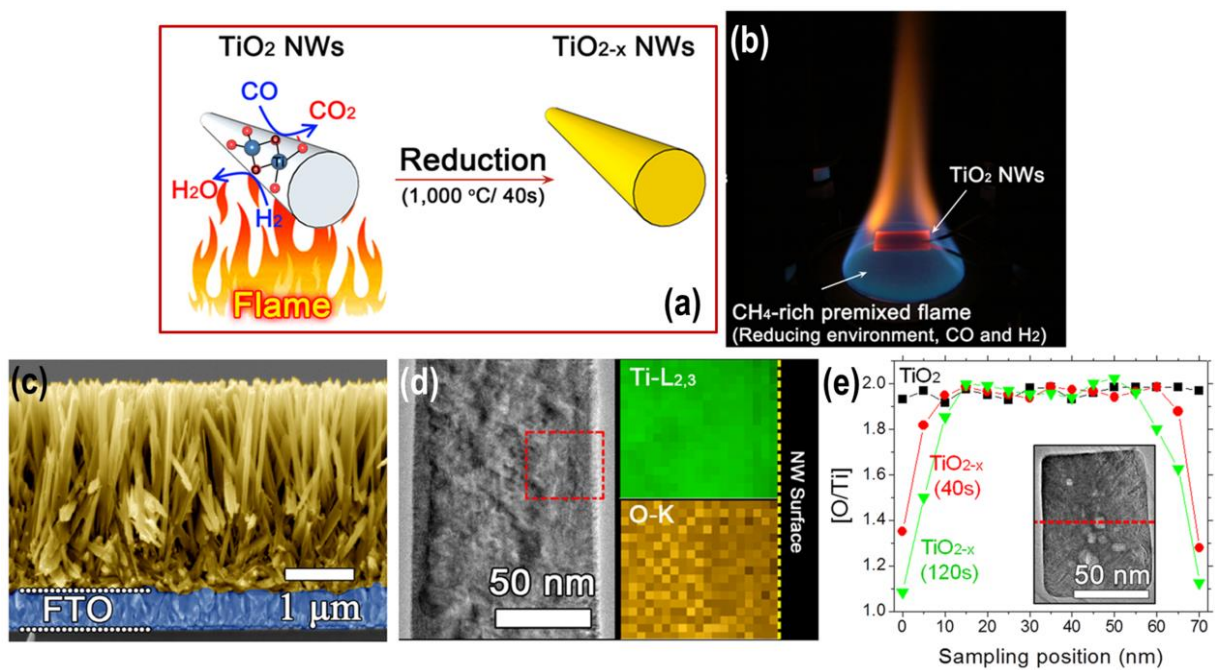

**Fig. 3.11** –Flame reduction process and characterization of oxygen vacancy distribution. (a) Schematic illustration of the flame reduction method. (b) A photograph of the flame reduction process. (c) Representative cross-sectional SEM image of the flame-reduced $TiO_2$ NWs. (d) Bright-field STEM image and the corresponding electron energy loss spectroscopy (EELS) elemental mapping of Ti-L2,3 and O−K edges inside the rectangular box, showing an oxygen deficiency near the NW surface. Darker color: lower concentration of oxygen. (e) O/Ti molar ratio distribution along the NW diameter (as-synthesized $TiO_2$ NW: black rectangles and flame-reduced $TiO_2$ NWs: red circles: 40 s, green triangles: 120 s), which is estimated using EELS spectra taken from a cross-line shown in the inset of a cross-sectional TEM image. Flame reduction conditions: T = 1000 °C, Φ = 1.4, and t = 40 or 120 s. Reproduced from Ref. 94 with permission from the American Chemical Society.

**Figure 3.12**

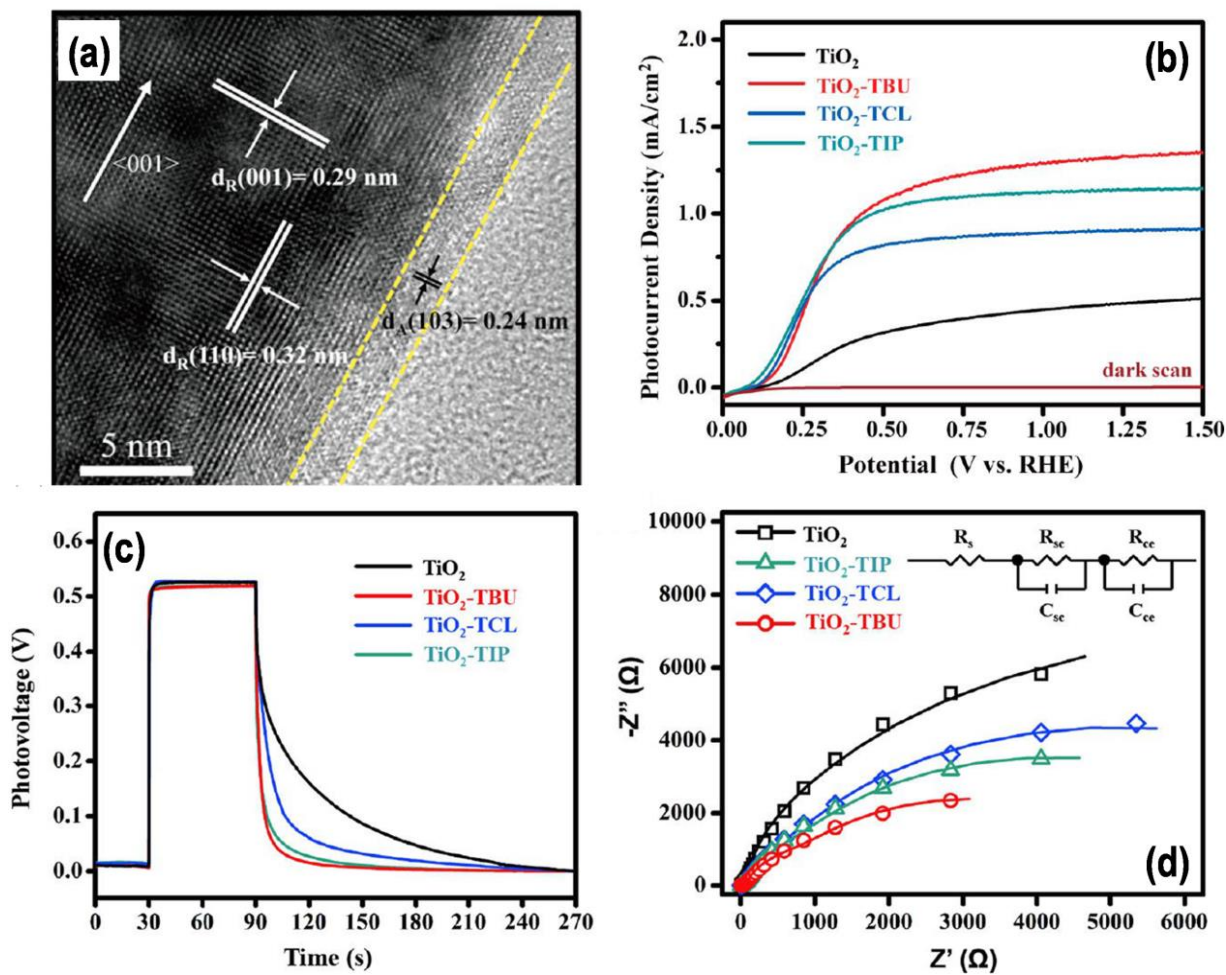

**Fig. 3.12** – a) HR-TEM image of TiO$_2$ nanowire array decorated with a TiO$_2$ anatase thin layer prepared from a Ti(OBu)$_4$ solution. b) Linear sweep voltammograms of pristine TiO$_2$ and variously modified TiO$_2$-based nanowire photoanodes recorded in 1 M KOH under AM 1.5 G (100 mW cm$^{-2}$) illumination. c) Photovoltage−time spectra collected for pristine TiO$_2$ and variously modified TiO$_2$-based nanowire photoanodes. d) EIS spectra of pristine TiO$_2$ and variously modified TiO$_2$-based nanowire photoanodes. The inset shows the equivalent circuit used to fit the spectra. In (b–d): TCL, TiCl$_4$; TIP, Ti(OiP)$_4$; TBU, Ti(OBu)$_4$. Reproduced from Ref. 95 with permission from the American Chemical Society.

**Figure 3.13**

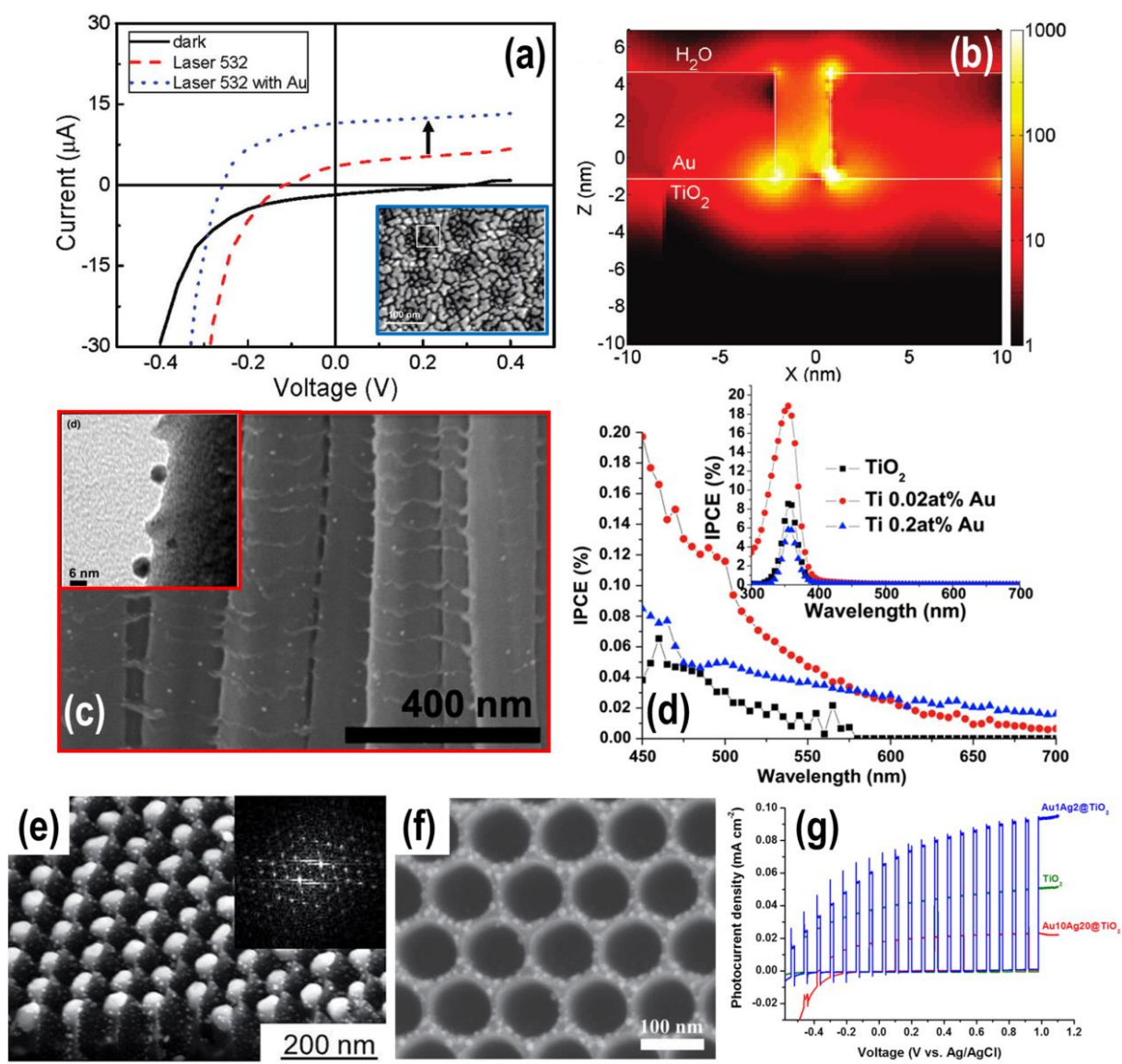

**Fig. 3.13** – a) Photocurrent versus bias voltage of anodic $TiO_2$ with and without Au nanoparticles irradiated with visible light (λ = 532 nm) – inset: SEM image of a 5 nm thick Au island film deposited on anodic $TiO_2$ NTs. b) Intensity of the plasmon-generated electric field at the interface of Au-$TiO_2$ calculated using FDTD. c) Cross-sectional SEM image of Au decorated $TiO_2$ nanotubes on 0.2 at.% of Au containing TiAu alloy (inset: TEM image of Au decorated $TiO_2$ nanotubes on Ti-(0.2 at.%)Au alloy); d) Photoresponse spectra of 12 μm thick $TiO_2$ nanotubes with different contents of Au in TiAu alloy. Photoelectrochemical spectra were measured in 0.1 M $Na_2SO_4$ at 500 mV (Ag/AgCl). e) Example of self-ordering

templated dewetting of Au film (20 nm-thick) on a highly ordered $TiO_2$ nanocavity array leading to total filling with exactly one metal nanoparticle per $TiO_2$ cavity (inset: FFT conversion of a top-view SEM image of array of highly-ordered Au NPs embedded in $TiO_2$ nanocavities). f) SEM image of $TiO_2$ NTs decorated with Au–Ag alloyed-dewetted NPs; g) Photocurrent density vs. potential curves under chopped solar light illumination under solar light illumination of $TiO_2$ NTs and $TiO_2$ NTs decorated with Au–Ag alloyed-dewetted NPs. Fig. (a,b) reproduced from Ref. 197 with permission from the American Chemical Society. Fig. (c,d) reproduced from Ref. 339 with permission from Elsevier. Fig. (e) reproduced from Ref. 342 with permission from John Wiley & Sons. Fig.(f,g) reproduced from Ref. 344 with permission from with permission from John Wiley & Sons.

**Figure 3.14**

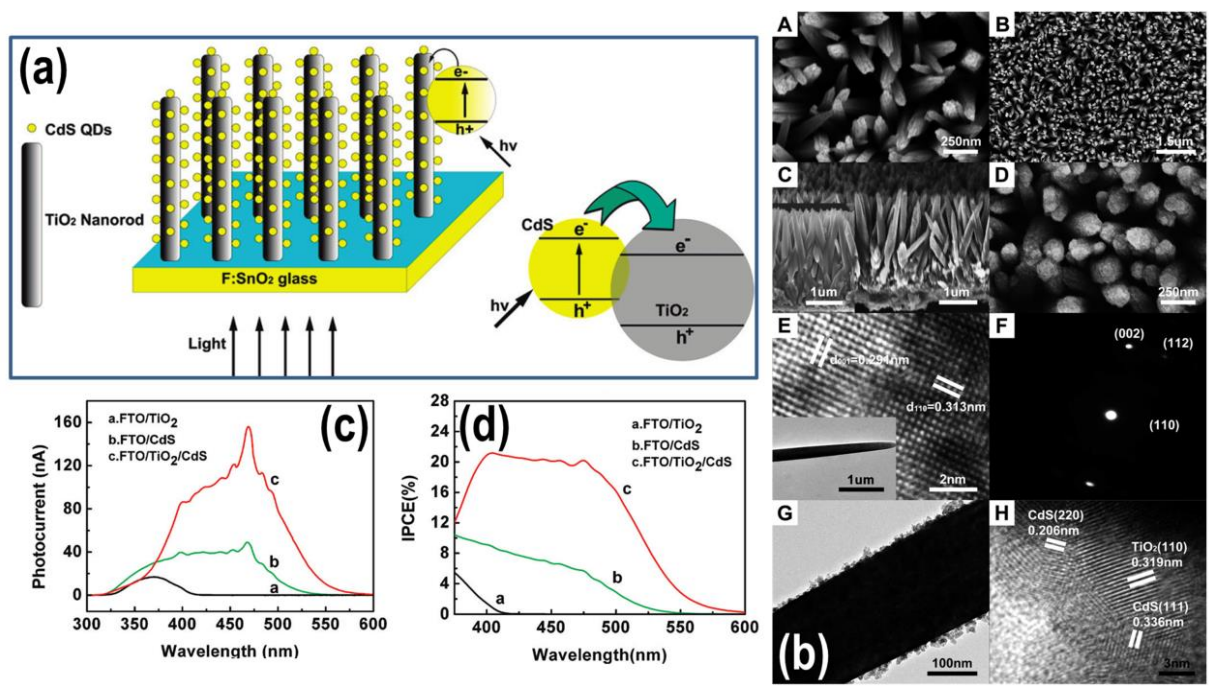

**Fig. 3.14** – a) Schematic representation of CdS QDs deposited on TiO$_2$ nanorods, relative energy band alignment, and photoinduced charge separation and injection mechanisms. b) Morphologies of FTO/TiO$_2$ and FTO/TiO$_2$/CdS electrodes: typical top view SEM images of FTO/TiO$_2$ electrode at (A) high and (B) low magnifications; (C) cross-sectional view of the well-aligned TiO$_2$ nanorod array; (D) FTO/TiO$_2$/CdS electrode; (E, F) HRTEM image and the corresponding selected area electron diffraction (SAED) pattern of bare TiO$_2$ nanorod array, respectively (inset in panel E: TEM image of a single bare TiO$_2$ nanorod); (G, H) TEM image and the corresponding HRTEM image of a single TiO$_2$ nanorod decorated with CdS QDs. c) Photocurrent action spectra and d) IPCE spectra of pristine TiO$_2$ and variously modified TiO$_2$ nanorod array. Photoelectrochemical spectra were measured under 475 nm (15 μW cm$^{-2}$) monochromatic irradiation. Reproduced from Ref. 346 with permission from the American Chemical Society.

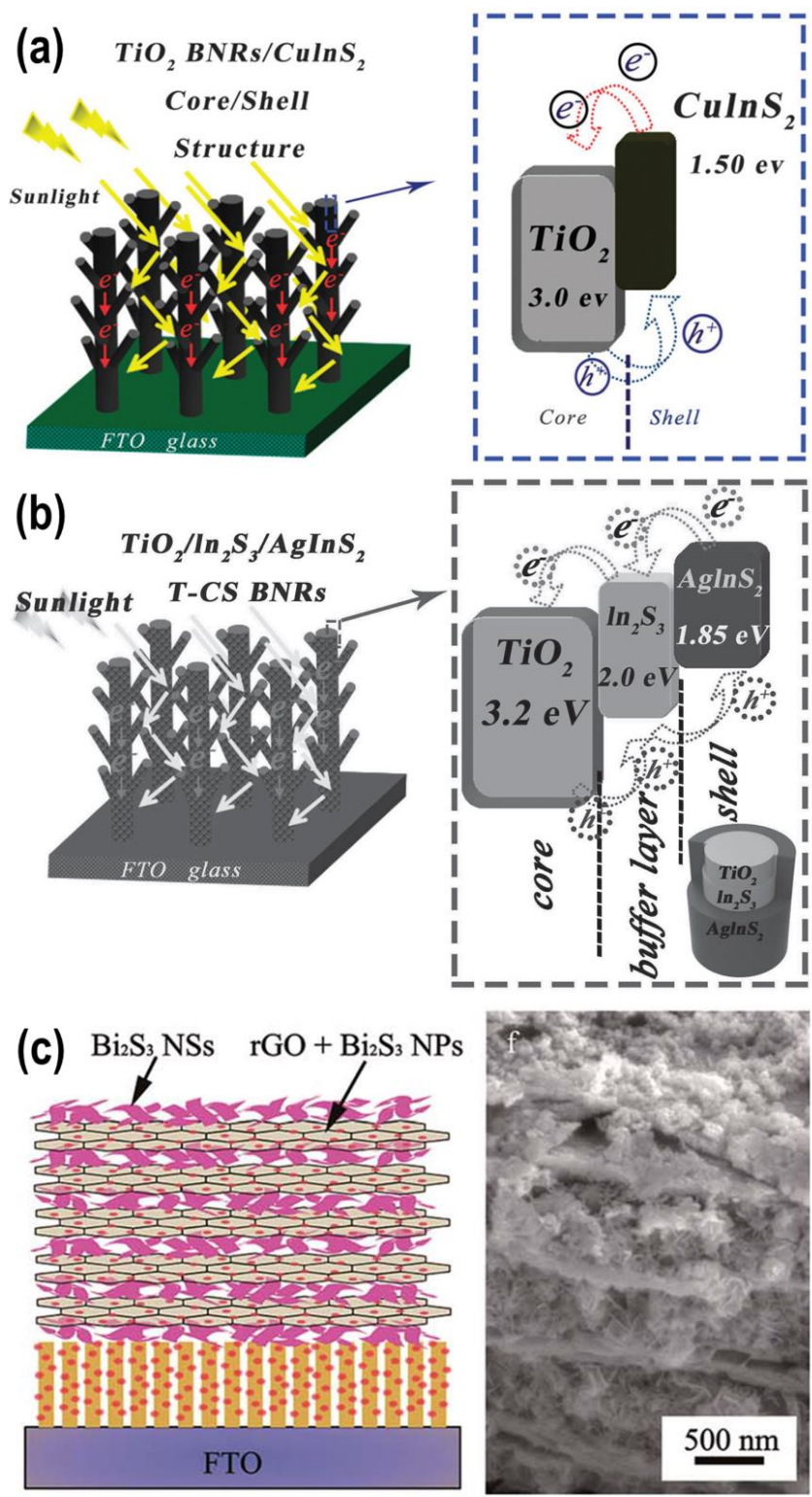

**Fig. 3.15** – Schematic representations of (a) the TiO$_2$ BNRs–CuInS$_2$ core–shell structure and (b) TiO$_2$/In$_2$S$_3$/AgInS$_2$ branched NRs electrode and relative band alignment. (c) Schematic

diagram and SEM image of five layers of $Bi_2S_3$/rGO (purple and light yellow) modified $Bi_2S_3$/$TiO_2$ NRs. Fig. (a) reproduced from Ref. 357 with permission from the Royal Society of Chemistry. Fig. (b) reproduced from Ref. 358 with permission from John Wiley & Sons. Fig. (c) reproduced from Ref. 359 with permission from the Royal Society of Chemistry.

**Figure 4.1**

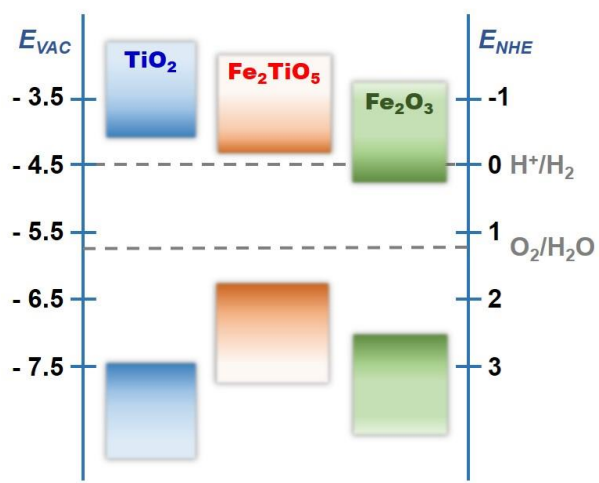

**Fig. 4.1** – Energy band structure of $TiO_2$, $Fe_2O_3$, and $Fe_2TiO_5$ relative to the vacuum level ($E_{VAC}$) and the normal hydrogen electrode ($E_{NHE}$).

**Figure 4.2**

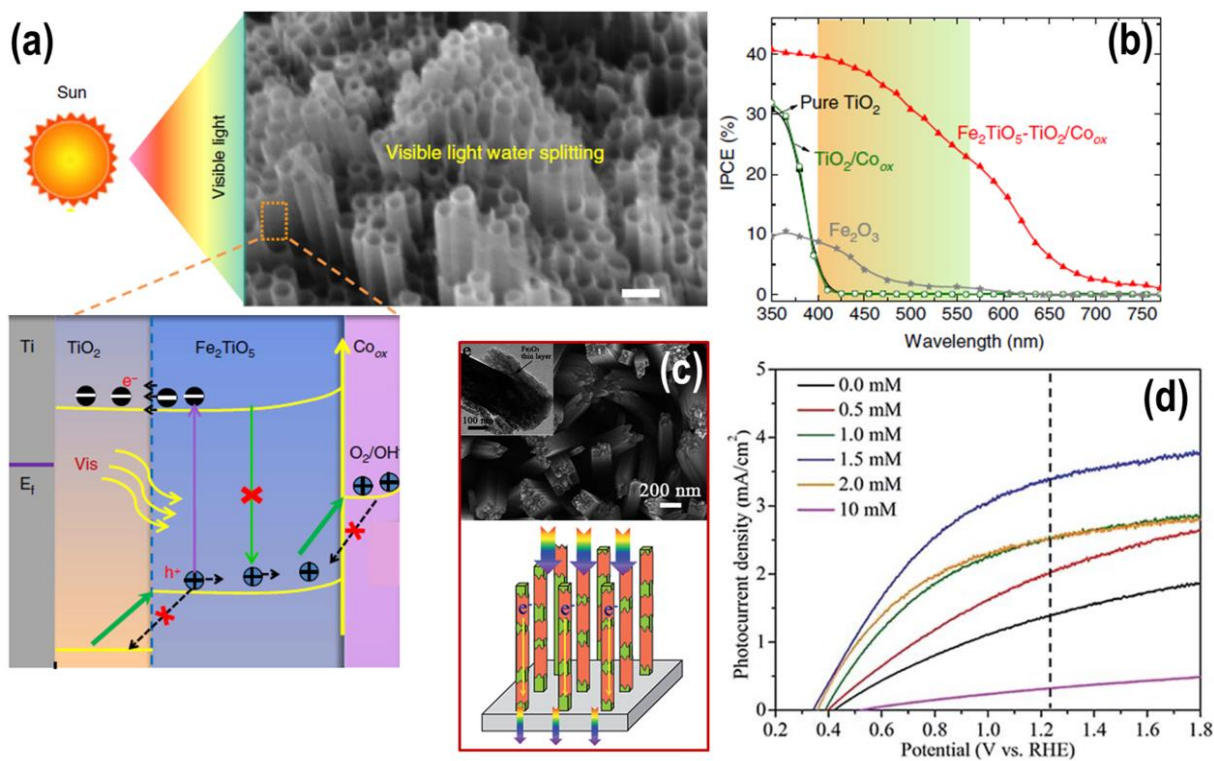

**Fig. 4.2** – (a) SEM image of $Fe_2TiO_5$-$TiO_2$/$Co_{ox}$ nanoarchitecture and schematic drawing of the photoanode design with most likely water oxidation mechanism considering the idealized electronic structure alignment of the composite photoanode, $E_f$ represents the Fermi level; (b) IPCE spectra of pure $TiO_2$, pure α-$Fe_2O_3$, $Co_{ox}$ decorated $TiO_2$ and $Co_{ox}$ decorated $Fe_2TiO_5$/$TiO_2$ layers. The spectra were measured in 1 M KOH at 1.23 V for α-$Fe_2O_3$, and at 0.4V vs. RHE for the other layers. (c) SEM and TEM (inset) images, and schematic representation of $TiO_2$ nanorod array covered by a thin α-$Fe_2O_3$ layer; (d) Linear sweep voltammetry plots of $Fe_2O_3$/$TiO_2$ nanorod photoanodes, containing different amounts of α-$Fe_2O_3$. The data were collected in 1.0 M NaOH, under Xe lamp irradiation (100 mW cm$^{-2}$). Fig. (a,b) reproduced from Ref. 363 with permission from Springer Nature. Fig. (c,d) reproduced from Ref. 364 with permission from the Royal Society of Chemistry.

**Figure 4.3**

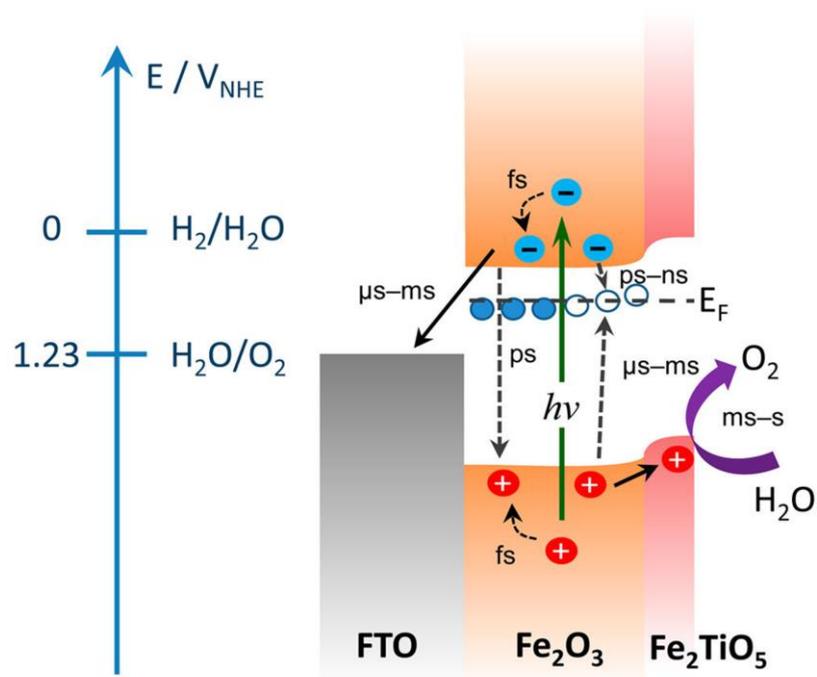

**Fig. 4.3** – Charge generation, recombination, and transfer in the nanocomposite photoelectrodes. Reproduced from Ref. 368 with permission from the American Chemical Society.

**Figure 4.4**

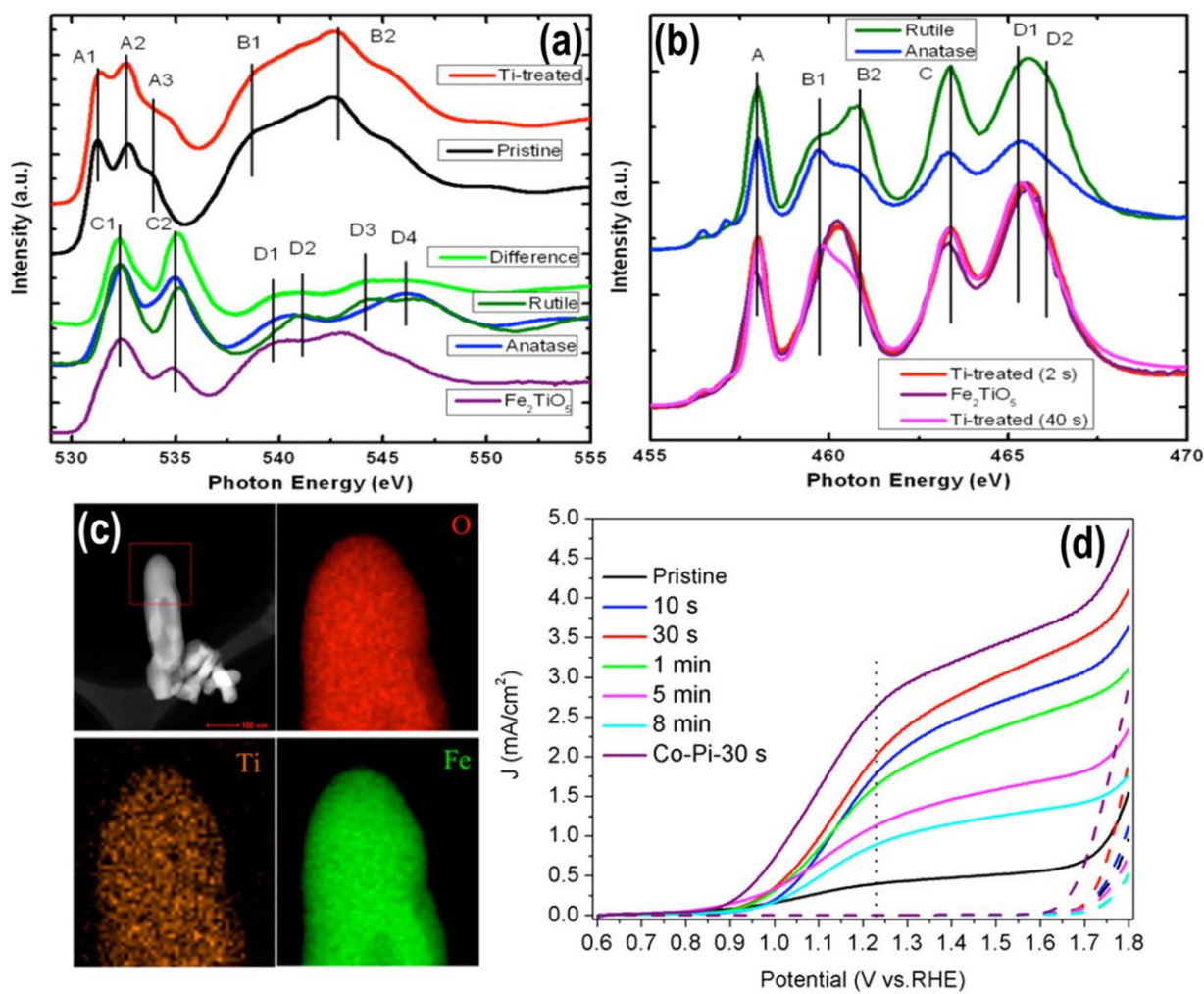

**Fig. 4.4** – (a) Illustration of the morphology and composition of Ti-treated hematite photoanode, and band edge positions of Fe$_2$TiO$_5$ and α-Fe$_2$O$_3$.; (b) J-V characteristics of the pristine hematite photoanode, HF-Ti treated hematite photoanodes with different treatment times (10 s, 30 s, 1 min, 5 min, and 8 min), and a Co-Pi deposited HF-Ti treated (30 s) hematite photoanode; (c) dark field TEM image and corresponding TEM elemental mappings of HF-Ti treated (30 s) hematite 1D nanostructured photoanode. Reproduced from Ref. 366 with permission from the American Chemical Society.

**Figure 4.5**

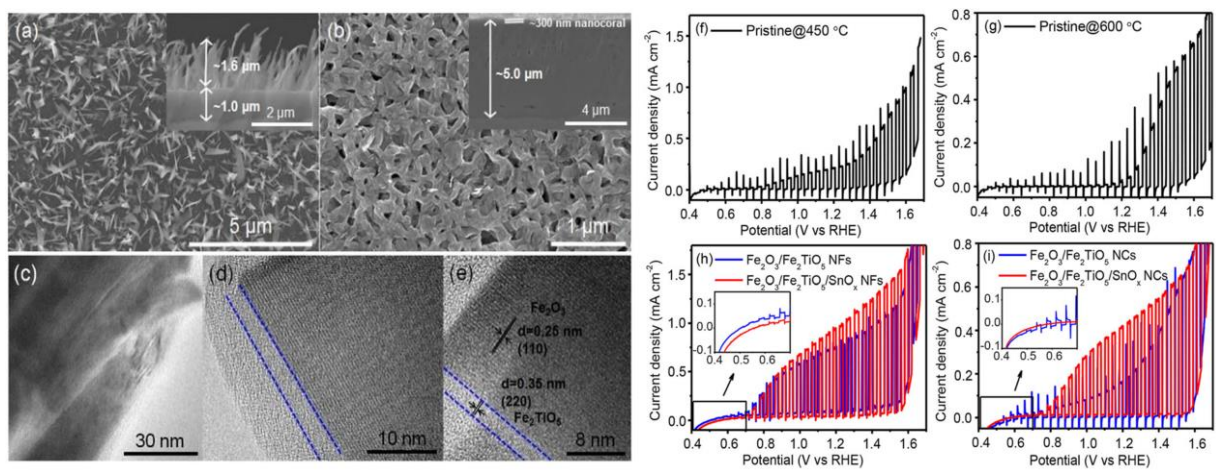

**Fig. 4.5** – SEM images of hematite $Fe_2O_3$ photoanodes prepared by thermal oxidation of Fe foils: (a) 1D nanoflakes (NFs) and (b) nanocorals (NCs); (c–e) TEM images of $Fe_2O_3/Fe_2TiO_5$ NFs; PEC polarization curves of hematite-based photoanodes: (f, h) NFs and (g, i) NCs. The PEC experiments were performed in 1 M KOH electrolyte AM 1.5 G simulated solar light irradiation (intensity 100 mW cm$^{-2}$) at a scan rate of 10 mV s$^{-1}$. Reproduced from Ref. 371 with permission from John Wiley & Sons.

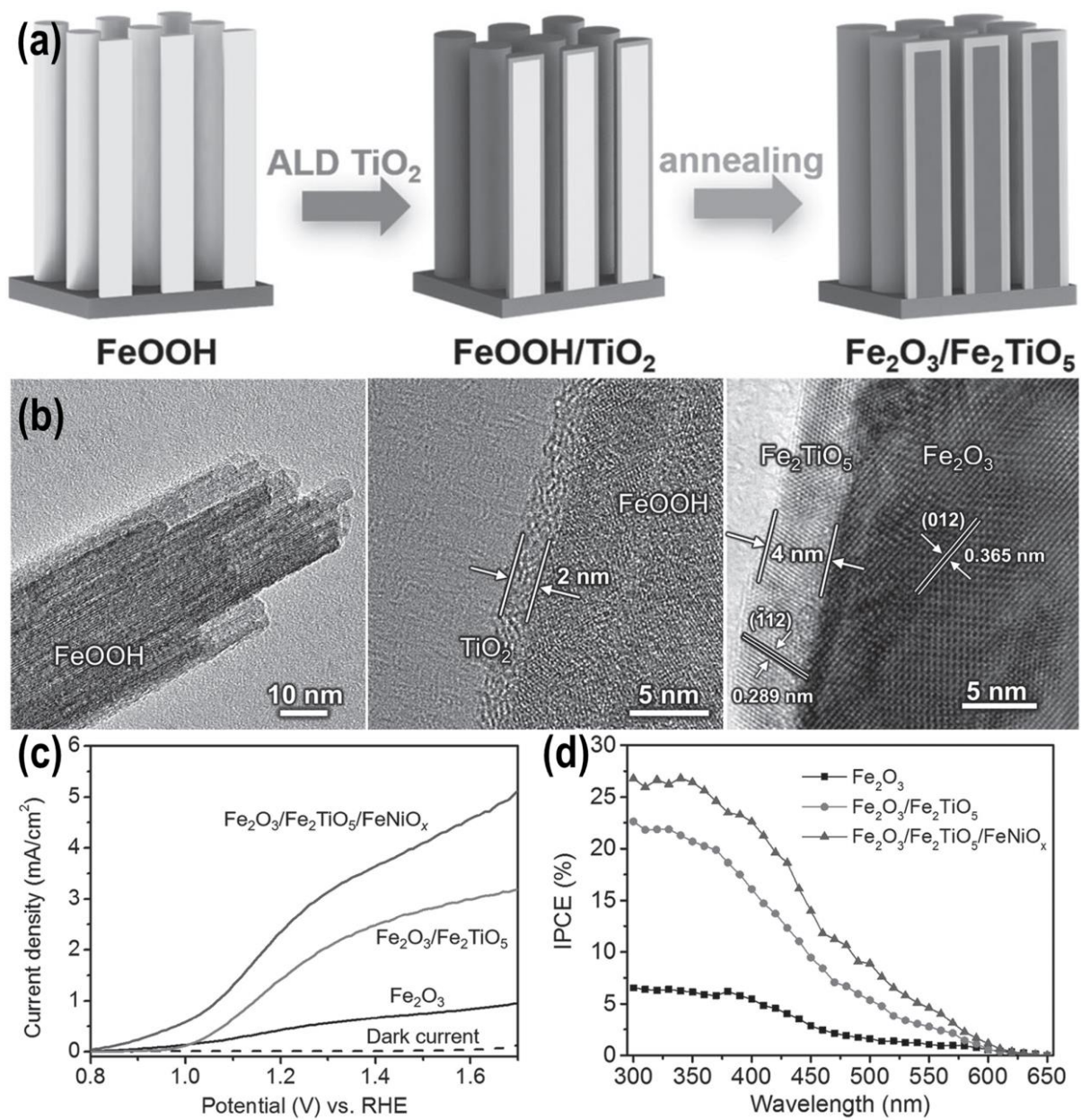

**Fig. 4.6** – (a) Representation of the process used to fabricate $Fe_2O_3/Fe_2TiO_5$ NR photoanodes; (b) HRTEM images of pristine FeOOH nanorod (left), after 90 cycles of $TiO_2$ ALD deposition (middle), and $Fe_2O_3/Fe_2TiO_5$ heterojunction NR after a two-step thermal treatment of $FeOOH/TiO_2$ (right); (c) linear sweep voltammetry (under AM 1.5 irradiation, in 1M KOH), and (d) IPCE spectra (under 150 W Xe lamp irradiation, in 1M KOH at 1.23V vs.

RHE) of pristine $Fe_2O_3$, $Fe_2O_3$/$Fe_2TiO_5$, and $Fe_2O_3$/$Fe_2TiO_5$/$FeNiO_x$ based photoanodes. Reproduced from Ref. 372 with permission from John Wiley & Sons.